\RequirePackage{fix-cm}
\pdfoutput=1
\documentclass[smallextended]{svjour3}       

\smartqed  
\usepackage{graphicx}
\usepackage{cite,amssymb,amsmath,amsfonts,graphicx,bm,dcolumn,mathrsfs}
\usepackage{amsmath,amssymb,amsfonts,graphicx,bm,dcolumn,mathrsfs}
\usepackage{color}
\usepackage{ulem}
\newcommand{\dd}{\mbox{d}}
\newcommand{\ii}{\mathrm{i}}

\begin{document}

\title{Relaxation dynamics in a long-range system with mixed Hamiltonian and non-Hamiltonian interactions}
\author{Alessandro Campa$^1$ and Shamik Gupta$^2$}
\institute{Alessandro Campa \at
alessandro.campa@iss.it
           \and
          Shamik Gupta \at
              shamikg1@gmail.com
              \and
              $^1$ National Center for Radiation Protection and
Computational Physics, Istituto Superiore di Sanit\`{a},
Viale Regina Elena 299, 00161 Roma, Italy  \at
\and
$^2$ Department of Theoretical Physics, Tata Institute of Fundamental Research, Homi Bhabha Road, Mumbai 400005, India
}
\date{Received: \today}
\authorrunning{A. Campa and S. Gupta}
\titlerunning{Relaxation dynamics in a long-range system with mixed interactions}
\maketitle

\begin{abstract}
It is sometimes the case that the dynamics of a physical system is described by equations of motion that do not derive from a Hamiltonian,
and additionally, the degrees of freedom constituting the system interact with each other via long-range interactions. A concrete 
example is that of particles interacting with light as encountered in free-electron laser and cold-atom experiments. In the last couple
of decades, long-range Hamiltonian systems have been found to present a peculiar relaxation dynamics, and in this work, we extend the study
of the relaxation dynamics to non-Hamiltonian systems, more precisely, to systems with interactions of both Hamiltonian
and non-Hamiltonian origin. Our model consists of $N$ globally-coupled particles moving on a circle of unit radius. Since every particle
is characterized by a single coordinate given by its location on the circle, the model is one-dimensional. We show
that in the infinite-size limit (the limit $N \to \infty$), the dynamics, similarly to the Hamiltonian case, is described by the Vlasov
equation for the one-particle distribution function. In the Hamiltonian case, the system eventually reaches an equilibrium state, even
though one has to wait for a long time diverging with $N$ for this to happen. By contrast, in the non-Hamiltonian case, there is no
equilibrium state that the system is expected to reach eventually; thus, the equations for the dynamical evolution
remain the only tool to analyze the state of the system. We characterize this state with its average magnetization. We find
that the relaxation dynamics depends strongly on the relative weight of the Hamiltonian and non-Hamiltonian contributions
to the interaction. When the non-Hamiltonian part is predominant, the magnetization attains a vanishing value, suggesting that
the system does not sustain states with constant magnetization, either stationary or rotating (for the fully
non-Hamiltonian case, we can prove this on the basis of the Vlasov equation). On the other hand, when the Hamiltonian part is predominant,
the magnetization presents long-lived strong oscillations, for which we provide a heuristic explanation. Furthermore, we
find that the finite-size corrections are much more pronounced than those in the Hamiltonian case; we justify this by showing
that the Lenard-Balescu equation, which gives leading-order corrections to the Vlasov equation, does not vanish, contrary to what occurs in
one-dimensional Hamiltonian long-range systems.

\keywords{Long--range interactions \and Non-Hamiltonian systems \and Vlasov equation \and Relaxation dynamics}
\PACS{05.20.-y \and 05.20.Dd \and 05.70.Ln}
\subclass{82C05 \and 82C22}
\end{abstract}

\section{Introduction}
\label{sec:intro}

Long-range interacting systems, both classical~\cite{Campa:2014} and quantum~\cite{Defenu:2021,Maity:2020}, are being investigated
extensively in recent years in the realm of statistical mechanics.  This surge in activity may be attributed
to the fact that such systems abound in nature, e.g., plasmas, self-gravitating systems, geophysical vortices,
wave-particle interacting systems such as free-electron lasers, and many more~\cite{Campa:2014}. From the perspective of
statistical physics, interests have largely stemmed from the observation of a spectrum of static and dynamic properties
exhibited by such systems that appears intriguing and unusual when viewed \textit{vis-\`{a}-vis} those of short-range
interacting systems~\cite{Campa:2009,Bouchet:2010,Gupta:2017}.  A particular issue that has generated a lot of interest
in the community is that of relaxation dynamics, whereby one is interested in how macroscopic observables of a system behave
as a function of time while starting from a given initial condition, and whether in the spirit of equilibrium statistical
mechanics there are time-independent or stationary values to which such observables relax in the limit of long times.
It has been revealed for classical long-range interacting systems described by a Hamiltonian that although macroscopic
observables do attain equilibrium state,  one has in fact to wait for a very long time (a time that diverges with the system
size) for this state to be observed~\cite{Campa:2014}. This phenomenon implies that for a very large system, macroscopic
observables remain trapped in quasistationary states for a very long time,  and this can actually be measured in laboratory
experiments.  

Most work exploring the aforementioned theme of slow relaxation in the classical setting has been devoted to many-body
dynamics derived from a long-range Hamiltonian. Our primary objective in the current work is to extend such studies to the case
in which the dynamical equations for a system of long-range-interacting particles do not derive from an underlying
Hamiltonian, but which nevertheless model bona fide dynamics of experimentally-realizable physical systems. The equations of
motion of our model contain both Hamiltonian and non-Hamiltonian contributions, and in this sense, the dynamics may be referred
to as mixed dynamics.  Indeed, the equations of motion have additive contributions of both; the dynamics  with solely the
Hamiltonian~\cite{Campa:2014,Dauxois:2002} and the non-Hamiltonian contribution~\cite{Bachelard:2019} has been studied separately
in the past, and signatures of slow relaxation have been identified in both, albeit with important differences. It is then
evidently of interest to study how a competition between the two contributions manifests in the relaxation of the system, an
issue we take up for a detailed investigation in the present work. 

The study that we present in this work involves considering a system of $N$ all-to-all-interacting particles of unit mass that
are moving on a circle of unit radius. Denoting by $\theta_i$ and $p_i$ the angular coordinate and the angular momentum,
respectively, of the $i$-th particle, $i=1,2,\ldots,N$, the time evolution is given by the following coupled equations of motion:
\begin{align}
\dot{\theta}_i=p_i,~~\dot{p}_i=\frac{1}{N}\sum_{j=1}^N~g(\theta_i-\theta_j).
\end{align}  
Here, $g(\theta)$ is a periodic function of $\theta$: $g(\theta+2\pi)=g(\theta)$, and the dot denotes derivative with respect to
time.  We may develop $g(\theta)$ in a Fourier series in $\theta$, as 
\begin{align}
g(\theta)=\frac{a_0}{2}+\sum_{n=1}^\infty \left[a_n \cos (n\theta)+b_n \sin(n\theta)\right].
\label{eq:Fourier-gtheta}
\end{align}

In this work, we will consider the Fourier expansion~(\ref{eq:Fourier-gtheta}) with truncation at the lowest order $n=1$.
Consequently, the equations of motion read
 \begin{align}
\dot{\theta}_i=p_i,~~\dot{p}_i=\frac{1}{N}\sum_{j=1}^N~\left[a_1\cos(\theta_i-\theta_j)+b_1 \sin(\theta_i-\theta_j)\right],
\label{eq:eom0}
\end{align} 
where we have dropped the contribution of the $\theta$-independent term in Eq.  (\ref{eq:Fourier-gtheta}) to $\dot{p}_i$ as it
cannot be interpreted as arising due to an interaction between the particles.  
In order to investigate the relative importance of the two terms in the sum in Eq.~(\ref{eq:eom0}) in dictating the dynamics of
the system, we will in this work choose the respective two coefficients to be $a_1=C$ and $b_1=C-1$, with $0 \le C \le 1$ a
given parameter. Thus, the equations of our study in this paper are 
\begin{align}
&\dot{\theta}_i=p_i, ~~\dot{p}_i=\frac{1-C}{N}\sum_{j=1}^N
\sin(\theta_j-\theta_i)+\frac{C}{N}\sum_{j=1}^N \cos(\theta_j-\theta_i).
\label{eq:eom} 
\end{align}
For $C=0$, the equations of motion define the so-called Hamiltonian mean-field (HMF) model~\cite{Campa:2014}, while
substituting $C=1$ relates the model to the one studied in
Ref.~\cite{Bachelard:2019}. In the former case, the equations~(\ref{eq:eom}) are the Hamilton equations of motion corresponding
to the HMF Hamiltonian~\cite{Campa:2014}
\begin{align}
H=\sum_{i=1}^N \frac{p_i^2}{2}+\frac{1}{2N}\sum_{i,j=1}^N [1-\cos(\theta_i-\theta_j)].
\end{align}
As such,  we refer to the case $C=0$ as the Hamiltonian limit of the model.  The equations of motion for $0 < C \le 1$ do not
derive as Hamilton equations corresponding to an underlying Hamiltonian.  Indeed,  writing $\dot{p}_i=\sum_j F_{ij}$, with $F_{ij}$
interpreted as the force on the $i$-th particle due to the $j$-th particle,  one observes from Eq.~(\ref{eq:eom}) and for
$0<C \le 1$ that $F_{ij} \ne - F_{ji}~\forall~i\ne j$, contrary to the case with Hamilton equations. The case $C=1$ models entirely
non-Hamiltonian dynamics, while $0 < C < 1$ models mixed dynamics with both Hamiltonian and non-Hamiltonian contributions in
the equations of motion.

In terms of the so-called magnetization components 
\begin{align}
(m_x,m_y)\equiv \frac{1}{N}\sum_{j=1}^N \left(\cos \theta_j,\sin \theta_j\right),
\label{magdefin}
\end{align}
Eq.~(\ref{eq:eom}) may be rewritten in a form that makes it evident the mean-field nature of the dynamics: each particle
evolves in presence of the mean fields $m_x$ and $m_y$ generated due to interaction between all the particles:
\begin{align}
\dot{\theta}_i=p_i, ~~\dot{p}_i=(m_y(1-C)+m_x C)\cos
\theta_i+(m_yC-m_x(1-C))\sin \theta_i.
\label{eq:eom1} 
\end{align}
In passing, let us define the magnetization as 
\begin{align}
m(t) \equiv \sqrt{m_x^2(t)+m_y^2(t)}.
\label{eq:mt}
\end{align}

We now discuss how the dynamics~(\ref{eq:eom}) is realized in experiments~(for details, see~\cite{Bachelard:2019} and references
therein). To this end,  consider the typical set-up of $N \gg 1$ particles interacting with light as encountered in free-electron
laser and cold-atom experiments,  in which the particles behave as pendula coupled by the common radiation field. The position
$\theta_i$ and the momentum $p_i$ of the $i$-th particle and the amplitude $A$ of the cavity field evolve in time as
\begin{align}
&\dot{\theta}_i=p_i,~~\dot{p}_i=-g~(Ae^{{\rm i}\theta_i}+\mathrm{c.c.}),\label{eq:thetai-pi}\\
&\dot A=\frac{g}{N}\sum_{i=1}^N e^{-{\rm i}\theta_i}-(\kappa-{\rm i}\Delta)A,
\end{align}
where c.c. stands for complex conjugate,  $g$ describes the coupling between the particles and the field,
the parameter $\kappa$ models cavity losses, and $\Delta$ is the frequency mismatch between the cavity and the atomic transition.
Effecting an adiabatic elimination of the field amplitude, which corresponds to assuming that the amplitude $A$ denotes a fast
variable with respect to the positions and the momenta of the particles and consequently attains stationary values on the time
scale of variation of the latter,  one obtains $A \approx g/(N(\kappa - {\rm i}\Delta))\sum_{i=1}^N e^{{\rm i}\theta_i}$.
Substituting this result in Eq.~(\ref{eq:thetai-pi}) yields 
\begin{align}
&\dot{\theta}_i=p_i, \nonumber \\ \label{eq:laser-dyn}\\
&\dot{p}_i=-\frac{2g^2\kappa}{\kappa^2+\Delta^2}\frac{1}{N}\sum_{j=1}^N \cos(\theta_j-\theta_i)-
\frac{2g^2\Delta}{\kappa^2+\Delta^2}\frac{1}{N}\sum_{j=1}^N \sin(\theta_j-\theta_i).\nonumber 
\end{align}
The above dynamics has the same form as the one we study in this paper, namely,  the dynamics~(\ref{eq:eom}), with the
coefficients of the two interaction terms in our case representing in contrast to Eq.~(\ref{eq:laser-dyn}) the situation in
which they add up to unity.

It is pertinent to state our main results right at the outset. While the system with $C=0$ relaxes at long times to the Boltzmann-Gibbs
equilibrium state, it is not known a priori the long-time state of the system as soon as the dynamics becomes non-Hamiltonian, i.e.,
for $0< C \le 1$. In the latter case,  one has to resort to equations determining the evolution of e.g.  the so-called one-particle
distribution function $f(\theta, p,t)$ (namely, the Vlasov equation, which describes the evolution in the limit $N\to \infty$, and the
Lenard-Balescu equation, which describes the leading-order correction to the Vlasov equation) in order to judge the form of the state the
system relaxes to at long times.  Our investigations reveal that the evolution of the system in time while starting from a homogeneous
state shows a change of character as one varies the parameter $C$ that determines the relative weight of the non-Hamiltonian force.
While for $C< 1/2$, the magnetization $m(t)$ presents strong oscillations, a different picture emerges for $C>1/2$,  in which case the
magnetization, after an initial transient, gets more or less rapidly to a practically vanishing value, without any signature of clear
oscillations. What we find is that for $1/2 < C <1$, when the non-Hamiltonian part of the interaction is dominant,  the repelling nature
of the interaction when two particles are close by does not allow formation of a clustered state that is stable in time. For $C$ between
$0$ and $1/2$, when the Hamiltonian part of the interaction is dominant, the particles of the system separate in two or more groups, with
each group composed of particles that are clustered to varying degrees and different groups having different average momentum. A major
theoretical result emerging from our analysis is the behavior of the Lenard-Balescu corrections to evolution of Vlasov-stable homogeneous
states. For $C=0$,  this correction is known to vanish for one-dimensional systems of the sort we are considering. On the other hand, we
have shown that with a non-Hamiltonian part in the interaction, the Lenard-Balescu correction does not vanish. This explains why in the case
of initial distributions that are Vlasov stable, the system for $C \ne 0$ evolves quite rapidly in time, unlike the situation for $C=0$.
However, we did not attempt to make any quantitative estimation arising from the Lenard-Balescu equation. 

The paper is structured as follows. In section \ref{sec:Vlasov-equation}, we introduce the Vlasov equation for the one-particle distribution
function $f(\theta,p,t)$, which describes the evolution of the system in the limit of a very large number of particles ($N \to \infty$).
In section \ref{sec:avermom}, we analyze the dynamics of the center-of-mass momentum, which is not conserved for the non-Hamiltonian model.
In section \ref{sec:Vlasov-stability}, we study the Vlasov stability of homogeneous distribution functions given by a distribution that is
uniform in position $\theta$ and with an arbitrary distribution for the momentum $p$; this is then applied to the case of Lorentzian and
Gaussian distribution functions for $p$. Although the simulations, presented in the following section, concern only Gaussian distribution
functions, we have chosen to give the stability results also for the Lorentzian distribution. This is done to highlight the different behavior,
as regards dependence on the relative weight of the Hamiltonian and non-Hamiltonian parts in the equations of motion, between the Lorentzian
and the Gaussian case. In section \ref{sec_results}, we present and discuss extensive numerical results obtained from $N$-body simulations
of the dynamics, obtained by integrating numerically the dynamical equations of motion. In section \ref{discuss}, we present our discussion
and conclusions. Derivation of a few technical details, including derivation of the Lenard-Balescu corrections, is relegated to the appendices.

\section{$N \to \infty$ limit and the Vlasov equation}
\label{sec:Vlasov-equation}

To characterize the system (\ref{eq:eom1}) in the limit $N \to \infty$, we consider the one-particle distribution $f(\theta,p,t)$, defined
such that $f(\theta,p,t){\rm d}\theta {\rm d}p$ gives the probability at time $t$ to find a particle with coordinate between $\theta$ and
$\theta+{\rm d}\theta$ and with momentum between $p$ and $p+{\rm d}p$. The distribution is normalized as 
\begin{align}
\int_{-\infty}^{+\infty} {\rm d}p \int_0^{2\pi} {\rm
d}\theta~f(\theta,p,t)=1~\forall~t.
\end{align}
Moreover, $f(\theta,p,t)$ is $2\pi$-periodic in $\theta$:
\begin{align}
f(\theta+2\pi,p,t)=f(\theta,p,t).
\end{align}
The time evolution of $f(\theta,p,t)$ is given by the Vlasov equation~\cite{Bachelard:2019}
\begin{align}
\frac{\partial f}{\partial t}+p\frac{\partial f}{\partial
\theta}+F[f](\theta,t)\frac{\partial f}{\partial p}=0,
\label{eq:vlasov}
\end{align}
with $F[f](\theta,t)$, a functional of $f$, defined as
\begin{align}
\label{forcefunctional}
F[f](\theta,t)\equiv (m_y(1-C)+m_x C)\cos
\theta+(m_yC-m_x(1-C))\sin \theta,
\end{align}
and
\begin{align}
(m_x,m_y)[f]\equiv \int {\rm
d}p~{\rm d}\theta~(\cos \theta,\sin \theta)f(\theta,p,t).
\end{align}

It is easily seen that any state 
\begin{align}
\label{distf0p}
f_0(p)=\frac{1}{2\pi}P(p),
\end{align}
with an arbitrary normalized distribution $P(p)$ for the momentum, and which is homogeneous in $\theta$, is a stationary
solution of Eq.~(\ref{eq:vlasov}). In Section \ref{sec:Vlasov-stability}, we will study its stability.

\section{The dynamics of the average momentum}
\label{sec:avermom}

From Eq. (\ref{eq:eom}), we see that for $C\neq 0$, the equation of motion for $p_i$ contains a sort of self-interaction of the particles
represented by the term with $i=j$ in the second sum on the right hand side of the equation. One may argue that such a term should be excluded.
However, we are going to show now that in the context of the dynamical behavior that we want to study in this paper, such a term is actually
not relevant.

The exclusion of the aforementioned term with $i=j$ in the equation of motion for $p_i$ can be done by adding to its right hand side the
term $-C/N$,  which then would obviously also appear in the equations of motion written in the form of Eq. (\ref{eq:eom1}).
This term can then be removed by viewing the dynamics in a frame that is uniformly accelerated with respect to an inertial frame, which is
equivalent to performing the transformation $\theta_i(t) \to \theta_i'(t) \equiv \theta_i(t) - Ct^2/(2N)$. The difference in the equations
of motion with and without the self-interaction term is then only related to a uniform and constant force on all particles given by $-C/N$.
In particular, such a force does not influence the dynamics of the magnetization~(\ref{eq:mt}), being determined by one-time observation of
$\theta$-values that will appear the same when viewed from either the inertial or the uniformly-accelerated frame. Concerning the momenta,
if $p_i(t)$ and $p_i'(t)$ are the momentum of the $i$-th particle with and without, respectively, the term $-C/N$ on the right hand side of
Eq. (\ref{eq:eom}), then we have $p_i'(t) = p_i(t) - Ct/N$. This means that the only difference will be in the distribution of the momenta,
which at any given time $t$ will be uniformly shifted by $C/N$ between the two cases of the inertial and the uniformly-accelerated frame.
Thus, for our purposes, the inclusion of the self-interaction term, i.e., the absence of a term $-C/N$ on the right hand side of
Eq. (\ref{eq:eom}), is not important, since it does not change the relevant physical properties of the system that constitute the object of
our study, namely, the magnetization. Therefore, making the choice between the two frames is a matter of practical convenience. Given the
above reasons for the presence of the self-interaction term being irrelevant for the analysis we want to perform, we now argue why it is
more convenient to include it. Let us consider the dynamics of the average momentum
\begin{equation}
\label{avermomentum}
{\cal P} \equiv \frac{1}{N} \sum_{i=1}^N p_i \, .
\end{equation}
It is a straightforward computation to obtain from Eq. (\ref{eq:eom}) by using the definition (\ref{magdefin}) that
\begin{equation}
\label{eqavermom}
\dot{\cal P} = C \left[ m_x^2(t) + m_y^2(t) \right] \equiv C m^2(t) \, .
\end{equation}
As expected,  for $C \neq 0$, i.e., when the system is non-Hamiltonian, the conservation of the average (or of the total) momentum is
not verified. Excluding self-interaction, an additional constant term equal to $-C/N$ would appear on the right hand side of
Eq.~(\ref{eqavermom}). This implies that with a vanishing magnetization, the average momentum is conserved only when self-interaction is
included. Furthermore, without self-interaction the definition of the functional $F[f](\theta,t)$, appearing in the Vlasov equation
(\ref{eq:vlasov}) and given in (\ref{forcefunctional}), would require the same additional term $-C/N$, implying that a distribution depending
only on the momentum $p$, as in (\ref{distf0p}), would not be stationary for a finite system (although such a term, in the spirit of the
Vlasov equation, describing the system in the infinite size limit, should not appear in this equation). So the choice of including
self-interaction is more convenient, since in this case a distribution uniform in $\theta$ will be stationary. A drifting distribution will
then be due only to a non-vanishing $m(t)$.

\section{Vlasov stability}
\label{sec:Vlasov-stability}

To study the linear stability of the homogeneous distribution (\ref{distf0p}), we write
\begin{align}
f(\theta,p,t)=f_0(p)+\delta f(\theta,p,t);~~|\delta f|\ll 1,
\label{eq:deltaf}
\end{align}
where we may expand $\delta f$ as
\begin{align}
\delta f(\theta,p,t)=\sum_{k=-\infty}^{+\infty} \widetilde{\delta f}_k(p)e^{\ii(k\theta-\omega t)}.
\label{eq:deltaf-expansion}
\end{align}

Substituting Eq.~(\ref{eq:deltaf}) in the Vlasov equation~(\ref{eq:vlasov}) and keeping terms to lowest order in $\delta f$ yield the
linearized Vlasov equation
\begin{align}
\frac{\partial \delta f}{\partial t}+p\frac{\partial \delta f}{\partial
\theta}+F[\delta f](\theta,t)f_0'(p)=0;~~f_0'(p)\equiv \frac{\partial f_0(p)}{\partial p}.
\label{eq:Vlasov-linear}
\end{align}
Plugging the expansion~(\ref{eq:deltaf-expansion}) in the above equation,  using the results
$m_x[\delta f]=\pi e^{-\ii \omega t}\int {\rm d}p~(\widetilde{\delta f}_{-1}(p)+\widetilde{\delta f}_1(p))$ and
$m_y[\delta f]=(\pi/\ii) e^{-\ii \omega t}\int {\rm d}p~(\widetilde{\delta f}_{-1}(p)-\widetilde{\delta f}_1(p))$, and equating the
coefficient of $e^{\ii(\theta-\omega t)}$ and $e^{\ii(-\theta-\omega t)}$ to zero, one obtains respectively that
\begin{align}
&\ii(\omega-p)\widetilde{\delta f}_1(p)=\pi[C+\ii (1-C)]f_0'(p)\int {\rm d}p~\widetilde{\delta f}_1(p),\\
&\ii(\omega+p)\widetilde{\delta f}_{-1}(p)=\pi[C-\ii (1-C)]f_0'(p)\int {\rm d}p~\widetilde{\delta f}_{-1}(p).
\end{align}
Integrating both sides of the above equations with respect to $p$ and noting that $\int {\rm d}p~\widetilde{\delta f}_{\pm 1}(p)\ne 0$, we
obtain for the Fourier modes $k=\pm 1$ that one has 
\begin{align}
&1+\pi [(1-C) -\ii C]\int_{-\infty}^{+\infty} \dd p \, \frac{f_0'(p)}{p-\omega} = 0;~~~~k=+1,
\label{eq:dispersion-1}\\
&1+\pi [(1-C) +\ii C]\int_{-\infty}^{+\infty} \dd p \, \frac{f_0'(p)}{p+\omega} = 0;~~~~k=-1.
\label{eq:dispersion-2}
\end{align}
We will be concerned with even $f_0(p)$, that is, functions for which $f_0(-p)=f_0(p)$. From now on, we will consider
only this case. Since we then have $f_0'(-p)=-f_0'(p)$, we may write the equation given above for $k=-1$ in such a way as to have in the
integrand the same denominator as in the equation for $k=+1$. This will prove to be more convenient for our discussions given below. Thus,
we rewrite the two equations as
\begin{align}
\epsilon_+(\omega) \, \equiv&\, 1+\pi [(1-C) -\ii C]\int_{-\infty}^{+\infty} \dd p \, \frac{f_0'(p)}{p-\omega} = 0;~~~~k=+1,
\label{eq:dispersion-1-evenf0}\\
\epsilon_-(\omega) \, \equiv&\, 1+\pi [(1-C) +\ii C]\int_{-\infty}^{+\infty} \dd p \, \frac{f_0'(p)}{p-\omega} = 0;~~~~k=-1,
\label{eq:dispersion-2-evenf0}
\end{align}
where we have introduced the functions $\epsilon_+(\omega)$ and $\epsilon_-(\omega)$.
The above equations are the dispersion relations for $k=+1$ and $k=-1$. Solving them
yields the frequencies $\omega=\omega_\pm$ of the Fourier modes $k=\pm 1$, respectively.
Obtaining $\omega$ with a positive imaginary part (respectively, a negative imaginary part) implies
that the corresponding mode grows in time (respectively, decays in time).

A little later, we will discuss how to interpret the quantities $\epsilon_+(\omega)$ and $\epsilon_-(\omega)$ for real $\omega$; but, 
before we do that, we remark on the properties of the complex solutions of $\epsilon_+(\omega)=0$ and $\epsilon_-(\omega)=0$.
From Eqs. ~(\ref{eq:dispersion-1-evenf0}) and (\ref{eq:dispersion-2-evenf0}), it is not difficult to see that
the following holds: if $\omega=\omega_0$ satisfies $\epsilon_+(\omega_0)=0$, then we have $\epsilon_+(-\omega_0)=0$ and
$\epsilon_-(\omega_0^*)=\epsilon_-(-\omega_0^*)=0$. Here, $*$ denotes complex conjugation.  Similarly,  if $\omega=\omega_0$ satisfies
$\epsilon_-(\omega_0)=0$, then we have $\epsilon_-(-\omega_0)=0$ and $\epsilon_+(\omega_0^*)=\epsilon_+(-\omega_0^*)=0$. For the particular
case of a Hamiltonian system ($C=0$), $\epsilon_+(\omega)$ and $\epsilon_-(\omega)$ are the same, and thus if $\omega$ is a solution, then
$\omega$, $-\omega$, $\omega^*$ and $-\omega^*$ are all solutions of the dispersion relation for both $k=1$ and $k=-1$. From these relations
between the solutions of the dispersion relations for $k=1$ and $k=-1$, we deduce that to study the stability of $f_0(p)$, it is sufficient
to study only one of them.

To consider real $\omega$, we first note that each of the two functions $\epsilon_+(\omega)$ and
$\epsilon_-(\omega)$ actually defines two different analytic functions in the upper half and in the lower half
of the complex $\omega$-plane,  which have different limits as $\omega$ approaches the real axis. Taking for definiteness
$\epsilon_+(\omega)$, using the well-known Plemelj formula
\begin{equation}
\label{pleme}
\lim_{\eta \to 0^+} \frac{1}{x\pm \ii \eta} = P \frac{1}{x} \mp \ii \pi \delta(x),
\end{equation}
where $P$ denotes the principal value, we have that
\begin{equation}
\label{limomegareal}
\lim_{{\rm Im}(\omega) \to 0^{\pm}} \epsilon_+(\omega) = 1+\pi [(1-C) -\ii C]
\left[ P \int_{-\infty}^{+\infty} \dd p \, \frac{f_0'(p)}{p-\omega} \pm \ii \pi f_0'(\omega) \right] \, ,
\end{equation}
where on the right hand side, it is understood that $\omega$ is real. Thus, in the limit of real $\omega$, the equation
$\epsilon_+(\omega)=0$ becomes equivalent as ${\rm Im}(\omega) \to 0^+$ to the following couple of equations, obtained by equating the
real and the imaginary part of the above equation to zero: 
\begin{eqnarray}
\label{dispomerealp1}
1 + \pi (1-C) P \int_{-\infty}^{+\infty} \dd p \, \frac{f_0'(p)}{p-\omega} + \pi^2 C  f_0'(\omega) &=& 0, \\
\label{dispomerealp2}
- C P \int_{-\infty}^{+\infty} \dd p \, \frac{f_0'(p)}{p-\omega} + \pi (1-C)  f_0'(\omega) &=& 0.
\end{eqnarray}
The equivalent couple of equations as ${\rm Im}(\omega) \to 0^-$ are
\begin{eqnarray}
\label{dispomerealm1}
1 + \pi (1-C) P \int_{-\infty}^{+\infty} \dd p \, \frac{f_0'(p)}{p-\omega} - \pi^2 C  f_0'(\omega) &=& 0, \\
\label{dispomerealm2}
C P \int_{-\infty}^{+\infty} \dd p \, \frac{f_0'(p)}{p-\omega} + \pi (1-C)  f_0'(\omega) &=& 0.
\end{eqnarray}
The two couples have different solutions; in fact, one may see that if $\omega$ is a solution of the first couple, $-\omega$ is a solution
of the second couple.

We end this brief summary of the properties of the dispersion relations by stressing that if
solving $\epsilon_+(\omega)=0$ and $\epsilon_-(\omega)=0$ yields a complex solution, then necessarily there
will be solutions with both signs of the imaginary part. This implies that a necessary condition for the stability
of the distribution $f_0(p)$ is the absence of complex solutions of the dispersion relations; namely, $f_0(p)$
can be stable only if the dispersion relations have either only real solutions or no solutions at all.
It can be shown that when there are no complex solutions of the dispersion relations, the density fluctuations,
i.e., the integral over $p$ of $\delta f(\theta,p,t)$, as determined by the linearized Vlasov equation, decay
exponentially in time. This issue is related to the so called Landau damping. In our case we are studying the
Fourier components with $k=\pm 1$ of $\delta f(\theta,p,t)$, since they are the only components that could have
complex solutions of the dispersion relation. Therefore, in the following we will consider the absence of complex
solutions of the dispersion relations as characterizing a stable $f_0(p)$. A concise
but clear and detailed description of the mathematical reason of this issue can be found in Ref. \cite{case1959}.

In the remaining of this section we will consider the dispersion relation for two concrete cases of $f_0(p)$, i.e. a
Lorentzian distribution and a Gaussian distribution. The solutions, for given $C$, of the dispersion relations
$\epsilon_+(\omega)=0$ and $\epsilon_-(\omega)=0$ will be functions of the width of $f_0(p)$, and then the distribution
will be stable or not depending on the width. As a direct consequence of what remarked in the previous paragraph,
the distributions will be stable only when the dispersion relations do not have complex solutions.
The stability threshold value of the width will be obtained as the value for which the complex solutions tend to the
real axis. Even if our simulation concern only the Gaussian case, we have included in this section also the
Lorentzian distribution, since the integral appearing in the dispersion relations in this case can be easily solved in closed form,
so that the solutions $\omega$ can be written in closed form as a function of $C$. For the Gaussian case, on the other hand, this
is not possible, but nevertheless it is not difficult, as we will show, to obtain, as a function of $C$, the threshold value
of the width of the Gaussian for Vlasov stability.

\subsection{Vlasov stability for the Lorentzian distribution}

\begin{figure}[!ht]
\centering
\includegraphics[width=8cm]{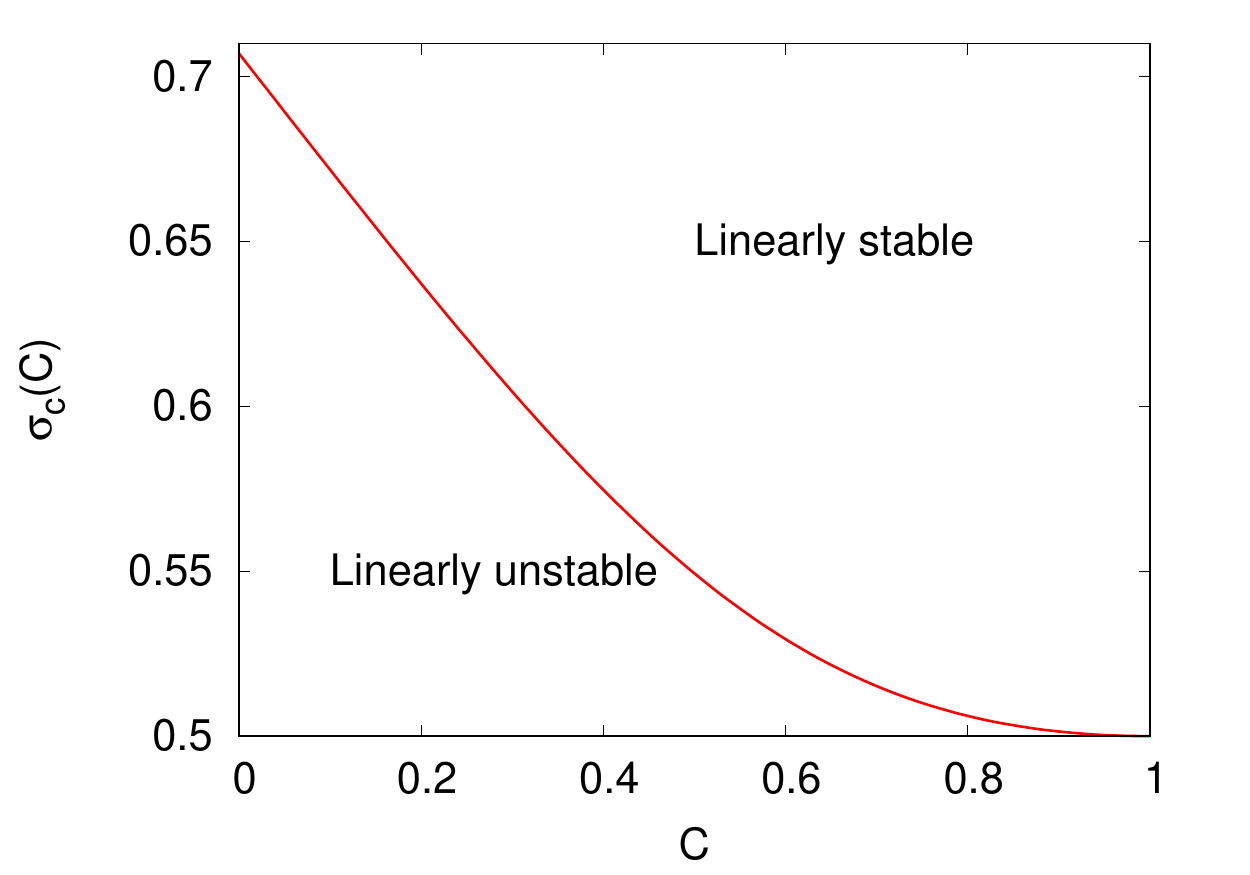}
\caption{Stability threshold for the state~(\ref{eq:f0-Lorentzian}); the threshold is given by Eq.~(\ref{eq:Lorentzian-stability-threshold}).}
\label{fig:lor-stability}
\end{figure}

As explained above, for the stability of an even distribution $f_0(p)$ it is sufficient to study only one of the two dispersion relations
$\epsilon_+(\omega)=0$ and $\epsilon_-(\omega)=0$. We will therefore focus on $\epsilon_+(\omega)$, studying Eq. (\ref{eq:dispersion-1-evenf0}).
Our first concrete case for $f_0(p)$ corresponds to a Lorentzian $P(p)$ in Eq. (\ref{distf0p}) so that we have 
\begin{align}
f_0(p)=\frac{1}{2\pi}\frac{\sigma}{\pi(p^2+\sigma^2)};~~\,\,\,\,\,\,\,\,\, \sigma >0.
\label{eq:f0-Lorentzian}
\end{align}
With this form of $f_0(p)$ it is easy to perform, using the residue theorem, the integral appearing in Eq. (\ref{eq:dispersion-1-evenf0})
when ${\rm Im}(\omega) \ne 0$. In particular, if there exist a solution with ${\rm Im}(\omega)>0$, it must be given by
\begin{equation}
1 + \frac{(1-C) - \ii C}{2(\omega + \ii \sigma)^2} = 0 \, .
\label{eq:dispersion-1-evenf0-lor}
\end{equation}  
Posing, for a lighter notation, $\omega_R \equiv {\rm Re}(\omega)$ and $\omega_I \equiv {\rm Im}(\omega)$, separating the real and the
imaginary parts of the last equation we get
\begin{align}
2\omega_R^2-2(\omega_I+\sigma)^2 + (1-C) = 0,~~4\omega_R(\omega_I+\sigma) - C = 0 \, .
\end{align}
Then, the solution with $\omega_I>0$, if it exists, is such that
\begin{align}
\omega_I=-\sigma +\frac{1}{2}\sqrt{1-C+\sqrt{(1-C)^2+C^2}} \, ,
\label{eq:threshold-lor}
\end{align}
and, as we have learned above, there will be another solution equal to $-\omega$, thus with a negative imaginary part. Therefore,
a complex solution of the dispersion relation exists only when the right hand side of Eq. (\ref{eq:threshold-lor}) is positive, and in this
case the Lorentzian distribution will be unstable. In conclusion, the threshold value for the Lorentzian width is
\begin{align}
\sigma_c(C)= \frac{1}{2}\sqrt{1-C+\sqrt{(1-C)^2+C^2}}.
\label{eq:Lorentzian-stability-threshold}
\end{align}
For $\sigma >\sigma_c(C)$ the state~(\ref{eq:f0-Lorentzian}) will be stable. In particular, we have $\sigma_c(0)=1/\sqrt{2}$ and
$\sigma_c(1)=1/2$. The stability diagram is shown in Fig.~\ref{fig:lor-stability}.

\subsection{Vlasov stability for the Gaussian distribution}

\begin{figure}[!ht]
\centering
\includegraphics[width=8cm]{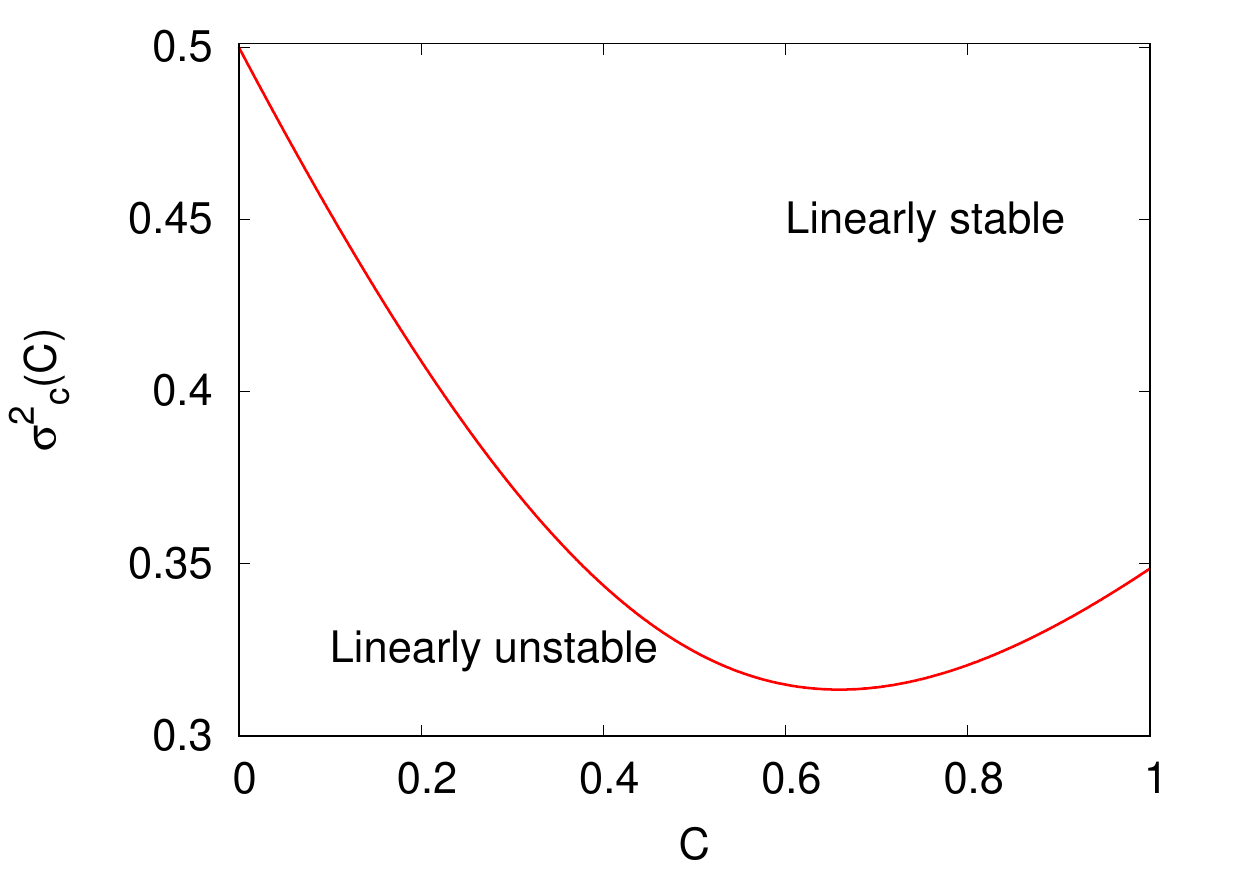}
\caption{Threshold temperature vs. $C$ for Gaussian initial conditions.}
\label{fig:gauss-stability}
\end{figure}

The next example that we will consider for $f_0(p)$ corresponds to a Gaussian $P(p)$, so that we have 
\begin{align}
f_0(p)=\frac{1}{2\pi}\frac{e^{-\frac{p^2}{2\sigma^2}}}{\sqrt{2\pi \sigma^2}};~~\,\,\,\,\,\,\,\,\, \sigma >0.
\label{eq:f0-Gaussian}
\end{align}
Like before, we consider the dispersion relation $\epsilon_+(\omega)=0$. Substituting the Gaussian in Eq. (\ref{eq:dispersion-1-evenf0})
we obtain, after some straightforward passage,
\begin{equation}
1-\frac{1}{2 \sigma^2} [(1-C) -\ii C]\left( 1+ \frac{\omega}{\sqrt{2\pi \sigma^2}}\int_{-\infty}^{+\infty} \dd p \,
\frac{e^{-\frac{p^2}{2\sigma^2}}}{p-\omega} \right) = 0 \, .
\label{eq:dispersiongauss}
\end{equation}
To find the threshold value for stability, $\sigma_c(C)$, we adopt the following strategy. We assume that there is a solution
of Eq. (\ref{eq:dispersiongauss}) with $\omega_I > 0$, and we look for a solution with $\omega_I \to 0^+$. This will provide a relation that
allows, for a given value of $C$, to obtain $\sigma_c(C)$. For this purpose we write the integral in Eq. (\ref{eq:dispersiongauss})
in the limit $\omega_I \to 0^+$. Using the Plemelj formula (\ref{pleme}) we have
\begin{equation}
\label{eqforometo0}
\lim_{\omega_I \to 0} \int_{-\infty}^{+\infty} \dd p \, \frac{e^{-\frac{p^2}{2\sigma^2}}}{p-\omega}
= P \int_{-\infty}^{+\infty} \dd p \, \frac{e^{-\frac{p^2}{2\sigma^2}}}{p-\omega_R} + \ii \pi e^{-\frac{\omega_R^2}{2\sigma^2}} \, .
\end{equation}
Substituting in (\ref{eq:dispersiongauss}) we get two relations equating to zero, respectively, the real and the imaginary parts;
they provide, as a function of $C$, both the threshold value $\sigma_c$, which is the quantity we are interested in,
and $\omega_R$. The derivation can be found in Appendix A. Here we only give the two relations, which are:
\begin{eqnarray}
\label{eqforometo0R}
\left( 1-C \right) \left[ 1 -2\nu e^{-\nu^2}\int_0^{\nu} \dd t \, e^{t^2} \right] + C\nu\sqrt{\pi}e^{-\nu^2}
&=& 2 \sigma^2
\\
\label{eqforometo0I}
C \left[ 1 -2\nu e^{-\nu^2}\int_0^{\nu} \dd t \, e^{t^2} \right] - \left( 1-C\right)\nu\sqrt{\pi}e^{-\nu^2}
&=& 0 \, ,
\end{eqnarray}
where $\nu = \frac{\omega_R}{\sqrt{2\sigma^2}}$. These are two equations in the unknown $\nu$ and $\sigma$, but the second equation
contains only $\nu$, and can be (numerically) solved to get $\nu$ as a function of $C$. Then, by substituting in the first equation,
we get $\sigma$ as a function of $C$. If one is interested also in the value of $\omega_R$, this will be given simply by
$\sqrt{2\sigma^2}\nu$. For the particular case of a Hamiltonian system, $C=0$, it is immediately seen that the solution of the above
relations is given by the known result $\sigma^2 = 1/2$ and $\nu=0$ (i.e., $\omega_R=0$) \cite{Campa:2009}. In
Fig. \ref{fig:gauss-stability} we plot the square of the threshold as a function of $C$. In this case we have chosen to plot
$\sigma_c^2$ since we identify $\sigma^2$ with the temperature $T$.
It is interesting to note that the threshold is not a monotonic function of the parameter $C$, contrary to what is found for the
Lorentzian distribution.

\section{Simulation results}
\label{sec_results}

The results will be presented and analyzed by showing two different types of plots obtained from the simulations. The first type is the
plot of the magnetization vs. time, $m(t)$, where $m(t) = \sqrt{m_x^2(t)+m_y^2(t)}$, with $m_x$ and $m_y$ defined in Eq. (\ref{magdefin}).
The second type of plots concerns the distribution at various times of the angular momentum of the particles, that in terms of the
one-particle distribution function $f(\theta,p,t)$ appearing in the Vlasov equation (see Section \ref{sec:Vlasov-equation}), is defined by
\begin{equation}
\label{defveldistr}
P(p,t) = \int_0^{2\pi} \dd \theta \, f(\theta,p,t) \, .
\end{equation}
We use the same symbol $P$ for the momentum distribution as in the homogeneous distribution $f_0(p)$ defined in Eq.~(\ref{distf0p}),
since clearly by plugging $f_0(p)$ on the right hand side of Eq.~(\ref{defveldistr}), we obtain $P(p)$.

We have studied the dynamics of the model with $5$ different values of $C$, i.e., $C=0$, $C=0.25$, $C=0.5$, $C=0.75$ and $C=1$. The first value,
as noted above, corresponds to the Hamiltonian model, the last value to the fully non-Hamiltonian model, and the other, intermediate values
of $C$, to the mixed cases. The system is initially prepared in a homogeneous state, with the angles uniformly distributed between $0$ and
$2\pi$, while the momenta are distributed according to a Gaussian. For each value of $C$ we have studied two situations: one in which the
width of the Gaussian (or the temperature) is above the stability threshold, and one in which it is below.

\subsection{The Hamiltonian and the fully non-Hamiltonian cases}

\begin{figure}[!htp]
\hspace{-2cm}\includegraphics[width=15cm]{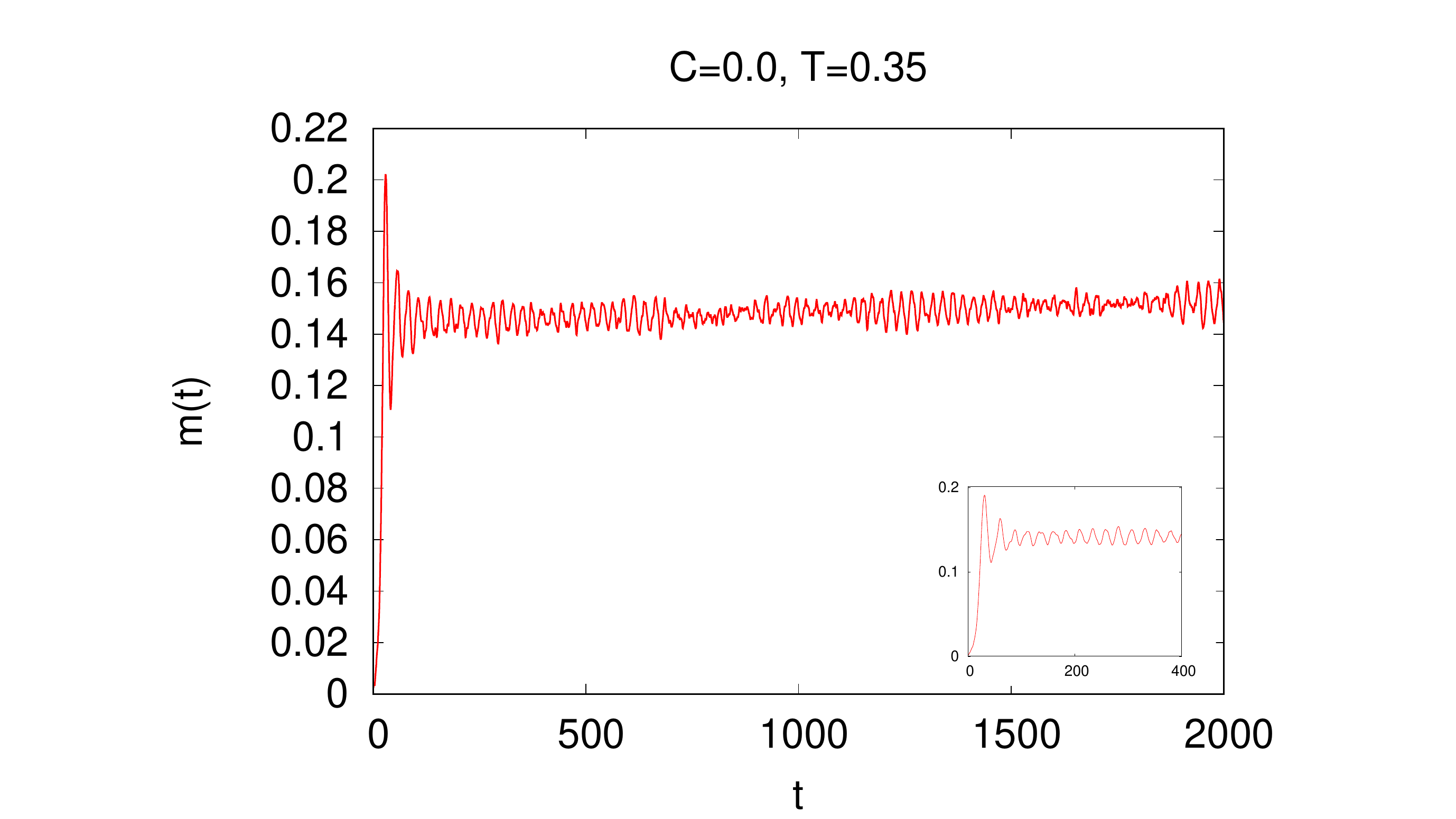}

\hspace{-2cm}\includegraphics[width=15cm]{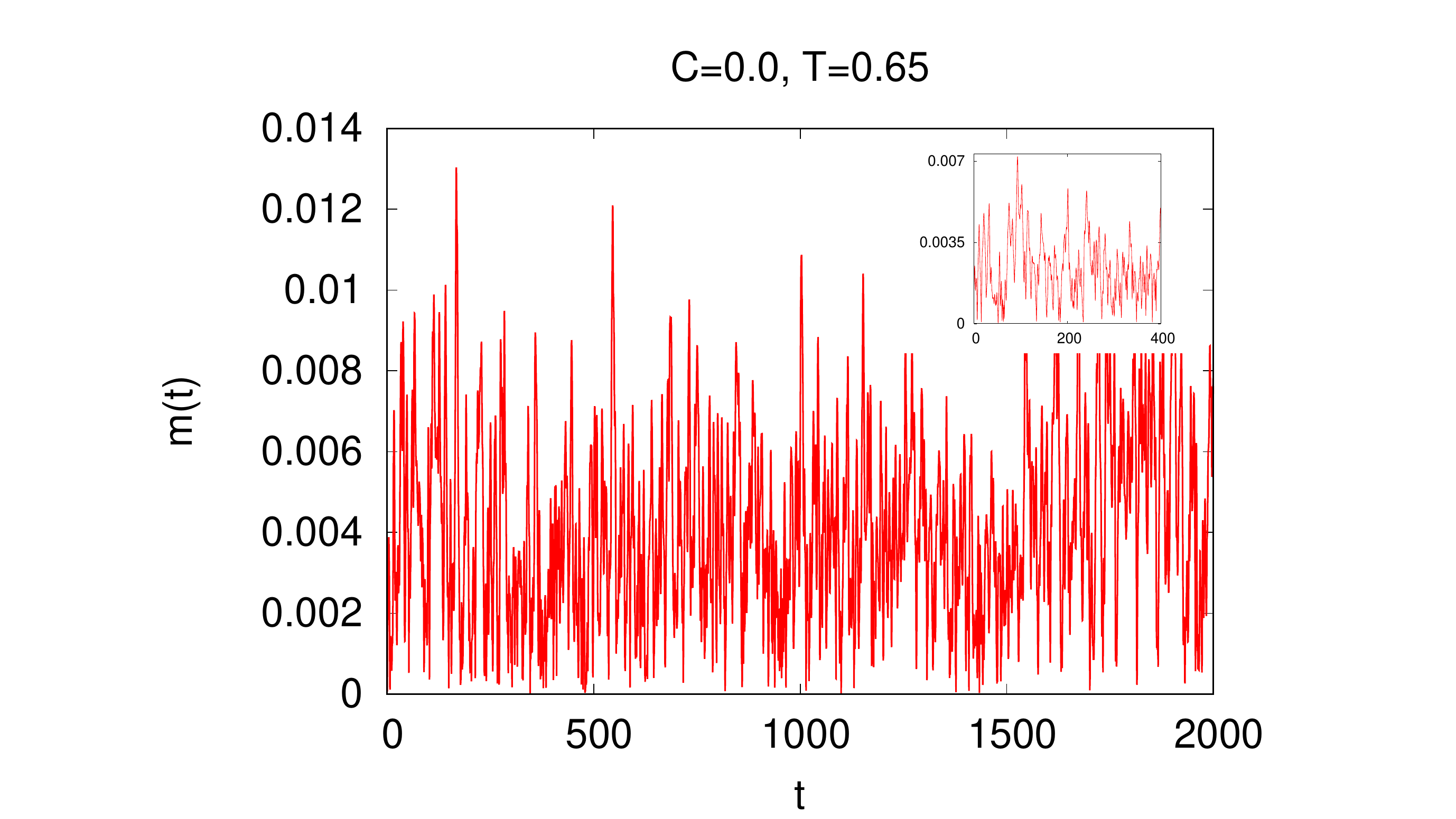}
\caption{Magnetization $m(t)$ as a function of time under the mixed dynamics~(\ref{eq:eom}) with $C=0.0$ and with the dynamics initiated
with the coordinates $\theta_i$ distributed independently and uniformly in $[-\pi,\pi]$ and with the momenta $p_i$ sampled independently
from the Gaussian distribution~(\ref{eq:f0-Gaussian}). The values of the parameter $T\equiv \sigma^2$ are $T=0.35$ (upper panel) and
$T=0.65$ (lower panel). The insets show the behavior at very short times. The system size is $N=2\times 10^5$ for the data in the main
plots and $N=5\times 10^5$ for the data in the insets.}
\label{fig:C0p0-m}
\end{figure}

\begin{figure}[!htp]
\includegraphics[width=12cm]{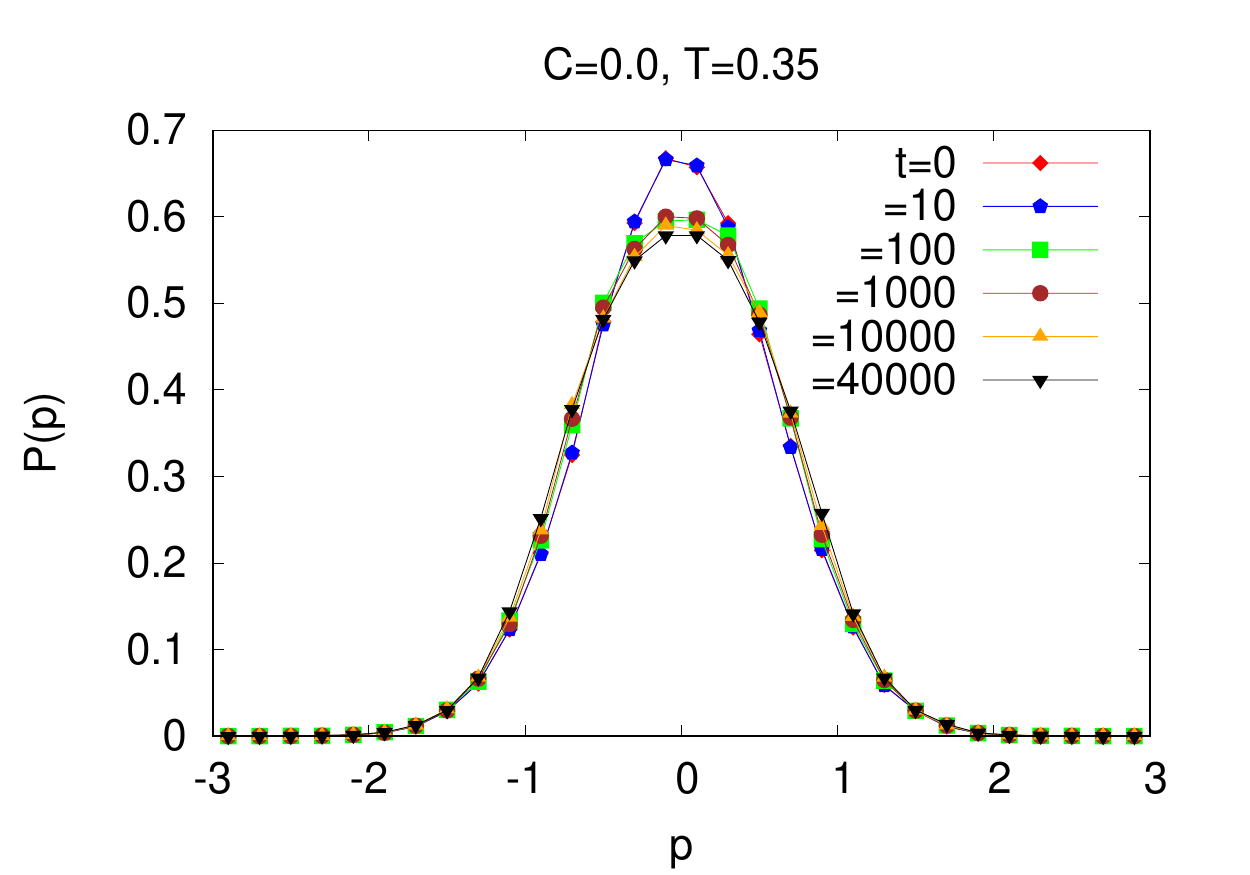}
\includegraphics[width=12cm]{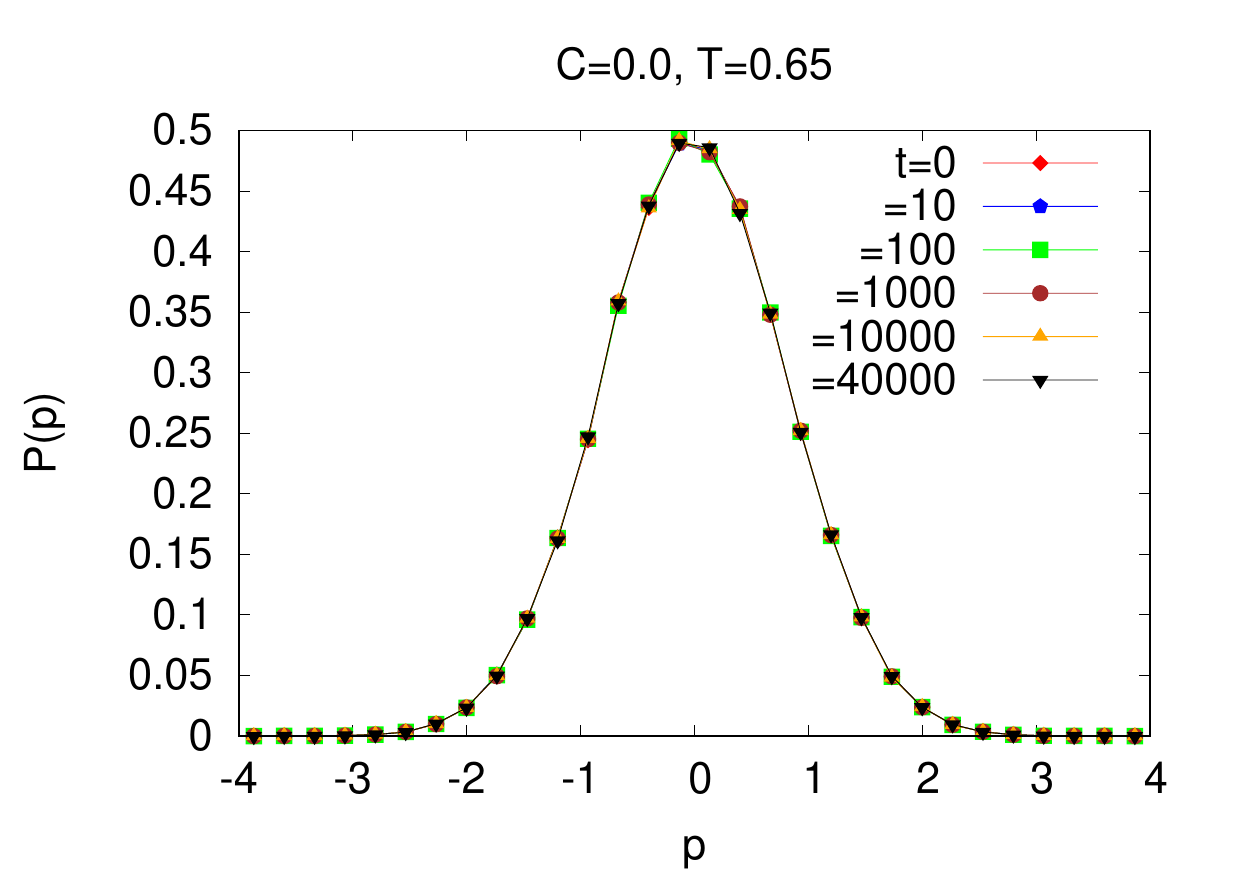}
\caption{Momentum distribution $P(p)$ at different times under the mixed dynamics~(\ref{eq:eom}) with $C=0.0$ and with the dynamics
initiated with the coordinates $\theta_i$ distributed independently and uniformly in $[-\pi,\pi]$ and with the momenta $p_i$ sampled
independently from the Gaussian distribution~(\ref{eq:f0-Gaussian}). The values of the parameter $T\equiv \sigma^2$ are $T=0.35$
(upper panel) and $T=0.65$ (lower panel). The system size is $N=10^5$.}
\label{fig:C0p0-Pv}
\end{figure}

We begin by showing a comparison of results obtained for the Hamiltonian case, $C=0$, and the fully non-Hamiltonian case, $C=1$.
In Fig. \ref{fig:C0p0-m}, we show the magnetization versus time for the Hamiltonian system, $C=0$, for two values of the temperature
$T\equiv \sigma^2$ characterizing the initial Gaussian momentum distribution,  namely,  $T=0.35$, which is below the stability threshold,
and $T=0.65$,  which is above the threshold. In Fig. \ref{fig:C0p0-Pv}, we give the plots of the momentum distribution $P(p)$ at time $t=0$
and at several times $t$ chosen at uniform logarithmic separation (except for the last value, corresponding to the final time of the
simulation run). We find it useful to provide now the information concerning the normalization in the plots of the
momentum distributions in Fig. ~\ref{fig:C0p0-Pv} and in the analogous ones in the following figures, which obviously are shown at discrete
values of the momentum $p$. We have chosen the normalization $\Delta p \sum_i P(p_i) = 1$, where $p_i$ are the momentum values
shown in the plot and $\Delta p$ is the homogeneous interval between them. This is the normalization that tends to the one that holds in the
continuum limit. We also note that in the figures, we have not explicitly indicated the time dependence of the momentum distribution and so
have written $P(p)$ instead of $P(p, t)$. Figure~\ref{fig:C0p0-Pv} as well as the figures for the plots of the momentum distributions shown
later in the paper have been obtained with simulation runs with $N = 10^5$ particles, simulating a total time of $t=4\times 10^4$. For the
plots of $m(t)$, on the other hand, we have used a higher number $N$ of particles for the following reason. From the time course of $m(t)$ in
the simulations with $10^5$ particles, we have seen that the most interesting part of the dynamics (on which we comment in a moment) occurs at
early times. Therefore, we have chosen to perform runs simulating shorter times but with a higher number of particles, so as to have smaller
finite size effects. Then, in Fig. \ref{fig:C0p0-m} and in all the other figures of $m(t)$, we have plotted the result of a run with
$N=2\times 10^5$ particles simulating a total time of $t=2000$; furthermore, in the inset we give $m(t)$ from a run with $N=5\times 10^5$,
simulating a total time of $t=400$.

\begin{figure}[!htp]
\hspace{-2cm}\includegraphics[width=15cm]{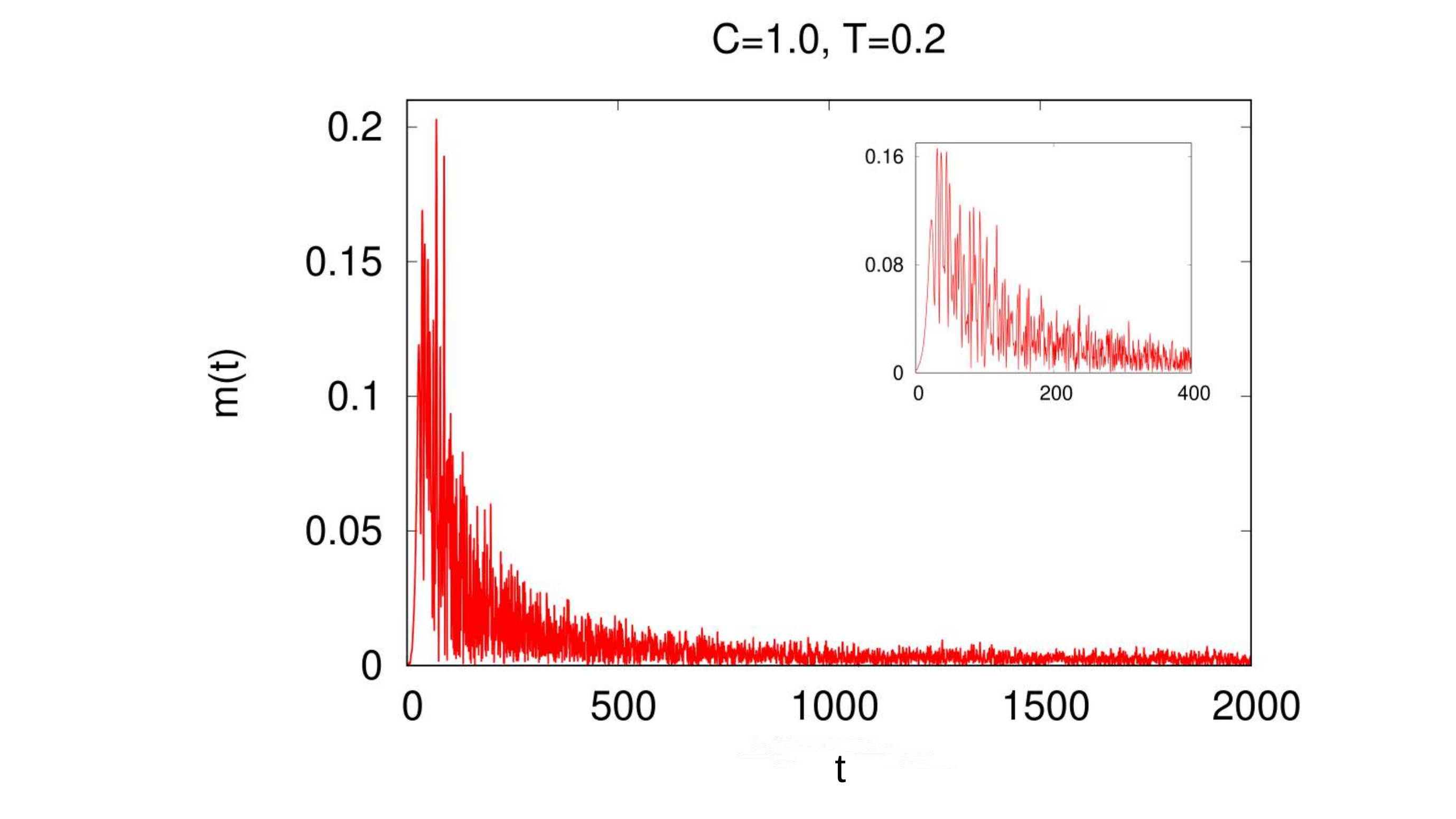}

\hspace{-2cm}\includegraphics[width=15cm]{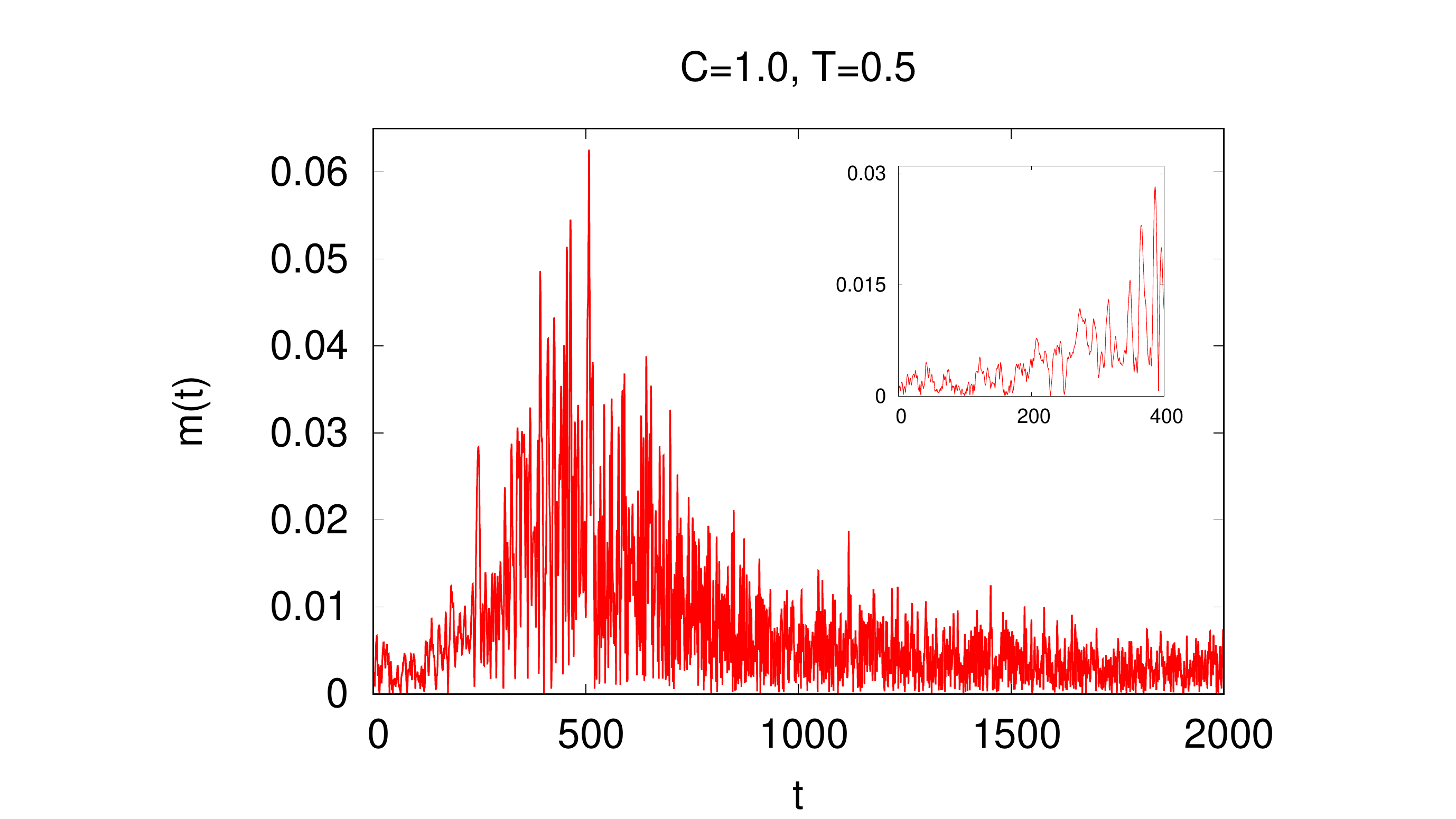}
\caption{Magnetization $m(t)$ as a function of time under the mixed dynamics~(\ref{eq:eom}) with $C=1.0$ and with the dynamics initiated
with the coordinates $\theta_i$ distributed independently and uniformly in $[-\pi,\pi]$ and with the momenta $p_i$ sampled independently
from the Gaussian distribution~(\ref{eq:f0-Gaussian}). The values of the parameter $T\equiv \sigma^2$ are $T=0.2$ (upper panel) and $T=0.5$
(lower panel). The insets show the behavior at very short times. The system size is $N=2\times 10^5$ for the data in the main plots and
$N=5\times 10^5$ for the data in the insets.}
\label{fig:C1-m}
\end{figure}

\begin{figure}[!htp]
\includegraphics[width=12cm]{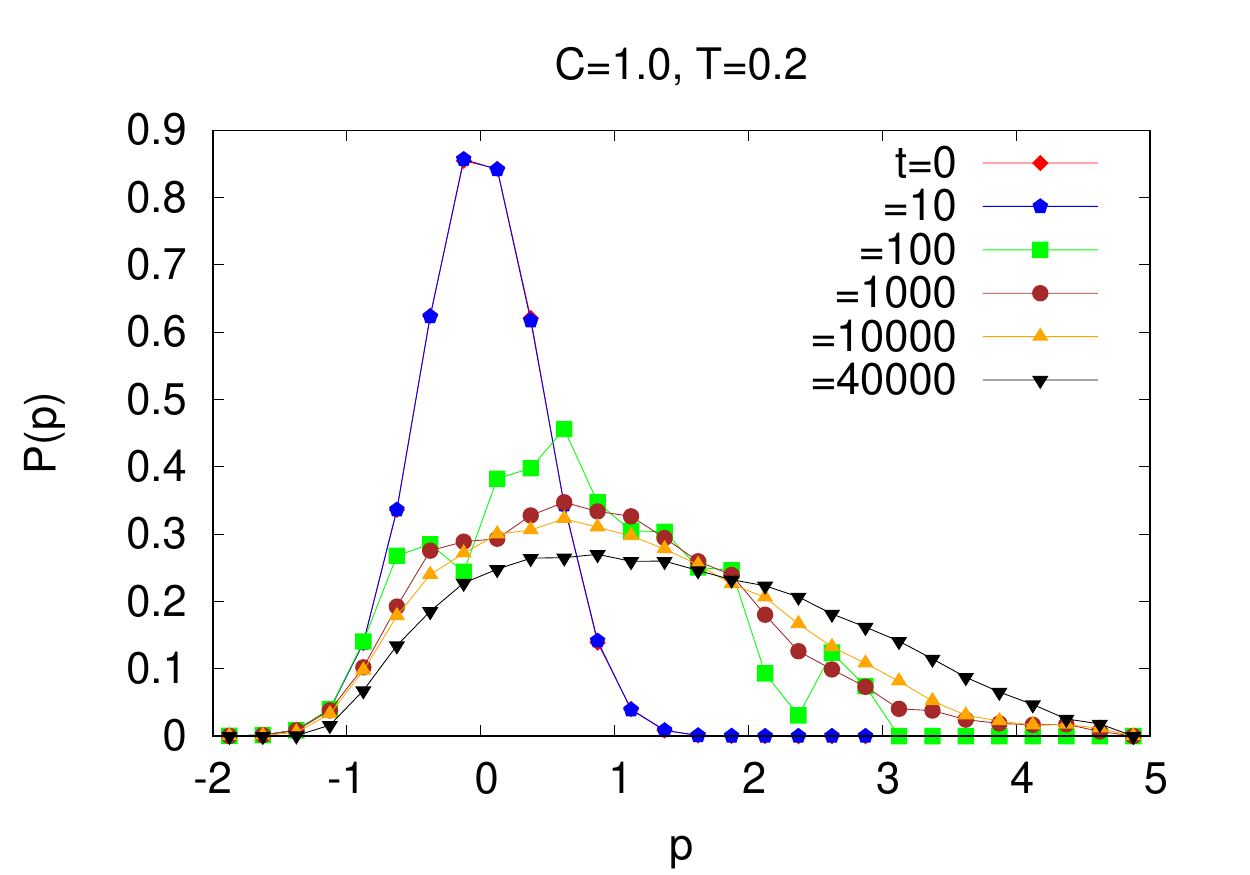}
\includegraphics[width=12cm]{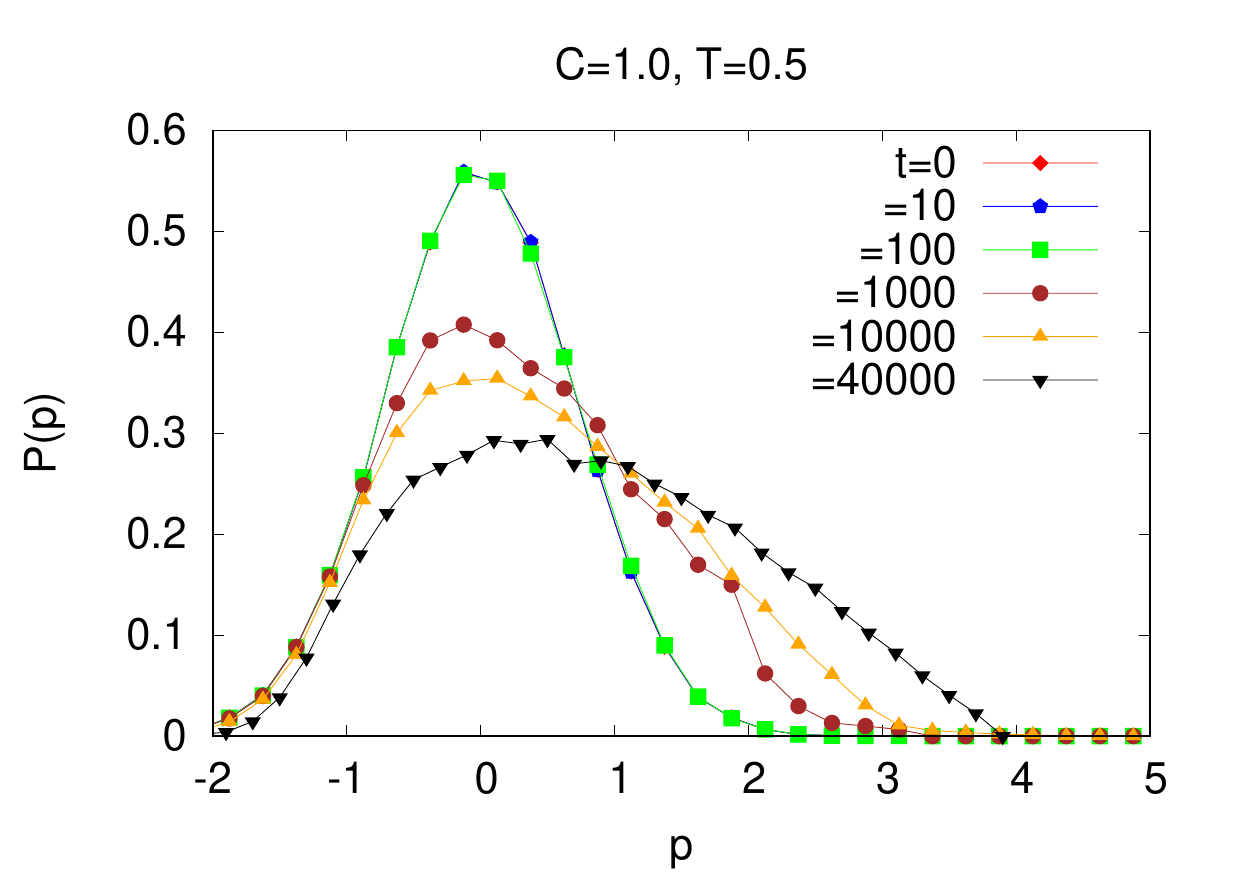}
\caption{Momentum distribution $P(p)$ at different times under the mixed dynamics~(\ref{eq:eom}) with $C=1.0$ and with the dynamics
initiated with the coordinates $\theta_i$ distributed independently and uniformly in $[-\pi,\pi]$ and with the momenta $p_i$ sampled
independently from the Gaussian distribution~(\ref{eq:f0-Gaussian}). The values of the parameter $T\equiv \sigma^2$ are $T=0.2$
(upper panel) and $T=0.5$ (lower panel). The system size is $N=10^5$.}
\label{fig:C1-Pv}
\end{figure}

In Figs. \ref{fig:C1-m} and \ref{fig:C1-Pv} we have the analogous plots for the fully non-Hamiltonian case, $C=1$. Now for the Vlasov
stable initial state we have chosen $T=0.5$, and for the Vlasov unstable state $T=0.2$. For the choice of $T$ of the stable and unstable
initial states in the various cases of $C$ we have adopted the criterion to have, roughly, similar distances above and below the threshold
value of $T$. 
We do not have to comment much on the plots of the Hamiltonian case, since this corresponds to the HMF system, that has long been studied
in the literature. The purpose of the plots for $C=0$ is to compare them with the other $C$ values. We just note that our simulations
reproduce the known fact that below threshold, $T=0.35$, the system develops very fast a magnetized state that slowly evolves towards
Boltzmann-Gibbs equilibrium (the equilibrium magnetization is reached on a time scale longer than that in the plot). Correspondingly, the
momentum distribution changes from the initial one (the differences in the $p$ range away from $p=0$ might seem small, but they are sufficient
to compensate the lower peak and have normalized distributions). Above threshold, $C=0.65$, the systems remains unmagnetized, and the
momentum distributions does not change: the system is already at $t=0$ in the Boltzmann-Gibbs equilibrium state.

For $C\ne 0$, there is no equilibrium state where the system is supposed to go. This makes a strong distinction with the Hamiltonian case.
In the latter, we know that, given the total conserved energy of the initial state, the system will evolve towards the corresponding
equilibrium state. If the initial state is Vlasov unstable, it will change very rapidly, while if it is Vlasov stable, in will remain in that
state for a time dependent on the system size, but eventually the finite size effects will drive it away from it (if it is not already the
equilibrium state). On the other hand, when $C\ne 0$, we can still predict the short time behavior from the Vlasov equation, depending on the
stability or instability of the initial state. But we do not have hints on where the system should head after instability (for Vlasov unstable
states) or finite size effects (for Vlasov stable states) have driven the system away from the initial state. Therefore in the following,
while commenting and trying to interpret the dynamics of the system, we will not have the support
of the knowledge of the final state (if any) as for the Hamiltonian case.

Looking at the plots for $C=1$, the most evident differences from the Hamiltonian case are the following. First, the unstable initial state
develops very rapidly a magnetization, as for $C=0$, but differently from this latter case, soon after it goes back to an unmagnetized state
(see Fig. \ref{fig:C1-m}, upper panel). Second, comparing the bottom panels of Fig. \ref{fig:C0p0-m} and of Fig. \ref{fig:C1-m},
we see that, although without attaining large values, for $C=1$ also in the Vlasov stable state the magnetization tends to increase somewhat
at rather early times, before going back practically to zero. Third, looking at the momentum distributions at various times in
Fig. \ref{fig:C1-Pv}, we clearly note the shift of the distributions as time passes. One can immediately argue that the last effect is related
to the nonconservation of the average momentum, as expressed in Eq. (\ref{eqavermom}); in the following we look at it under another
point of view. Let us then analyze these various effects.

Concerning the fact that for $C=1$ the magnetization seems to prefer to keep a vanishing value, it is not difficult to provide a physical
argument that can explain why the system with $C=1$ does not sustain a magnetized state. In fact, from the equations
of motion (\ref{eq:eom}) we see that for $C=1$ there is a repulsion between two particles when they are at the same angle. One can then infer
that, when the system is pushed away from the initial state due to the Vlasov instability and it builds a magnetization (since the instability
is in the first Fourier component of the distribution, the one related to the magnetization of the system), it is then driven quickly to
another unmagnetized state which is Vlasov stable. This suggests that not only a stationary magnetized state (where $m$, $m_x$, $m_y$ are all
time independent), but that also a non-stationary magnetized state (where $m$ is time independent, while $m_x$ and $m_y$ are not), is not
attained. As a matter of fact, from Eq. (\ref{eqavermom}), one can deduce that it is not possible to have a magnetized
Vlasov stationary state for a non-Hamiltonian system ($C \neq 0$), in particular for the fully non-Hamiltonian case ($C=1$), since for
$m \ne 0$, the average momentum, i.e., the expectation value of $p$, will not be constant. For the case $C=1$, it is interesting to derive
from the Vlasov equation that states of constant magnetization, either stationary or rotating, are not possible. Thus, let us begin by
considering possible magnetized stationary states of the Vlasov equation (\ref{eq:vlasov}) for the fully non-Hamiltonian system. For this
purpose, let us start from the stationary version of the Vlasov equation, i.e.
\begin{equation}
\label{stat_vlas}
p\frac{\partial f}{\partial \theta} + F[f](\theta)\frac{\partial f}{\partial p} = 0 \, ,
\end{equation}
trying to find a solution $f(\theta,p)$ with a given magnetization $m$. Redefining, if necessary, the axes orientation, we can take without
loss of generality the magnetization along the $x$ axis. Therefore, from Eq. (\ref{forcefunctional}) the force term in the above equation for
the fully non-Hamiltonian system is given by
\begin{equation}
\label{nonham_force}
F[f](\theta) = m \cos \theta \, ,
\end{equation}
with
\begin{equation}
\label{selfcons}
m = m_x = \int \dd p \, \dd \theta \, \cos \theta f(\theta,p) \, .
\end{equation}
We now make a Fourier expansion of $f(\theta,p)$, to have
\begin{equation}
\label{fourexp}
f(\theta,p) = \sum_{k=-\infty}^{+\infty} g_k(p) e^{\ii k \theta} \, .
\end{equation}
The constraints on the functions $g_k(p)$ are the following. The reality of $f(\theta,p)$
requires that $g_{-k}(p)=g_k^*(p)$ for $k \ne 0$ and that $g_0(p)$ is real; from its normalization we get
\begin{equation}
\label{normcons}
2 \pi \int \dd p \, g_0(p) = 1 \, ;
\end{equation}
finally from the self-consistency we obtain
\begin{equation}
\label{cpselfcons}
\pi \int \dd p \, [g_{-1}(p) + g_1(p)] = 2\pi \int \dd p \, {\rm Re}[g_1(p)] = m \, .
\end{equation}
Plugging the expansion (\ref{fourexp}) in Eq. (\ref{stat_vlas}), using the expressions (\ref{nonham_force}) of
$F[f](\theta)$, we have
\begin{equation}
\sum_{k=-\infty}^{+\infty} \left[ \ii pkg_k(p) + \frac{m}{2}(e^{\ii \theta} + e^{-\ii \theta})
\frac{\dd g_k}{\dd p} \right] e^{\ii k \theta} = 0 \, .
\end{equation}
This equation can be rewritten as
\begin{equation}
\label{sum_sol_vlas}
\sum_{k=-\infty}^{+\infty} \left[ \ii pkg_k(p) + \frac{m}{2}\left(\frac{\dd g_{k+1}}{\dd p}
+ \frac{\dd g_{k-1}}{\dd p}\right)\right] e^{\ii k \theta} = 0 \, .
\end{equation}
Each term in square bracket must then be equal to $0$. If we consider the term with $k=0$ we have
\begin{equation}
\label{sum_sol_vlas_k0}
\frac{m}{2} \left( \frac{\dd g_1}{\dd p} + \frac{\dd g_{-1}}{\dd p}\right) =
m \frac{\dd {\rm Re}[g_1]}{\dd p} = 0 \, .
\end{equation}
Since by hypothesis we are assuming $m\neq 0$, we get
\begin{equation}
\label{constg1}
{\rm Re}[g_1(p)] = cnst \, .
\end{equation}
We are forced to put ${\rm Re}[g_1(p)] = 0$ to have an integrable term; but then, the self-consistent
equation (\ref{cpselfcons}) cannot be satisfied with $m\ne 0$. Therefore, in the fully non-Hamiltonian case it is not
possible to have a stationary magnetized solution of the Vlasov equation. Applying the same procedure in the Hamiltonian
case, instead of the condition (\ref{constg1}) we find that what has to vanish is the imaginary part of $g_1(p)$, and this has
no consequence on the constraint (\ref{cpselfcons}), allowing, as of course we know, the existence of stationary magnetized
solutions.

We now pass to consider the possibility of a magnetized state that rotates. Since Eq. (\ref{eqavermom}) shows that
the center of mass has an acceleration equal to the square of magnetization, $m^2$, such a state could be represented
by a distribution function $f(\theta,p,t)= \widetilde{f}(\theta - \frac{1}{2}m^2 t^2,p-m^2 t)$, with $\widetilde{f}(\theta,p)$ a
function to be found and $m$ the constant value of the magnetization. In the accelerated frame defined by $\theta'=\theta-\frac{1}{2}m^2 t^2$,
$p'=p-m^2 t$, the solution $\widetilde{f}(\theta',p')$ would be given by the time independent Vlasov equation
\begin{equation}
\label{stat_vlas_acc}
p\frac{\partial \widetilde{f}}{\partial \theta'} + \left( m\cos \theta' - m^2\right)\frac{\partial \widetilde{f}}{\partial p'} = 0 \, ,
\end{equation}
different from (\ref{stat_vlas}) for the presence of the term proportional to $m^2$, due to the inertial force in the
accelerated frame. Expanding the function $\widetilde{f}(\theta',p')$ as in Eq. (\ref{fourexp}), we arrive at the analogous of
Eq. (\ref{sum_sol_vlas}), i.e.,
\begin{equation}
\label{sum_sol_vlas_acc}
\sum_{k=-\infty}^{+\infty} \left[ \ii p'kg_k(p') + \frac{m}{2}\left(\frac{\dd g_{k+1}}{\dd p'}
+ \frac{\dd g_{k-1}}{\dd p'}\right)-m^2 \frac{\dd g_k}{\dd p'} \right] e^{\ii k \theta'} = 0 \, ,
\end{equation}
which implies that for every $k$, we have
\begin{equation}
\label{sum_sol_vlas_acc-1}
 \ii p'kg_k(p') + \frac{m}{2}\left(\frac{\dd g_{k+1}}{\dd p'}
+ \frac{\dd g_{k-1}}{\dd p'}\right)-m^2 \frac{\dd g_k}{\dd p'}= 0 \, .
\end{equation}
Now equating to zero the term with $k=0$ we obtain
\begin{equation}
\label{constg1_acc}
{\rm Re}[g_1(p')] = m g_0(p') \, ,
\end{equation}
that satisfies exactly the self-consistency equation (\ref{cpselfcons}), keeping in mind the normalization condition
on $g_0$. On the other hand, the $k=1$ term gives
\begin{equation}
\label{sum_sol_vlas_acc_k1}
\ii p'g_1(p') + \frac{m}{2}\left(\frac{\dd g_2}{\dd p'}
+ \frac{\dd g_0}{\dd p'}\right)-m^2 \frac{\dd g_1}{\dd p'} = 0 \, .
\end{equation}
Let us now suppose that the functions $g_k(p')$ can be expanded in powers of $m$:
\begin{equation}
\label{expandgp}
g_k(p') = \sum_{s=0}^\infty h_{ks}(p')m^s \, .
\end{equation}
Plugging $m=0$ in Eq.~(\ref{sum_sol_vlas_acc-1}), we have that $h_{k0}=0$ for $k\ne 0$. Inserting now Eq.~(\ref{expandgp}) in
Eq. (\ref{sum_sol_vlas_acc_k1}), the term linear in $m$ gives
\begin{equation}
\label{eqforg_1}
\ii p' h_{11}(p') + \frac{1}{2} \frac{\dd h_{00}}{\dd p'} = 0 \, .
\end{equation}
Since $g_0$ is real, this equation implies that $h_{11}$ is purely imaginary, but this is incompatible with (\ref{constg1_acc}).
Thus, we deduce that also rotating states with constant magnetization $m\ne 0$ are not possible for $C=1$.

The other effects mentioned above show that for $C=1$ the dynamics deviates more markedly from what predicted by the linearized Vlasov equation.
There can be two reasons for such a deviation. One is related to the nonlinear corrections to the linearized Vlasov equation.
Since the Vlasov equation describes the dynamics in the limit $N\to \infty$, these nonlinear corrections to the linearized equation are
present also in this limit. The other cause of deviation is connected with the finite size effects. We can show that both corrections
are much more important for $C=1$, and more generally for $C\ne 0$, than for the Hamiltonian case, $C=0$. Let us begin with the nonlinear
corrections to the Vlasov equation. As in Section \ref{sec:Vlasov-stability}, we write $f(\theta,p,t)=f_0(p)+\delta f(\theta,p,t)$, with
$f_0(p)$ as in (\ref{distf0p}), and we expand $\delta f(\theta,p,t)$ in Fourier series as:
\begin{align}
\delta f(\theta,p,t)=\sum_{k=-\infty}^{+\infty} \widehat{\delta f}_k(p,t)e^{\ii k\theta} \, ,
\label{eq:deltaf-expansion-fou}
\end{align}
with $\widehat{\delta f}_{-k}(p,t) = \widehat{\delta f}_k^*(p,t)$ because of the reality of $\delta f(\theta,p,t)$.
Note that, differently from the expansion (\ref{eq:deltaf-expansion}), we leave the time dependence in the function
$\widehat{\delta f}_k(p,t)$, since we are going to consider nonlinear terms, and we are not going to compute a dispersion relation.
From Eq. (\ref{eqavermom}), expressing the time derivative of the average momentum, we guess that for $C=1$ one has
\begin{eqnarray}
\label{nonlineareq1}
\frac{\dd}{\dd t} \langle p \rangle &\equiv& \int_{-\infty}^{+\infty} \dd p \int_0^{2\pi} \dd \theta \, p
\frac{\partial}{\partial t}f(\theta,p,t) = m^2(t) \nonumber \\ &=&
\left| \int_{-\infty}^{+\infty} \dd p \int_0^{2\pi} \dd \theta \, e^{\ii \theta} f(\theta,p,t) \right|^2
= 4\pi^2 \left| \int_{-\infty}^{+\infty} \dd p \widehat{\delta f}_1(p,t) \right|^2 \, .
\end{eqnarray}
In Appendix B we give few details on the straightforward computation that shows that the last term in the second line of this equation
arises as a nonlinear correction to the linearized Vlasov equation.

The second source of deviation from the prediction of the linearized Vlasov equation stems from the finite size corrections. In systems
with long-range interactions these so called collisional corrections, when the system is initially prepared in a homogeneous Vlasov stable
state, are described by the Lenard-Balescu equation~\cite{Nicholson:1992}. It is known that for one-dimensional systems the correction
vanishes for Hamiltonian systems~\cite{Campa:2014}. This is the origin, e.g., of the long life, whose length increases more rapidly than
the system size $N$, of the homogeneous stable states of the HMF model. However, computing the finite size correction for the non-Hamiltonian
case, one finds that, on the contrary, the correction does not vanish. In Appendix C we provide a detailed computation of the Lenard-Balescu
equation for the general case in which the force in the equations of motion has all the Fourier components as in Eq. (\ref{eq:eom0}), then
specializing to our case. 

\subsection{The mixed cases}

In this section we present and analyze the results for the system where the force term has both components, one of Hamiltonian origin
and one which is non-Hamiltonian. As we anticipated, we focus on three such cases, with the values of $C$ given respectively
by $C=0.25$, $C=0.5$ and $C=0.75$.

\begin{figure}[!htp]
\hspace{-2cm}\includegraphics[width=15cm]{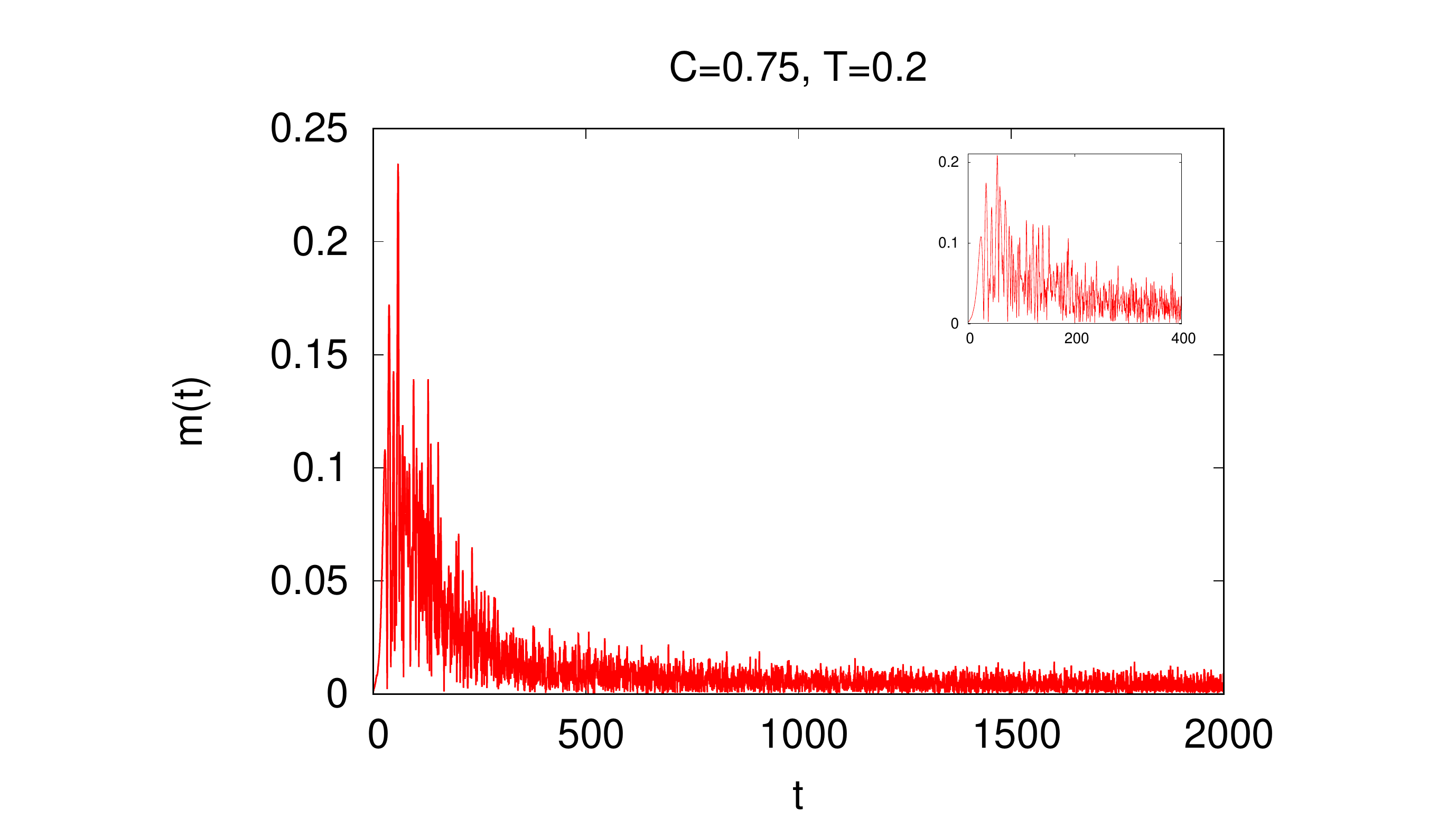}

\hspace{-2cm}\includegraphics[width=15cm]{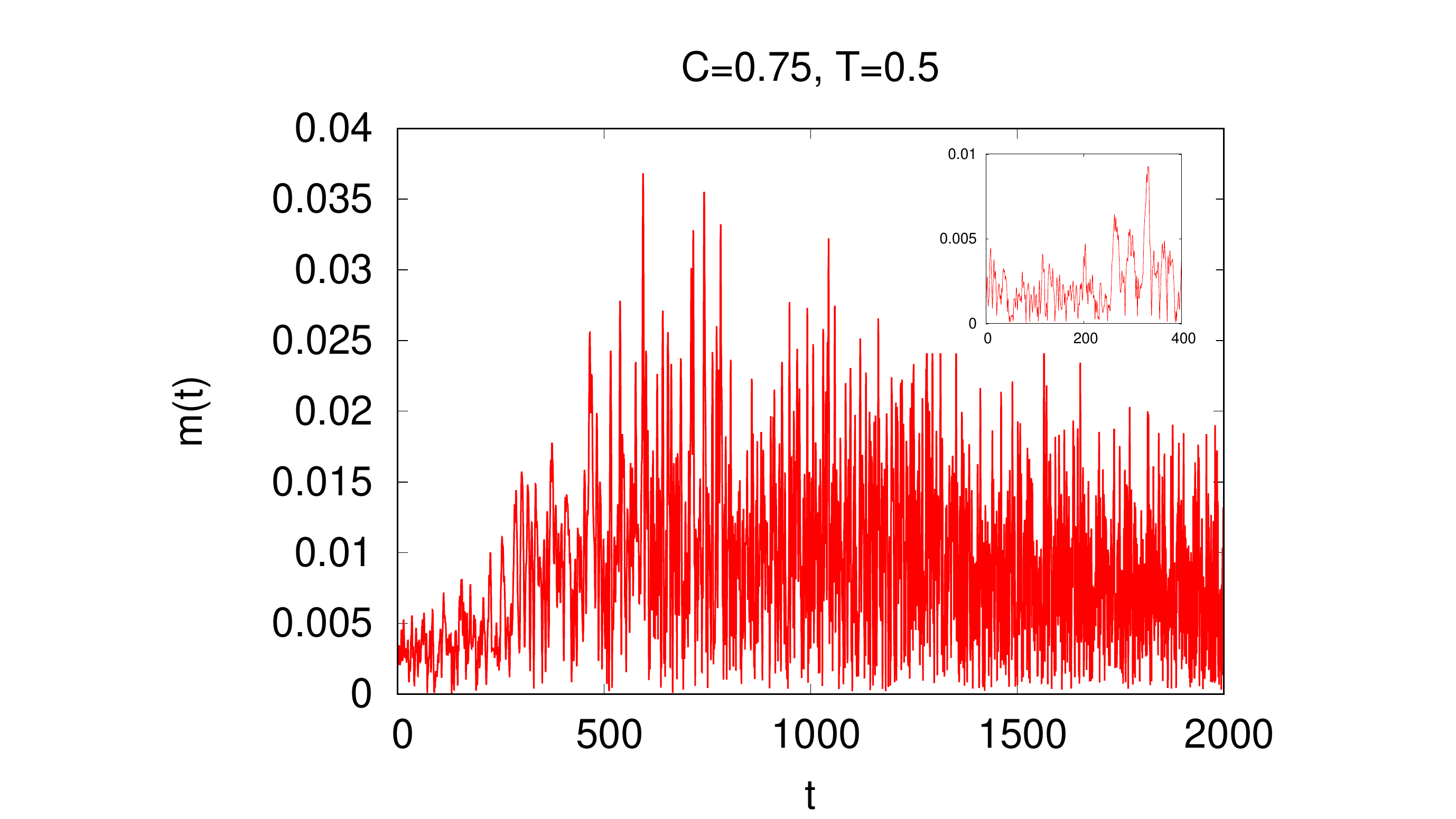}
\caption{Magnetization $m(t)$ as a function of time under the mixed dynamics~(\ref{eq:eom}) with $C=0.75$ and with the dynamics initiated
with the coordinates $\theta_i$ distributed independently and uniformly in $[-\pi,\pi]$ and with the momenta $p_i$ sampled independently from
the Gaussian distribution~(\ref{eq:f0-Gaussian}). The values of the parameter $T\equiv \sigma^2$ are $T=0.2$ (upper panel) and $T=0.5$
(lower panel). The insets show the behavior at very short times. The system size is $N=2\times 10^5$ for the data in the main plots and
$N=5\times 10^5$ for the data in the insets.}
\label{fig:C0p75-m}
\end{figure}

\begin{figure}[!htp]
\includegraphics[width=12cm]{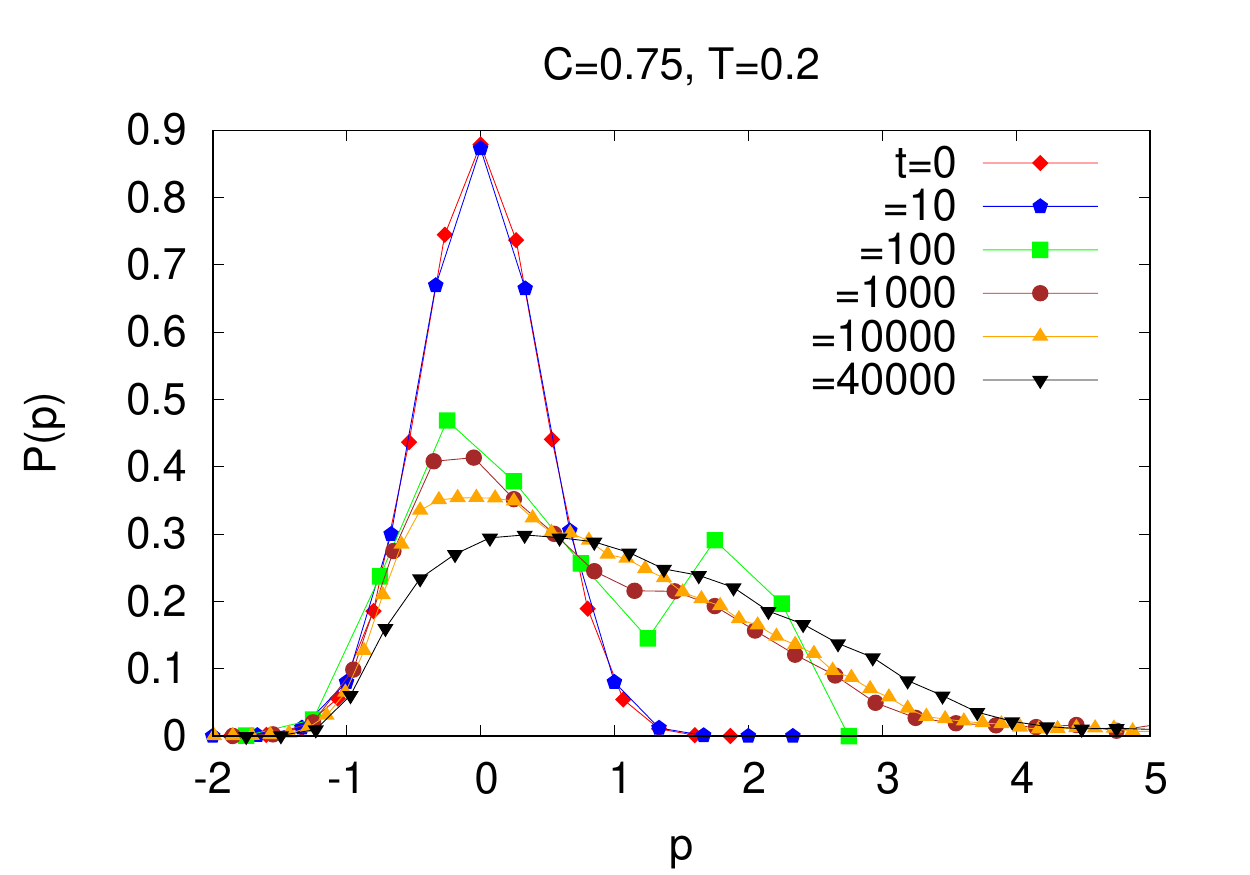}
\includegraphics[width=12cm]{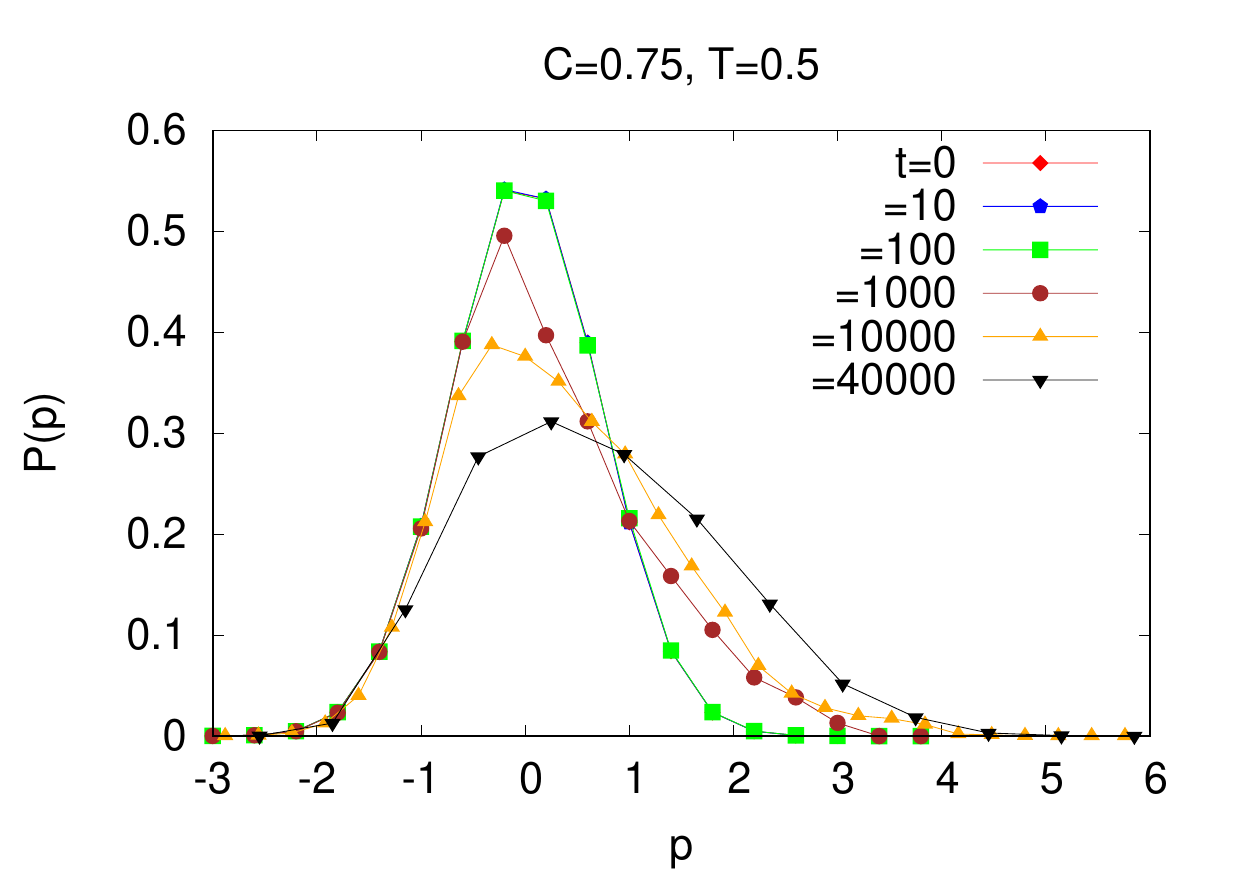}
\caption{Momentum distribution $P(p)$ at different times under the mixed dynamics~(\ref{eq:eom}) with $C=0.75$ and with the dynamics initiated
with the coordinates $\theta_i$ distributed independently and uniformly in $[-\pi,\pi]$ and with the momenta $p_i$ sampled independently
from the Gaussian distribution~(\ref{eq:f0-Gaussian}). The values of the parameter $T\equiv \sigma^2$ are $T=0.2$ (upper panel) and $T=0.5$
(lower panel). The system size is $N=10^5$.}
\label{fig:C0p75-Pv}
\end{figure}

We begin by considering the latter case, $C=0.75$. In Figs. \ref{fig:C0p75-m} and
\ref{fig:C0p75-Pv} we show the analogous of the plots presented above for the Hamiltonian and the fully non-Hamiltonian systems. The initial
temperature is $T=0.2$ for the Vlasov stable state and $T=0.5$ for the Vlasov unstable state.
We see that the behavior is not dissimilar from that occurring at $C=1$, with the magnetization increasing rapidly in the unstable case and then
going back to practically zero, and increasing somewhat also for the stable state at rather early times, before vanishing again. It can then
be argued that the same characteristics envisaged for $C=1$ should be valid also in this case.

\begin{figure}[!htp]
\hspace{-2cm}\includegraphics[width=15cm]{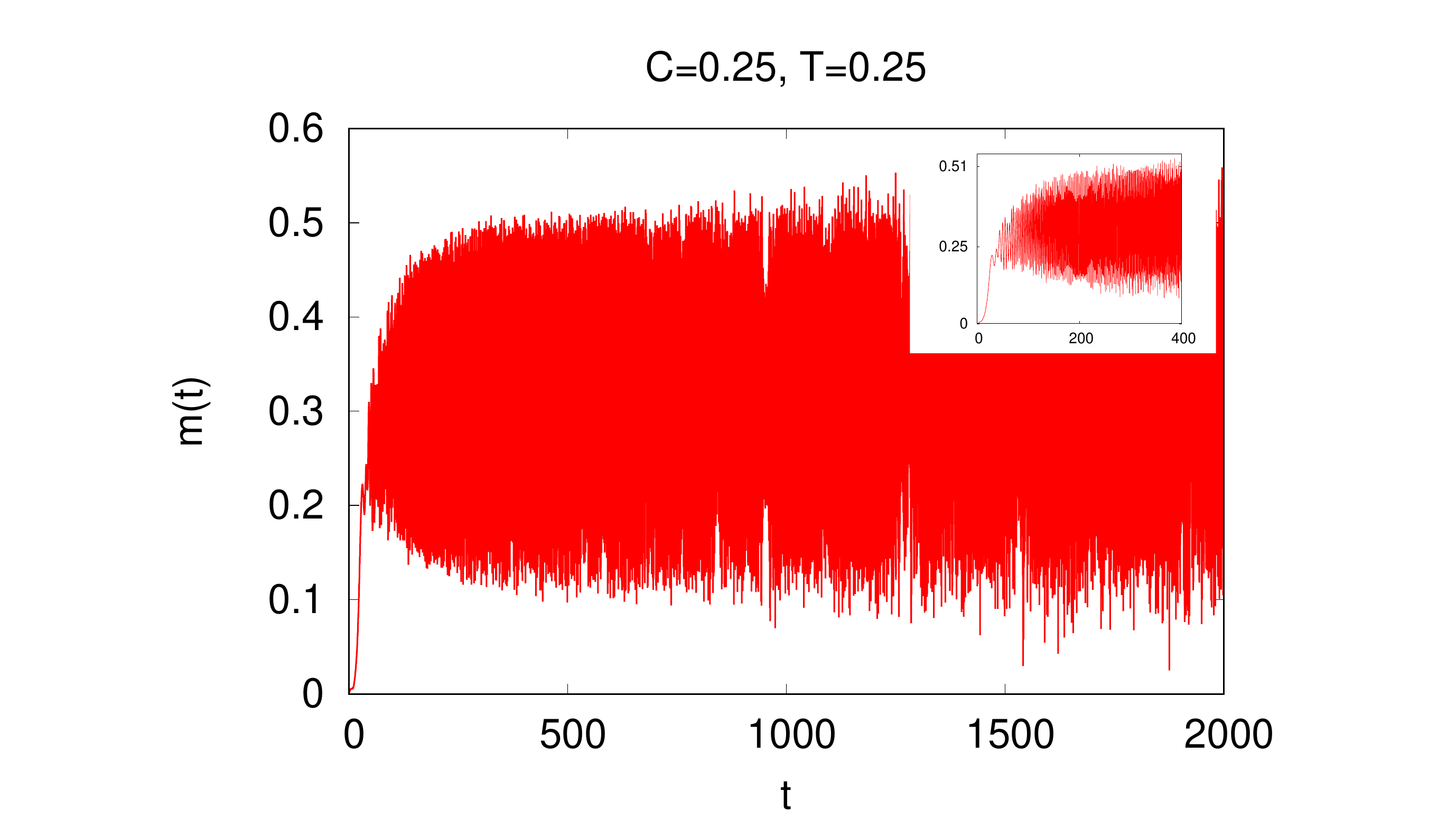}

\hspace{-2cm}\includegraphics[width=15cm]{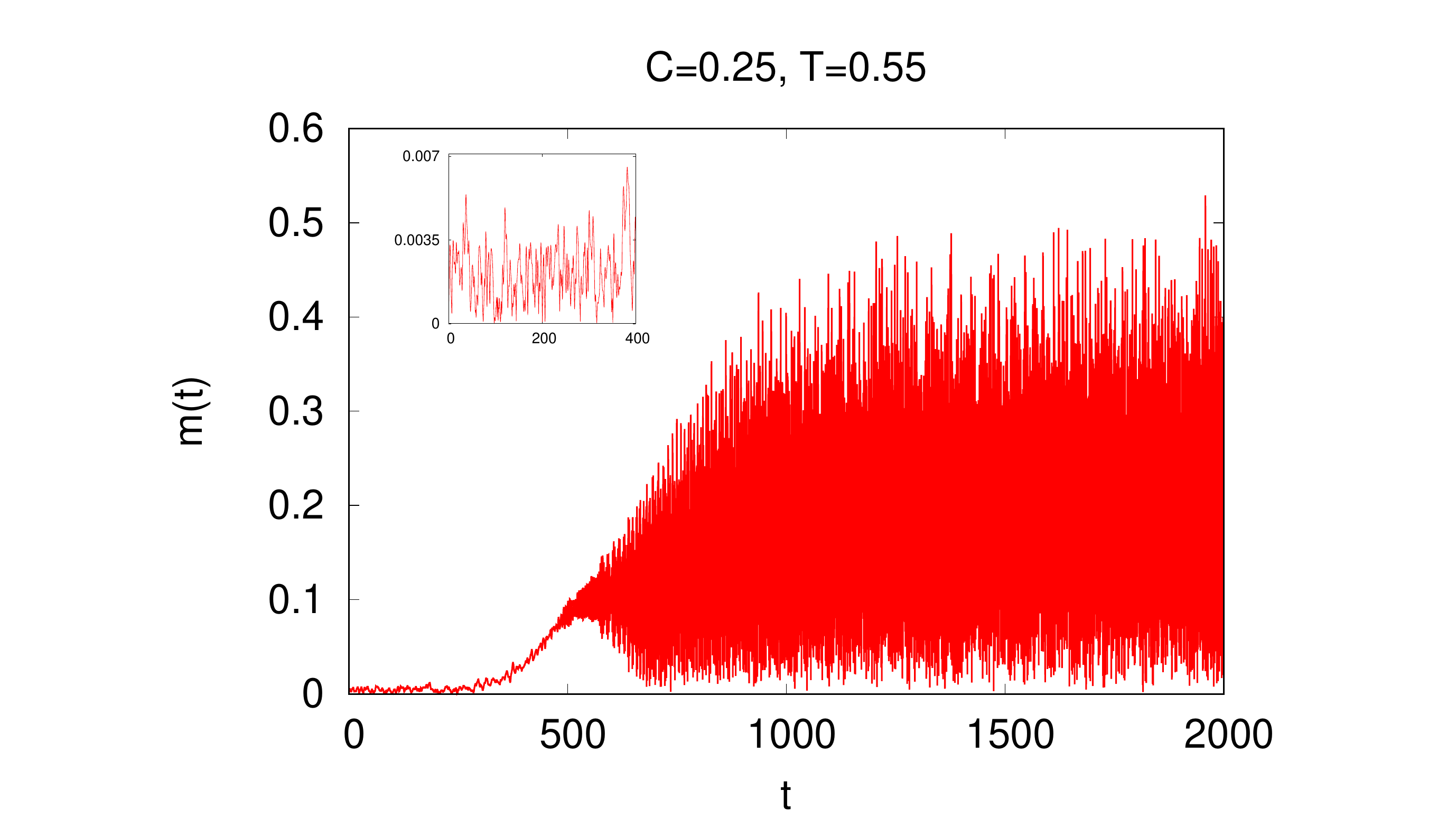}
\caption{Magnetization $m(t)$ as a function of time under the mixed dynamics~(\ref{eq:eom}) with $C=0.25$ and with the dynamics initiated with
the coordinates $\theta_i$ distributed independently and uniformly in $[-\pi,\pi]$ and with the momenta $p_i$ sampled independently from the
Gaussian distribution~(\ref{eq:f0-Gaussian}). The values of the parameter $T\equiv \sigma^2$ are $T=0.25$ (upper panel) and $T=0.55$
(lower panel). The insets show the behavior at very short times.  The system size is $N=2\times 10^5$ for the data in the main plots and
$N=5\times 10^5$ for the data in the insets.}
\label{fig:C0p25-m}
\end{figure}

\begin{figure}[!htp]
\hspace{-1.6cm}\includegraphics[width=15cm]{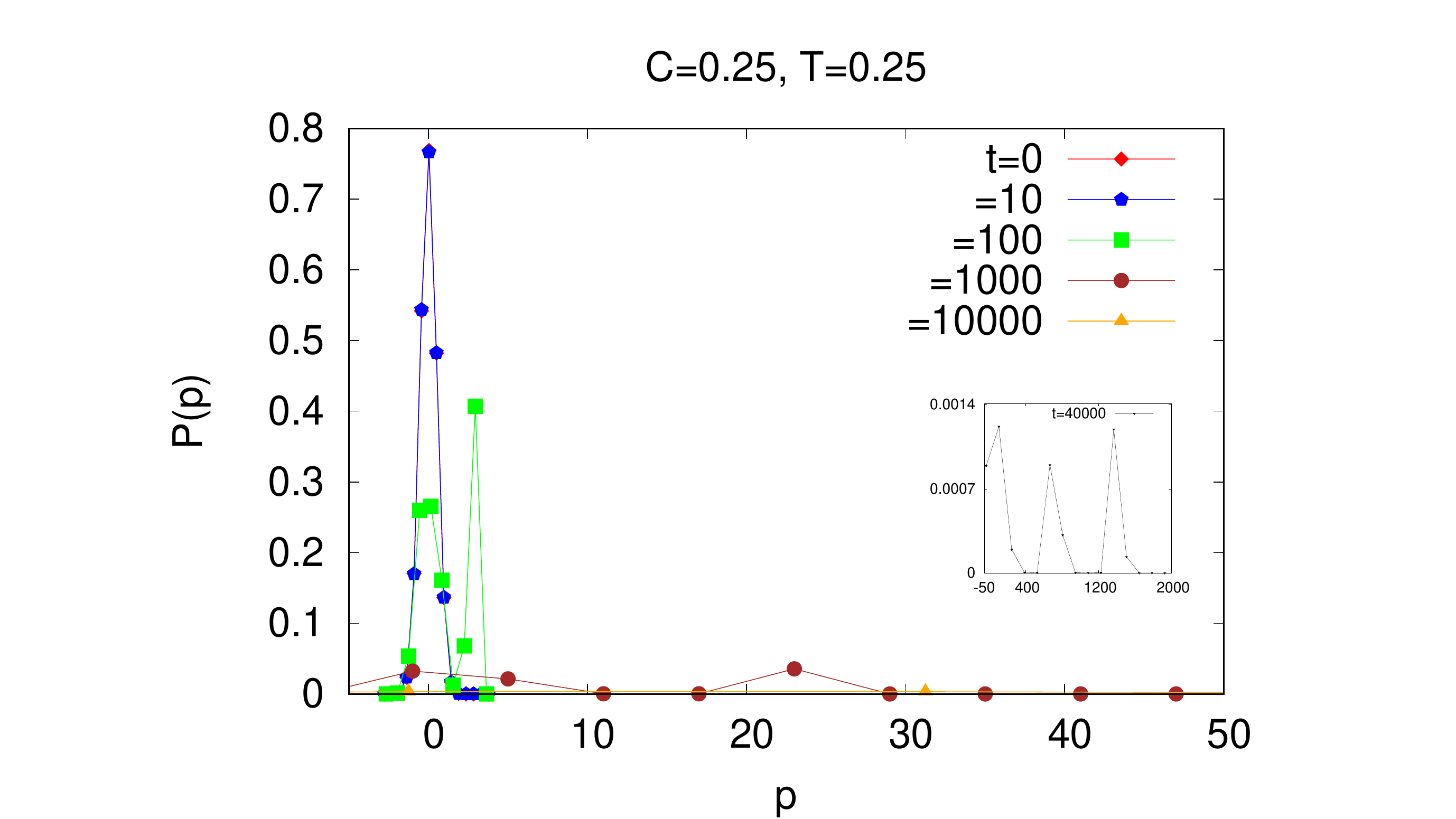}

\includegraphics[width=12cm]{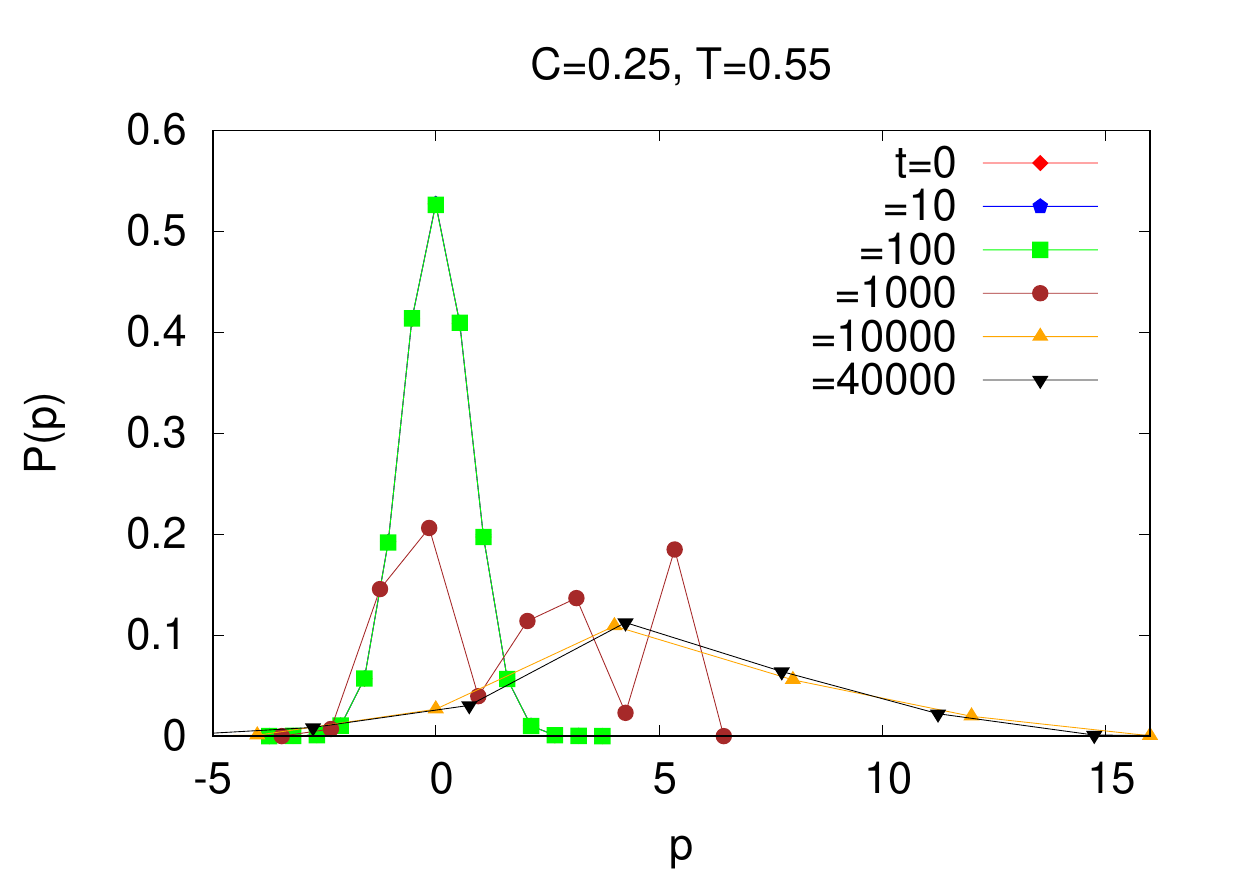}
\caption{Momentum distribution $P(p)$ at different times under the mixed dynamics~(\ref{eq:eom}) with $C=0.25$ and with the dynamics initiated
with the coordinates $\theta_i$ distributed independently and uniformly in $[-\pi,\pi]$ and with the momenta $p_i$ sampled independently from
the Gaussian distribution~(\ref{eq:f0-Gaussian}). The values of the parameter $T\equiv \sigma^2$ are $T=0.25$ (upper panel) and $T=0.55$
(lower panel). The system size is $N=10^5$. For the upper panel, the range of $p$ for $t=40000$ extends significantly beyond the range of the
main plot, and the data are therefore shown separately in the inset. }
\label{fig:C0p25-Pv}
\end{figure}

\begin{figure}[!htp]
\hspace{-2cm}\includegraphics[width=15cm]{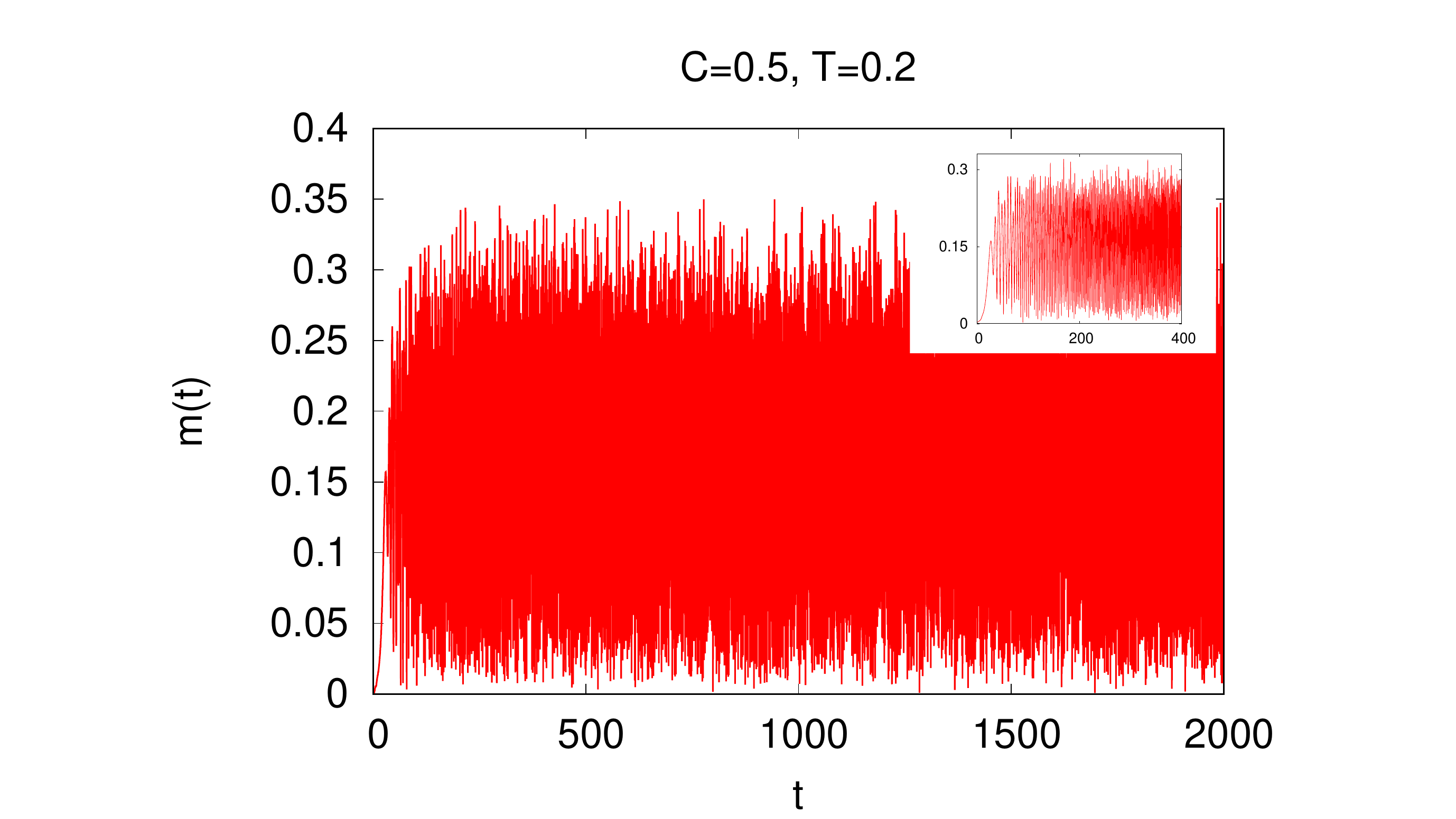}

\hspace{-2cm}\includegraphics[width=15cm]{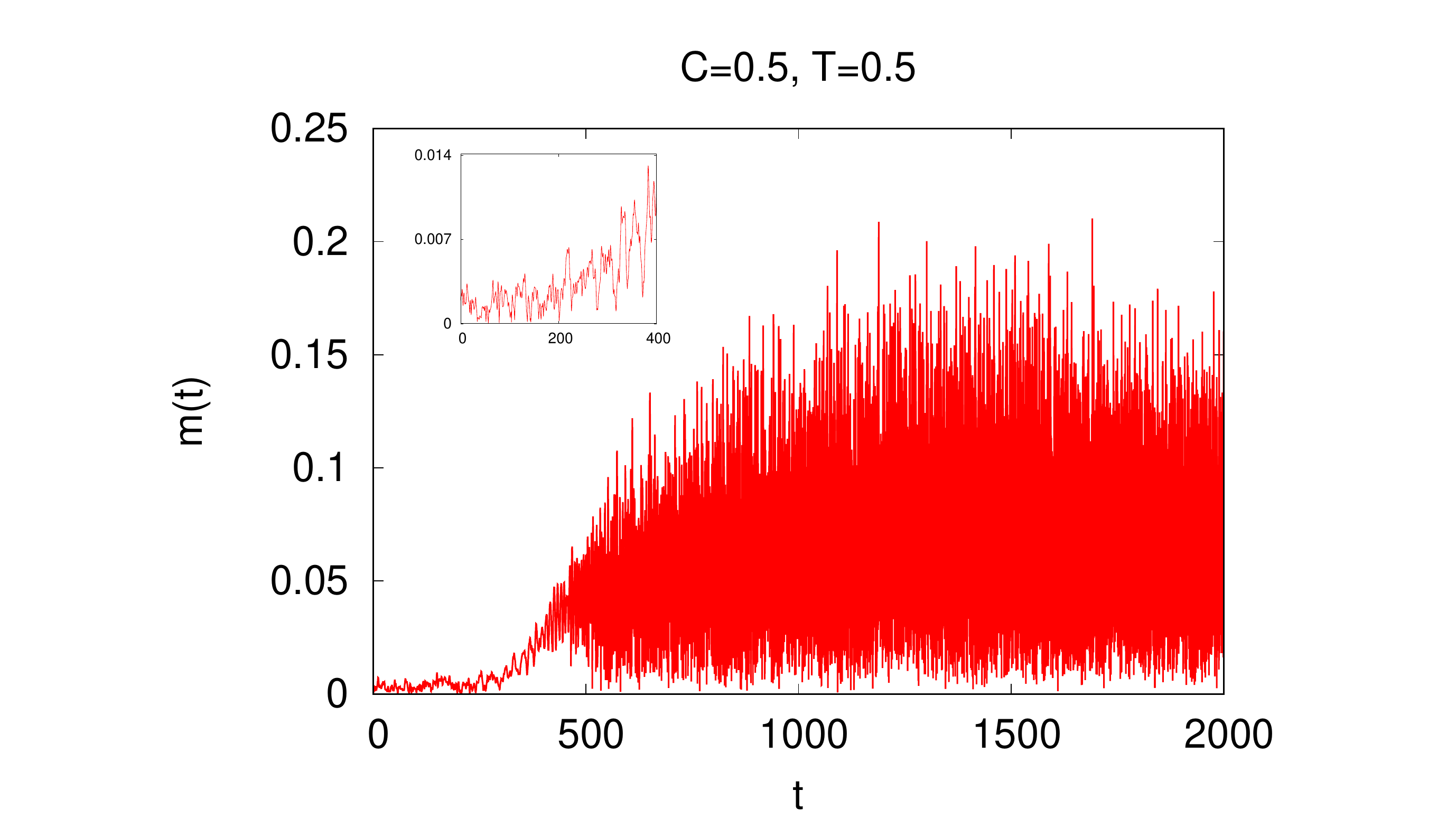}
\caption{Magnetization $m(t)$ as a function of time under the mixed dynamics~(\ref{eq:eom}) with $C=0.5$ and with the dynamics initiated with
the coordinates $\theta_i$ distributed independently and uniformly in $[-\pi,\pi]$ and with the momenta $p_i$ sampled independently from the
Gaussian distribution~(\ref{eq:f0-Gaussian}). The values of the parameter $T\equiv \sigma^2$ are $T=0.2$ (upper panel) and $T=0.5$ (lower panel).
The insets show the behavior at very short times. The system size is $N=2\times 10^5$ for the data in the main plots and $N=5\times 10^5$ for
the data in the insets.}
\label{fig:C0p5-m}
\end{figure}

\begin{figure}[!htp]
\includegraphics[width=12cm]{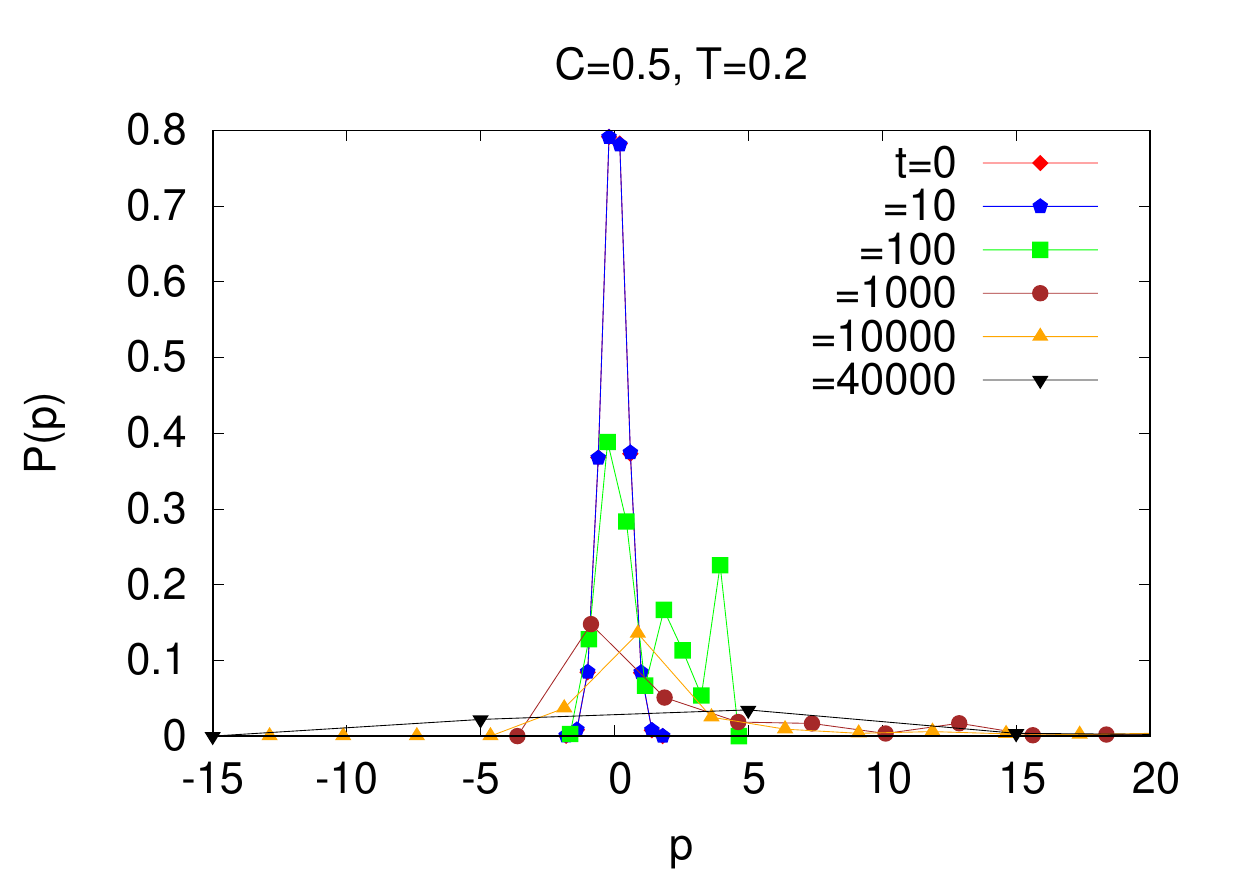}
\includegraphics[width=12cm]{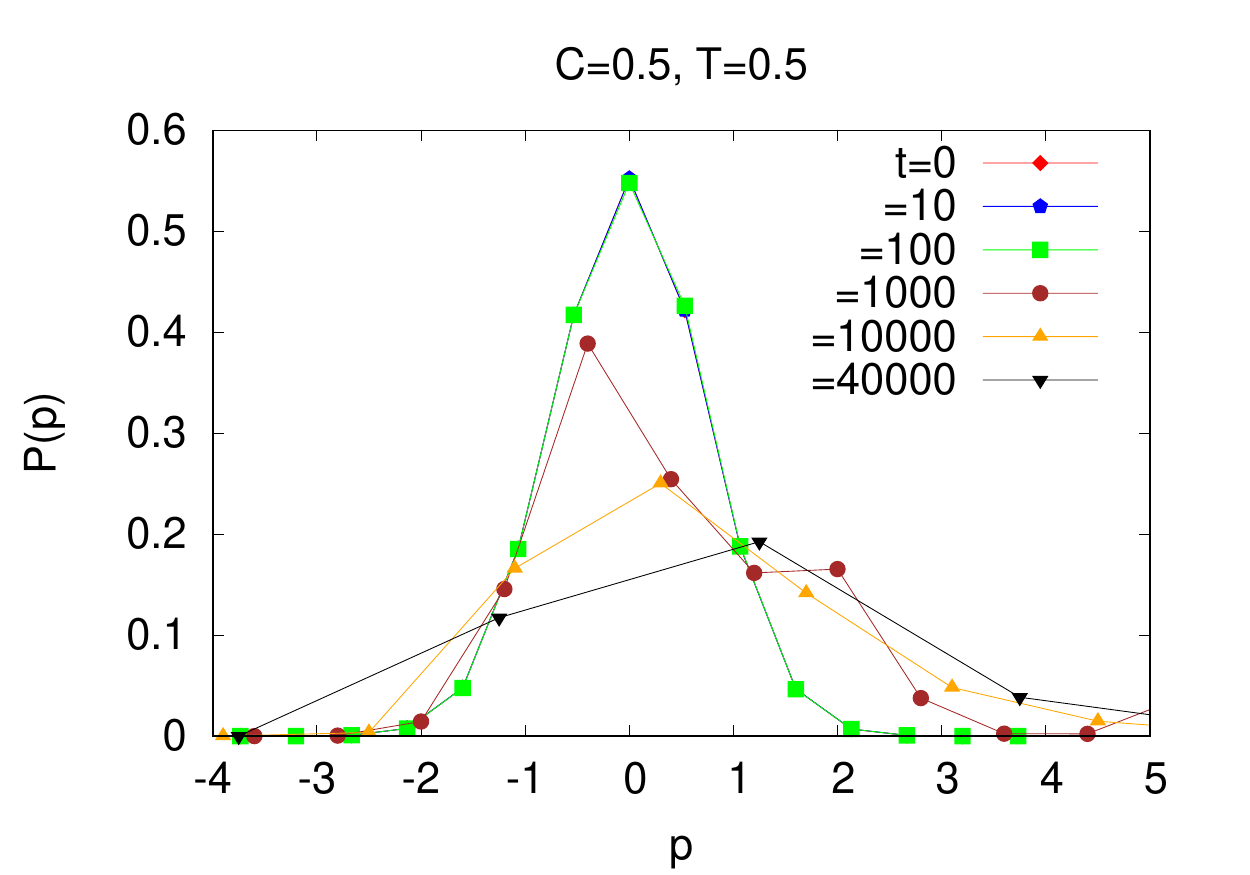}
\caption{Momentum distribution $P(p)$ at different times under the mixed dynamics~(\ref{eq:eom}) with $C=0.5$ and with the dynamics initiated
with the coordinates $\theta_i$ distributed independently and uniformly in $[-\pi,\pi]$ and with the momenta $p_i$ sampled independently from
the Gaussian distribution~(\ref{eq:f0-Gaussian}). The values of the parameter $T\equiv \sigma^2$ are $T=0.2$ (upper panel) and $T=0.5$
(lower panel). The system size is $N=10^5$.}
\label{fig:C0p5-Pv}
\end{figure}

\begin{figure}[!htp]
\includegraphics[width=12cm]{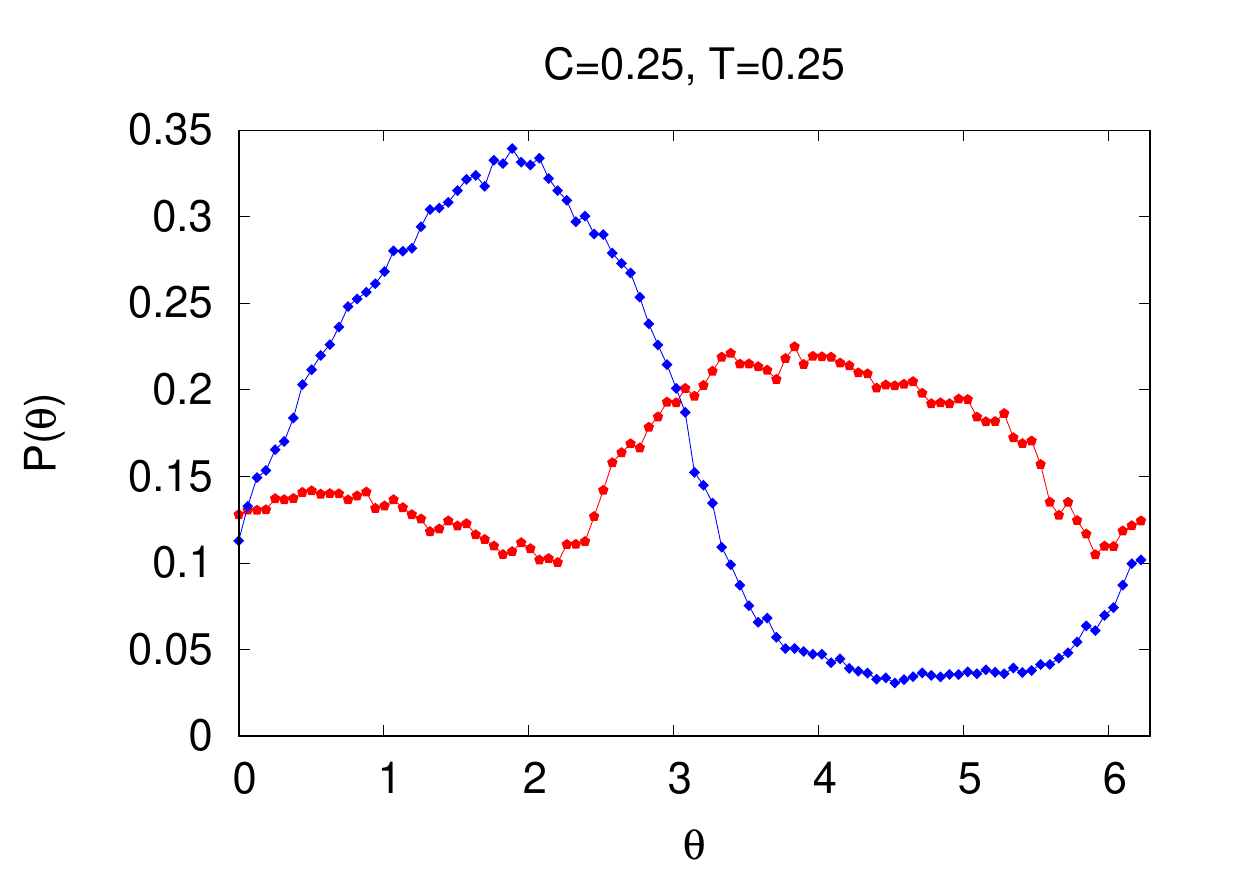}
\includegraphics[width=12cm]{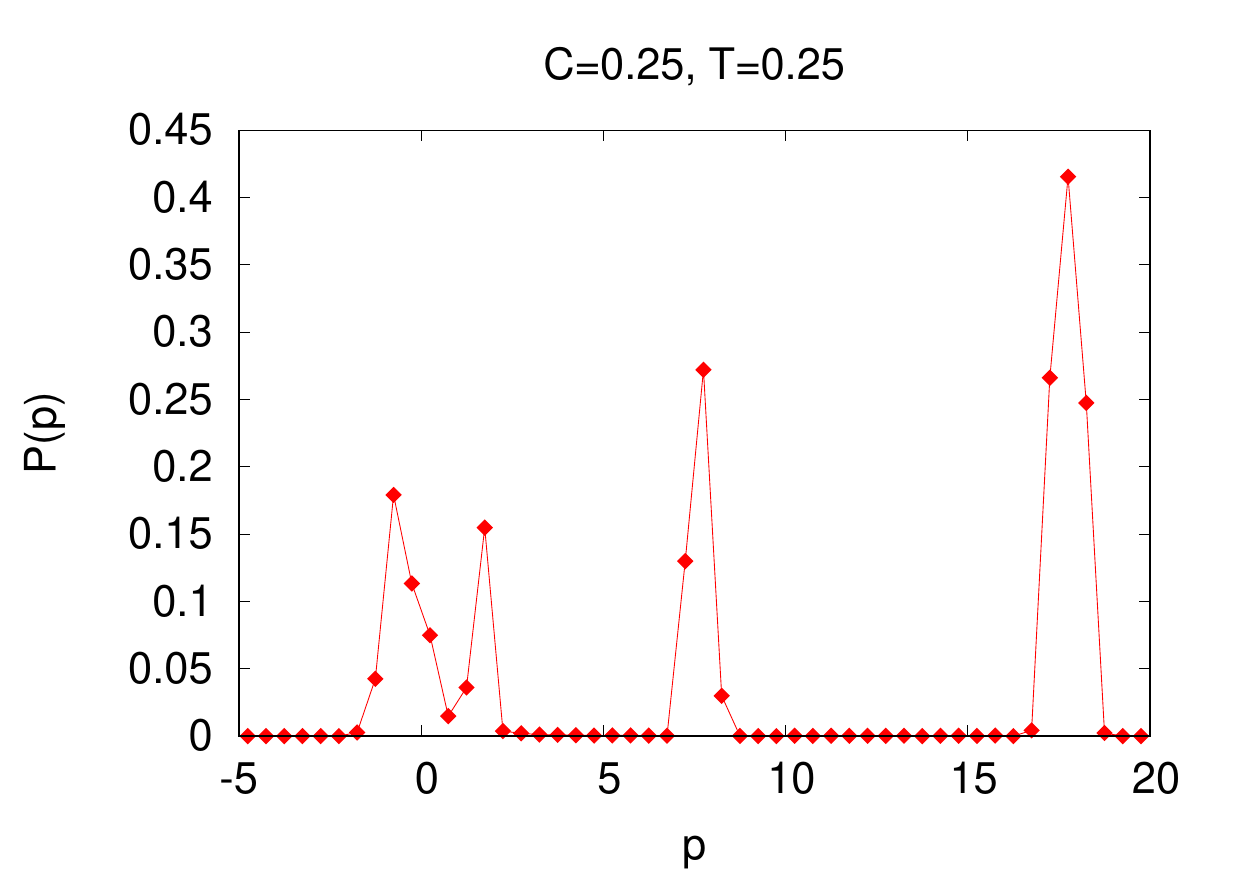}
\caption{As evident from the upper panel of Fig.~\ref{fig:C0p25-m}, the magnetization shows in time wild oscillations between small and
large values. For the same parameter values, here, the upper panel shows the $\theta$-distribution corresponding to a time at which the
magnetization has a low and a high value, with the corresponding distribution being respectively more uniform and highly non-uniform. The
lower panel shows the momentum distribution at time instant $t=400$. The system size is $N=5\times10^5$.}
\label{fig:C0p25-Ptheta}
\end{figure}

The remaining two mixed cases are presented, with plots analogous to those used before, in Figs. \ref{fig:C0p25-m} and \ref{fig:C0p25-Pv} for
$C=0.25$, and in Figs. \ref{fig:C0p5-m} and \ref{fig:C0p5-Pv} for $C=0.5$. We see a new phenomenon, that we are going to describe.
However, we first note the similarity concerning the early time behavior of the magnetization: like in the $C$ values analyzed up to now,
$m(t)$ rises almost immediately when the system starts in the Vlasov unstable state, characterized by $T=0.25$ for $C=0.25$ and $T=0.2$ for
$C=0.5$; starting in the Vlasov stable state, in our runs corresponding to $T=0.55$ for $C=0.25$ and $T=0.5$ for $C=0.5$, the rise of the
magnetization is also occurring quite early, due to finite size effects. But now the system does not settle to a quasi-stationary unmagnetized
state, with just a slow evolution characterized by a progressive shift of the momentum distribution due to the slow increase of the average
momentum. In fact, we see that the magnetization settles in a state with strong and rapid oscillations. We argue that these oscillations are
due to the separation of the oscillators in two or more groups, each one characterized by a different average momentum of its components. In
particular, we argue that there is a group of oscillators with a more or less vanishing average momentum, similarly to what happens for the
whole population of oscillators in the Hamiltonian case, where the average momentum is conserved and remains equal to the initial vanishing
value; besides this group of more or less standing oscillators, there is at least another group that moves with a finite average
momentum that keeps increasing, in order to satisfy Eq. (\ref{eqavermom}). If both groups are not uniformly distributed between $0$ and
$2 \pi$, then we deduce that we have an oscillatory behavior of $m(t)$, with the maximum when the peak of two groups coincide and the minimum
when they form an angle of $\pi$. 
Obviously this is a simplified and approximate explanation. We do not expect to have a bunch of oscillators circling around and having, at
any given time, the same velocity, but we do argue that it is a rough description that does not go very far from what happens. In support of
this, we present the plots in Fig. \ref{fig:C0p25-Ptheta}, showing the distribution of the angles $\theta$ in the upper panel and of that of
the momenta $p$ in the lower panel, taken during the simulation of the dynamics started in the Vlasov unstable state for $C=0.25$, i.e., the
one with initial temperature $T=0.25$, whose magnetization is given in the inset of Fig. \ref{fig:C0p25-m} for the run with
$N=5\times 10^5$ particles.  The momentum distribution $P(p)$ is defined in Eq.~(\ref{defveldistr}), while the angle distribution is
defined by the analogous expression:
\begin{equation}
P(\theta, t) = \int_{-\infty}^\infty {\rm d}p~ f(\theta, p, t). 
\end{equation}
The normalizing procedure for $P(\theta, t)$ is the same as that for $P(p, t)$ described before. The
$\theta$-distribution is taken at two different but close time values: one time in which $m(t)$ is at one of the peaks during its
oscillations, and the following close time in which $m(t)$ attains the lowest value before raising again. The momentum distribution is taken
at the time $t=400$. Like in the other plots,
in Fig.~\ref{fig:C0p25-Ptheta} the time dependence of the distributions $P(p, t)$ and $P(\theta, t)$ is not
explicitly indicated.
The fact that a higher magnetization value corresponds to a less uniform $\theta$-distribution than a lower magnetization is of course expected,
and in fact the two plots of Fig. \ref{fig:C0p25-Ptheta} have to be considered together. The momentum distribution in the lower panel gives
support to our picture of different group of oscillators with different average velocities, one of which being very small. The groups with
high velocities, then, produce a high magnetization when they are in phase with the almost still group, and a low magnetization when they are
in opposite phase. 

\section{Discussion and conclusions}
\label{discuss}

The dynamics of some physical systems can be modeled effectively by non-Hamiltonian equations of motion.  When these equations contain
long-range interactions, as is the case of the system described in the Introduction,  one may ask if, similarly to what occurs
in Hamiltonian systems, it is possible to witness some peculiar properties connected to the long-range character of the interaction.
In this work, we have studied the relaxation dynamics resulting from Eqs. (\ref{eq:eom}), in which there is both a Hamiltonian part and a
non-Hamiltonian part in the force, with the relative weight determined by the parameter $C$. The system with $C=1$, when only the
non-Hamiltonian force is present, has been considered in Ref. \cite{Bachelard:2019}, and it was found that, similarly to the Hamiltonian
model, the dynamics shows a slow relaxation, although with important differences. The aim of this work has been mainly an analysis of the
relaxation process as a function of the relative contribution of the Hamiltonian and the non-Hamiltonian interactions.

The first consideration one has to make, although it appears almost obvious, is that as soon as there is a non-Hamiltonian part in the
force, there is no more a Boltzmann-Gibbs equilibrium state to which the system is supposed to relax at large times.  Numerous
studies of long-range Hamiltonian systems have shown that, although it is often necessary to wait for a long time (a time that diverges with
the system size), these systems do relax to Boltzmann-Gibbs equilibrium. The fact that in some cases, e.g., for very large
self-gravitating systems, the required time could be larger than the age of the Universe, does not change this fact. When a non-Hamiltonian
component is present, we have to rely on the equations determining the evolution of the distribution function, like the Vlasov equation,
or the Lenard-Balescu equation at the following order of approximation to the Vlasov equation,  but we do not have any a priori clue as to
where the system relaxes at long times.

We have found that the slow relaxation of the system shows a change of character as one varies the parameter $C$ that determines
the relative weight of the non-Hamiltonian force. In fact, considering the values of $C$ that have been investigated, our simulations
have shown that for $C\leq 1/2$, the magnetization $m(t)$ presents
strong oscillations, implying that during the relaxation, the distribution function $f(\theta,p,t)$ is not quite uniform
in time, but that it has an almost periodic variation. On the other hand, for $C>1/2$, the magnetization, after an initial
transient, gets more or less rapidly to a practically vanishing value. We have proposed the following explanation for these
behaviors. When $C$ is close to $1$, i.e., when the non-Hamiltonian part of the interaction is dominant, its repulsive property
when two particles are close forbids a stable formation of a clustered state, which would be necessary to develop a finite magnetization.
When $C$ is smaller than $1/2$, and there is a strong contribution of the Hamiltonian attractive interaction, the particles of the
system separate in two or more groups; each group is composed of particles that are not uniformly distributed between $0$ and $2\pi$,
i.e.,  which presents a degree of clustering, but the different groups have different average momentum. This simple picture, for which
we have given support through the data obtained from simulations (see Fig. \ref{fig:C0p25-Ptheta}), explains
the rapid magnetization oscillations.

\begin{figure}[!htp]
\hspace{-2cm}\vspace{-0.9cm}\includegraphics[width=15cm]{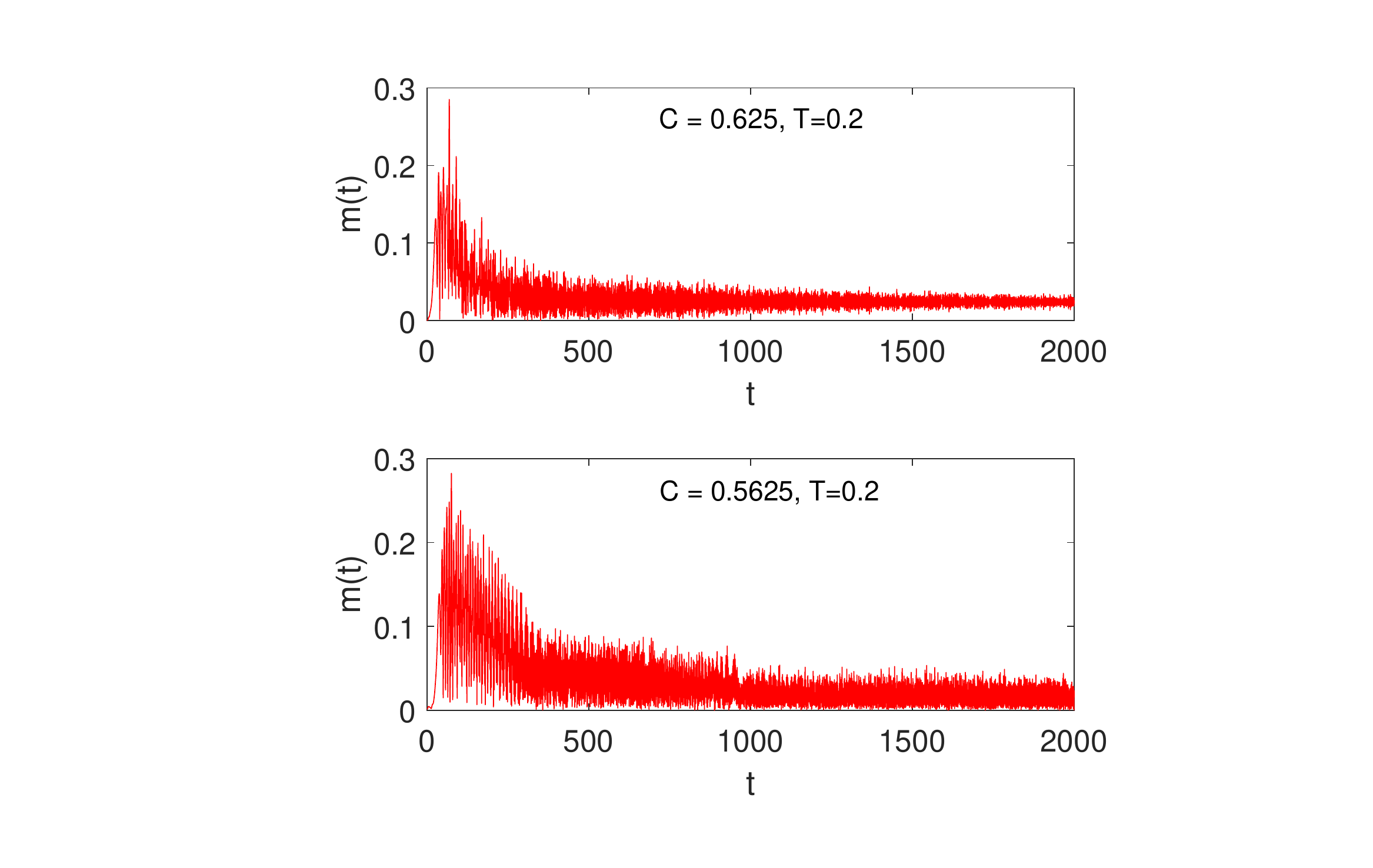}
\caption{Magnetization $m(t)$ as a function of time under the mixed dynamics~(\ref{eq:eom}) with, respectively, $C=0.625$ (upper panel)
and $C=0.5625$ (lower panel); the dynamics is initiated
with the coordinates $\theta_i$ distributed independently and uniformly in $[-\pi,\pi]$ and with the momenta $p_i$ sampled independently
from the Gaussian distribution~(\ref{eq:f0-Gaussian}). The values of the parameter $T\equiv \sigma^2$ is $T=0.2$ in both cases.
The system size is $N=2\times 10^5$.}
\label{fig:C0p625}
\end{figure}

A precise characterization of the crossover, as a function of $C$, of the behaviour of the magnetization (in particular, the determination
of whether such a crossover occurs in a very narrow range of $C$ that allows to interpret it as a sort of phase transition) would require
investigation at more closely spaced values of $C$ than the one presented here. We have performed a first step in this direction by studying
the dynamics of the magnetization for two further values of $C$, i.e., $C=0.625$ and $C=0.5625$, for the Vlasov-stable initial conditions,
$T=0.2$. The first value of $C$ is halfway between the two values $C=0.5$ and $C=0.75$ considered in the main
part of our analysis, and the second one is halfway between $C=0.5$ and the first value, $C=0.625$. The reason for this choice was the following.
We first checked if the strong oscillations present at $C=0.5$ and practically absent at $C=0.75$ were present, and in what measure, halfway between
these two values. Having found that the behavior at $C=0.625$ is much more similar to that at $C=0.75$ than at $C=0.5$, with very few oscillations
of an almost vanishing magnetization (see upper panel of Fig. \ref{fig:C0p625}), we have then considered the dynamics at $C=0.5625$. Also in this
latter case we have found that the magnetization value is quite small and with very few oscillations. In Table I, this analysis is supported with
the values of the standard deviation of the magnetization related to the plots in, respectively, the upper panel of Fig. \ref{fig:C0p25-m}
($C=0.25$, $T=0.25$), the upper panel of Fig. \ref{fig:C0p5-m} ($C=0.5$, $T=0.2$), the bottom and the upper panel of Fig. \ref{fig:C0p625}
($C=0.5625$, $T=0.2$ and $C=0.625$; $T=0.2$) and the upper panel of Fig. \ref{fig:C0p75-m} ($C=0.75$, $T=0.2$). The standard deviation has been
computed by considering the magnetization values between times $t=500$ and $t=2000$. The table shows that the standard deviation has a marked
decrease passing from $C=0.5$ to $C=0.5625$. This analysis is not sufficient, of course, to infer that the crossover is extremely sharp, but it puts
in evidence that the range of $C$ where it occurs is rather narrow and close to $C=0.5$. 

The fact that the crossover happens at or very close to $C=0.5$ is not due, in our opinion, to any particular property of
the equations of motion of the system, but to the fact that crossing that value one goes, as already emphasized, from a situation in which the
Hamiltonian part is predominant to one where the non-Hamiltonian part is predominant. We can argue that this change in the relaxation properties
should be present also in other models sharing with the model studied in this work the property that the non-Hamiltonian part of the force
tends to separate two particles, in contrast to the Hamiltonian force. Obviously, this anticipation should be supported with specific studies.
On the contrary, if also the non-Hamiltonian part of the force favors particle clustering, it is likely that the dependence of the relaxation
properties on the value of $C$ would be different than the one presented in this work.

\begin{table}[!htp]
\centering
\begin{tabular}{|c|c|}
\hline 
$C$  & $\mathrm{std}[m]$   \\ \hline 
0 25 & 0.110 \\ \hline 
0.5 &  0.073 \\ \hline 
0.5625 & 0.014  \\ \hline 
0.625 &  0.007 \\ \hline 
0.75 &  0.003  \\ \hline
\end{tabular}
\caption{Standard deviation of the magnetization as a function of the parameter $C$ (more details in the text).}
\label{Table}
\end{table}

A marked difference with respect to the Hamiltonian case is the presence of much more pronounced effects, when $C\ne 0$, of the finite
size effects. It is known that in one-dimensional systems, the Lenard-Balescu equation,  which determines the correction at order $1/N$,
with respect to the Vlasov equation,  for the evolution of the distribution function, shows that actually, this correction vanishes. With
a detailed derivation given in Appendix C, we have shown that with a non-Hamiltonian part in the interaction,
the Lenard-Balescu correction does not vanish. This should explain why also in the case of initial distributions that are Vlasov stable,
the system evolves quite rapidly, and to appreciate a significant difference in the early stages of the dynamics between Vlasov stable and
Vlasov unstable distributions it is necessary to consider systems with a rather large number of particles. We have also shown
in Appendix B that the time derivative of the total momentum, proportional to $Cm^2$, stems from the nonlinear
part of the Vlasov equation.

In one-dimensional systems, like the one studied in this work, we expect that finite-size effects will be relevant in general due to the
fact that the leading correction to the Vlasov equation, of order $1/N$, vanishes in the Hamiltonian case and does not vanish in the
non-Hamiltonian case, although the size of the correction in the case in which a non-Hamiltonian force is present might strongly vary among
the various models. What one should expect at higher dimension is not clear from the study of a one-dimensional system, since in two or more
dimensions, the Lenard-Balescu term does not vanish in general for Hamiltonian systems.

In the analysis of our simulations, we have focussed mainly on the features appearing at relatively early times, but we have to consider
the following if we want to discuss the implications of our results for the behavior of concrete systems. Any dynamical evolution develops
on time scales that increase with the size of the system, and therefore in a real system, what we have observed will occur at much larger
times, since the number of particles will be much larger than that employed in our simulations. Thus, e.g., the strong magnetization
oscillations that are present in a range of values of the parameter $C$, should be a relevant characteristic of a real system.

In Hamiltonian long-range systems, it has been found that, apart from some details that depend on the concrete form of the interaction,
there are many properties, both at equilibrium and out of equilibrium, that are shared by all of them. Here, as already remarked,
we do not have an equilibrium state and then we concentrate on the dynamical properties, but we can argue that also in the
non-Hamiltonian case, the most relevant peculiarities will be common. However, this is something that should be verified through
investigation with other forms of interaction; we feel that this line of research deserves to be further developed.

\begin{acknowledgements}
A.C. acknowledges financial support from INFN (Istituto Nazionale di Fisica Nucleare) through the projects DYNSYSMATH and ENESMA.
S.G. acknowledges support from the Science and Engineering Research Board (SERB), India, under SERB-TARE scheme
Grant No. TAR/2018/000023, SERB-MATRICS scheme Grant No. MTR/2019/000560, and SERB-CRG Scheme
Grant No. CRG/2020/000596. He also thanks ICTP – The Abdus Salam International Centre for Theoretical Physics,
Trieste, Italy, for support under its Regular Associateship scheme.
\end{acknowledgements}

\section*{Declarations}

{\bf Conflict of interest} $\,\,\,\,\,$ The authors declare that they have no conflict of interest.
\par\noindent
{\bf Data availability} $\,\,\,\,\,$ All data generated or analysed during this study are included in this article.

\section*{Appendix A: Derivation of Eqs. (\ref{eqforometo0R}) and (\ref{eqforometo0I})}
\label{append_threshgauss}
Here, we derive the two relations (\ref{eqforometo0R}) and (\ref{eqforometo0I}),  which are used to compute the threshold $\sigma_c(C)$
for the Gaussian distribution (\ref{eq:f0-Gaussian}). By using in Eq. (\ref{eq:dispersiongauss}) the expression (\ref{eqforometo0}), a 
consequence of the Plemelj formula (\ref{pleme}), and by separating the real and the imaginary part, we obtain
\begin{eqnarray}
\label{eqforometo0Rap}
\left( 1-C\right) \left[1 +\frac{\omega_R}{\sqrt{2\pi \sigma^2}}P\int_{-\infty}^{+\infty} \dd p \,
\frac{e^{-\frac{p^2}{2\sigma^2}}}{p-\omega_R} \right] +C\frac{\omega_R}{\sqrt{2\pi \sigma^2}}\pi e^{-\frac{\omega_R^2}{2\sigma^2}}
&=& 2\sigma^2,
\\
\label{eqforometo0Iap}
C \left[ 1+ \frac{\omega_R}{\sqrt{2\pi \sigma^2}}P\int_{-\infty}^{+\infty} \dd p \, \frac{e^{-\frac{p^2}{2\sigma^2}}}{p-\omega_R}
\right] -\left(1-C\right)\frac{\omega_R}{\sqrt{2\pi \sigma^2}}\pi e^{-\frac{\omega_R^2}{2\sigma^2}}
&=& 0 \, .
\end{eqnarray}
We need a manageable expression for the principal part of the integral.  For this purpose, with a simple change of variable in the integral,
i.e., $p=\omega_R + y$, we get 
\begin{equation}
\label{principal_1}
P\int_{-\infty}^{+\infty} \dd p \, \frac{e^{-\frac{p^2}{2\sigma^2}}}{p-\omega_R}=
e^{-\frac{\omega_R^2}{2\sigma^2}}P\int_{-\infty}^{+\infty} \dd y \, \frac{1}{y} e^{-\frac{y^2}{2\sigma^2}}e^{-\frac{\omega_R y}{\sigma^2}}
\, .
\end{equation}
The limiting procedure implied in the evaluation of the principal value can be performed by integrating from $0$ to $+\infty$ and by
substituting $e^{-\frac{\omega_R y}{\sigma^2}}$ with $e^{-\frac{\omega_R y}{\sigma^2}}-e^{\frac{\omega_R y}{\sigma^2}}
=-2\sinh\left(\frac{\omega_R y}{\sigma^2}\right)$.  We get
\begin{eqnarray}
P\int_{-\infty}^{+\infty} \dd p \, \frac{e^{-\frac{p^2}{2\sigma^2}}}{p-\omega_R}&=&
-2 e^{-\frac{\omega_R^2}{2\sigma^2}}\int_0^{+\infty} \dd y \, \frac{1}{y}\sinh\left(\frac{\omega_R y}{\sigma^2}\right)
e^{-\frac{y^2}{2\sigma^2}} \nonumber \\ &=&
-2 e^{-\nu^2}\int_0^{+\infty} \dd w \, \frac{1}{w}\sinh\left(2\nu w\right) e^{-w^2} \, ,
\label{principal_2}
\end{eqnarray}
where the right hand side of the second equality has been obtained with the change of variable $y=\sqrt{2\sigma^2}w$ in the integral and by
defining, as in the main text, $\nu$ by $\omega_R = \sqrt{2\sigma^2}\nu$. It is more convenient, for the search of the numerical solution, to
write the integral in the second equality of (\ref{principal_2}) in another form. To do this, we perform the derivative with respect
to $\nu$ of the integral, to get 
\begin{equation}
\frac{\partial}{\partial \nu} \int_0^{+\infty} \dd w \, \frac{1}{w}\sinh\left(2\nu w\right) e^{-w^2}
= 2 \int_0^{+\infty} \dd w \, \cosh\left(2\nu w\right) e^{-w^2} = \sqrt{\pi} e^{\nu^2}\, .
\label{principal_3}
\end{equation}
Therefore, noting that for $\nu=0$, the integral in the second equality of (\ref{principal_2}) vanishes, we have
\begin{equation}
P\int_{-\infty}^{+\infty} \dd p \, \frac{e^{-\frac{p^2}{2\sigma^2}}}{p-\omega_R}=
-2 \sqrt{\pi} e^{-\nu^2}\int_0^{\nu} \dd t \,  e^{t^2} \, .
\label{principal_4}
\end{equation}
Substituting the above result in Eqs. (\ref{eqforometo0Rap}) and (\ref{eqforometo0Iap}), and using the definition of $\nu$, we obtain
the equations in the main text, i.e., Eqs. (\ref{eqforometo0R}) and (\ref{eqforometo0I}).

\section*{Appendix B: Nonlinear terms in the Vlasov equation}

We consider the cases $C=0$ and $C=1$.
We write $f(\theta,p,t)=f_0(p)+\delta f(\theta,p,t)$, with $f_0(p)$ as in (\ref{distf0p}), and make the expansion
(\ref{eq:deltaf-expansion-fou}),  which for convenience we rewrite here
\begin{align}
\delta f(\theta,p,t)=\sum_{k=-\infty}^{+\infty} \widehat{\delta f}_k(p,t)e^{\ii k\theta} \, .
\label{eq:deltaf-expansion-fou_app}
\end{align}
Substituting in Eq. (\ref{eq:vlasov}), we obtain an equation for each Fourier component. The nonlinearity,  which clearly comes from the last
term in Eq. (\ref{eq:vlasov}), results in the coupling between different Fourier components $\widehat{\delta f}_k(p,t)$. We consider the
force term in the two cases $C=0$ and $C=1$. For $C=0$, we have
\begin{eqnarray}
\label{force_ceq0}
F[f](\theta,t) &=& \int_{-\infty}^{+\infty} \dd p' \int_0^{2\pi} \dd \theta' \, \sin (\theta' - \theta) f(\theta',p',) \nonumber \\
&=& \ii \pi \sum_{k=-\infty}^{+\infty} e^{\ii k \theta}\left( \delta_{k,1} - \delta_{k,-1}\right)
\int_{-\infty}^{+\infty} \dd p' \widehat{\delta f}_k(p',t) \, ,
\end{eqnarray}
while for $C=1$, we have
\begin{eqnarray}
\label{force_ceq1}
F[f](\theta,t) &=& \int_{-\infty}^{+\infty} \dd p' \int_0^{2\pi} \dd \theta' \, \cos (\theta' - \theta) f(\theta',p',) \nonumber \\
&=& \pi \sum_{k=-\infty}^{+\infty} e^{\ii k \theta}\left( \delta_{k,1} + \delta_{k,-1}\right)
\int_{-\infty}^{+\infty} \dd p' \widehat{\delta f}_k(p',t) \, ,
\end{eqnarray}
Plugging in the Vlasov equation (\ref{eq:vlasov}), we get the equations for the Fourier components $\widehat{\delta f}_k(p,t)$.
In a few steps, one obtains
\begin{eqnarray}
\label{vlasov-fou-nonlin}
&&\frac{\partial \widehat{\delta f}_k}{\partial t} + \ii k p \widehat{\delta f}_k +\pi \left(A_C \delta_{k,1} + B_C \delta_{k,-1}\right)
\frac{\partial f_0}{\partial p} \int_{-\infty}^{+\infty} \dd p' \, \widehat{\delta f}_k(p') \\
&+& \pi \left( A_C \frac{\partial \widehat{\delta f}_{k-1}}{\partial p}\int_{-\infty}^{+\infty} \dd p' \, \widehat{\delta f}_1(p')
+B_C \frac{\partial \widehat{\delta f}_{k+1}}{\partial p}\int_{-\infty}^{+\infty} \dd p' \, \widehat{\delta f}_{-1}(p') \right) = 0 \, ,
\nonumber
\end{eqnarray}
where $A_0 = \ii$, $B_0 = -\ii$, $A_1=B_1=1$. The linear term in the force, i.e. , the last term in the first row, appears only
for $k=\pm 1$, and ignoring the nonlinear term, we get the linearized equation for $\widehat{\delta f}_{\pm 1}$ from
which one can obtain the dispersion relations treated in the main text. Here, we are interested in the equation for $k=0$, which is
\begin{equation}
\label{vlasov-fou-nonlin_0}
\frac{\partial \widehat{\delta f}_0}{\partial t}
+ \pi \left( A_C \frac{\partial \widehat{\delta f}_{-1}}{\partial p}\int_{-\infty}^{+\infty} \dd p' \, \widehat{\delta f}_1(p')
+B_C \frac{\partial \widehat{\delta f}_{1}}{\partial p}\int_{-\infty}^{+\infty} \dd p' \, \widehat{\delta f}_{-1}(p') \right) = 0 \, .
\end{equation}
We see that the zero-th Fourier component is acted upon only by the nonlinear term. The equation shows that the integral of
$\widehat{\delta f}_0(p,t)$ is conserved, and it is identically $0$, because of the normalization of $f(\theta,p,t)$. However, a
difference between $C=0$ and $C=1$ arises when we compute the integral of this equation multiplied by $p$. From the expansion
(\ref{eq:deltaf-expansion-fou_app}), taking into account the definition (\ref{distf0p}), we obtain
\begin{equation}
\label{intdeltaf0}
\int_{-\infty}^{+\infty} \dd p \int_0^{2\pi} \dd \theta \, p f(\theta,p,t) =
\int_{-\infty}^{+\infty} \dd p \, p \left( P(p) + 2\pi \widehat{\delta f}_0(p,t) \right) \, .
\end{equation}
Thus, from Eq. (\ref{vlasov-fou-nonlin_0}), we get
\begin{eqnarray}
\label{eq_for_deltaf0}
\frac{\dd}{\dd t} \langle p \rangle &\equiv& \int_{-\infty}^{+\infty} \dd p \int_0^{2\pi} \dd \theta \, p
\frac{\partial}{\partial t}f(\theta,p,t) = 2\pi \int_{-\infty}^{+\infty} \dd p \, p \frac{\partial \widehat{\delta f}_0}{\partial t}
\\ &=&
2\pi^2 \left[ A_C \left(\int_{-\infty}^{+\infty} \dd p \, \widehat{\delta f}_{-1}(p)\right) \left(\int_{-\infty}^{+\infty} \dd p' \,
\widehat{\delta f}_1(p')\right) \right. \nonumber 
\\ &&\left. \,\,\,\,\,\,\,\,\,\,\, +B_C \left( \int_{-\infty}^{+\infty} \dd p \,
\widehat{\delta f}_{1}(p)\right) \left(\int_{-\infty}^{+\infty} \dd p' \, \widehat{\delta f}_{-1}(p')\right) \right] \, ,\nonumber
\end{eqnarray}
where an integration by parts has been used. Since $\widehat{\delta f}_{-1}=\widehat{\delta f}_1^*$, we can rewrite the last member,
obtaining
\begin{equation}
\label{nonlineareq1b}
\frac{\dd}{\dd t} \langle p \rangle
= 2\pi^2 (A_C + B_C) \left| \int_{-\infty}^{+\infty} \dd p \widehat{\delta f}_1(p,t) \right|^2 \, .
\end{equation}
From the values of $A_C$ and $B_C$ we see that for $C=0$, the right hand side vanishes, while for $C=1$, it is equal to the last member
of Eq. (\ref{nonlineareq1}). We thus see that the variation of the average momentum is obtained as a nonlinear correction to the
linearized Vlasov equation.

\section*{Appendix C: Lenard-Balescu equation for non-Hamiltonian systems}
\label{append_lbnonham}

In the main text, we have shown that systems with long-range interactions, even if of non-Hamiltonian origin, share with the more common
Hamiltonian systems the property of being described, in the thermodynamic limit $N \to \infty$, by the Vlasov equation. When one considers
the corrections in the dynamics due to the collisional effects, which are of order $1/N$ with respect to the mean-field interaction embodied
in the Vlasov equation, a difference arises.  As will be shown below,  this difference is essentially related to the different properties of
the Fourier coefficients of the two-body interaction; the difference is particularly relevant for 1D systems like ours.

The kinetic equation describing the slow evolution, due to the collisional effects, of Vlasov-stable homogeneous one-particle distribution
functions for Hamiltonian systems is the Lenard-Balescu equation \cite{Nicholson:1992}. It is known that for 1D systems, the right-hand side
of the Lenard-Balescu equation, which defines the evolution operator, vanishes.  This implies that collisional effects are of higher order
than $1/N$. We will see that this is not the case when the interactions are non-Hamiltonian. Therefore, in this case, the corrections to the
Vlasov equation are considerably more important.

It is possible to adopt more than one procedure to derive the Lenard-Balescu equation. One procedure starts from the first two equations of
the so-called Bogoliubov-Born-Green-Kirkwood-Yvon (BBGKY) hierarchy.  Then, by introducing suitable approximations for the dynamical evolution
of the two-particle distribution function, this set of equations is expressed in terms of the one-particle distribution function, thus
obtaining a closed equation for the latter. Here, we have chosen to follow an alternative route \cite{Campa:2014,Campa:2009}, in which one
starts from the so-called Klimontovich equation \cite{Campa:2009,Nicholson:1992}. We have tried to make this description as self-contained as
possible, by writing explicitly also some expressions and definitions that are well established in the literature and text books.

In view of the application to our 1D model of rotators, in our derivation we employ, as canonical coordinates of the particles, the angle
$\theta_i$ and the angular momentum $p_i$. The Klimontovich equation describes the time evolution of the following one-particle density
function:
\begin{equation}
\label{def_klim}
f_{{\rm d}}(\theta,p,t) = \frac{1}{N}\sum_{i=1}^N \delta(\theta - \theta_i(t)) \delta(p-p_i(t)) \, .
\end{equation}
Here, the set $(\theta_i,p_i)$ are the canonical coordinates of the $N$ particles, while $(\theta,p)$ without subscript are the Eulerian
coordinates of the two-dimensional one-particle phase space. In spite of the fact that due to the presence of the Dirac delta function,
$f_{{\rm d}}$ is a singular function, its dynamical evolution is perfectly defined through that of the canonical coordinates of the particles.
For Hamiltonian systems with Hamiltonian
\begin{equation}
\label{gen_ham}
H = \sum_{i=i}^N \frac{p_i^2}{2} + U(\{\theta_i\}) =  \sum_{i=i}^N \frac{p_i^2}{2} + \sum_{i<j=1}^N V(\theta_i - \theta_j) \, ,
\end{equation}
and equations of motion given by
\begin{equation}
\label{gen_equ}
\frac{\dd \theta_i}{\dd t} = p_i \, ,\,\,\,\,\,\,\,\,\,\,\,\,\,\,\,\,\,\,\, \frac{\dd p_i}{\dd t} = -\frac{\partial U}{\partial \theta_i} \, ,
\end{equation}
it is not difficult to show that the evolution of $f_{{\rm d}}(\theta,p,t)$ is governed by the Klimontovich equation:
\begin{equation}
\label{klim_eq}
\frac{\partial f_{{\rm d}}}{\partial t} + p \frac{\partial f_{{\rm d}}}{\partial \theta} - \frac{\partial v(\theta,t)}{\partial \theta}
\frac{\partial f_{{\rm d}}}{\partial p} = 0 \, ,
\end{equation}
where the potential $v(\theta,t)$ is given by
\begin{equation}
\label{pot_klim}
v(\theta,t) = N \int \dd \theta' \dd p \, V(\theta - \theta') f_{{\rm d}}(\theta',p,t) \, .
\end{equation}
The Klimontovich equation (\ref{klim_eq}) is exact, but is useless in practice, since its solution requires the solution of the equations of
motion. In fact, as it is clear from its definition (\ref{def_klim}), there is an implicit dependence of $f_{{\rm d}}$ on time through that
of the canonical coordinates of the $N$ particles, a dependence that is defined by the solution of the equations of motion. The Klimontovich
equation can become useful when used to derive approximate equations, i.e., kinetic equations for the one-particle distribution function.
Before proceeding in this direction, we have to generalize the Klimontovich equation to non-Hamiltonian systems. In this case, a potential
energy $U$ does not exist, and the Hamilton equations of motion are substituted by
\begin{equation}
\label{gen_equ_non}
\frac{\dd \theta_i}{\dd t} = p_i \, ,\,\,\,\,\,\,\,\,\,\,\,\,\,\,\,\,\,\,\, \frac{\dd p_i}{\dd t} = \sum_{j=i}^N
{\cal F}(\theta_i - \theta_j) \, ,
\end{equation}
where ${\cal F}(\theta_i - \theta_j)$ is the force exerted on the $i$-th particle by the $j$-th particle. The
Klimontovich equation (\ref{klim_eq}) is substituted by\footnote{Actually, there is a caveat that depends on whether in the equations
of motion of the model the term with $j=i$ appears or not on the right hand side of the second equation in (\ref{gen_equ_non}). This will be
clarified at the end of the procedure, whose development does not depend crucially on this issue.}
\begin{equation}
\label{klim_eq_non}
\frac{\partial f_{{\rm d}}}{\partial t} + p \frac{\partial f_{{\rm d}}}{\partial \theta} +F_{{\rm d}}(\theta,t)
\frac{\partial f_{{\rm d}}}{\partial p} = 0 \, ,
\end{equation}
where now
\begin{equation}
\label{force_klim}
F_{{\rm d}}(\theta,t) = N \int_0^{2\pi} \dd \theta' \int_{-\infty}^{+\infty} \dd p \, {\cal F}(\theta - \theta') f_{{\rm d}}(\theta',p,t) \, .
\end{equation}

By a smoothing procedure, realized by averaging over a distribution function $\rho_N(\theta_1,p_1,\theta_2,p_2,\dots,\theta_N,p_N)$, one can
obtain from $f_{{\rm d}}(\theta,p,t)$ a smooth function $f(\theta,p,t)$:
\begin{equation}
\label{smooth}
f(\theta,p,t) = \int \dd \theta_1\dots\dd \theta_N \dd p_1 \dots \dd p_N \, \rho_N(\theta_1,p_1,\theta_2,p_2,\dots,\theta_N,p_N)
f_{{\rm d}}(\theta,p,t) \, .
\end{equation}
The explicit form of the function $\rho_N$ is not relevant for our procedure. Physically, it may be taken to represent the $N$-particle
distribution function associated to some given macroscopic state of the system\footnote{The function $\rho_N$ is assumed to be normalized to
unity. We note that this implies that $f(\theta,p,t)$ is normalized to unity,  as is also the case for $f_{{\rm d}}(\theta,p,t)$.}. The
smoothing by itself does not provide a real simplification of the dynamical problem, since for each point $(\theta_1,p_1,\dots,\theta_N,p_N)$
of the $N$-dimensional dynamical phase space of the system, one has to consider the implicit dependence on time of $f_{{\rm d}}(\theta,p,t)$,
in the integral of Eq. (\ref{smooth}), coming from the equations of motion with that point as initial conditions. However, the simplification
can be obtained as follows. First, one defines the deviation $\delta f(\theta,p,t)$ from $f_{{\rm d}}(\theta,p,t)$ by
\begin{equation}
\label{deltadef}
f_{{\rm d}}(\theta,p,t) = f(\theta,p,t) +\delta f(\theta,p,t) \, .
\end{equation}
Next we substitute $f_{{\rm d}} = f + \delta f$ in the Klimontovich equation (\ref{klim_eq_non}) and in Eq. (\ref{force_klim}). The latter
substitution defines $F_{{\rm d}}(\theta,t) = F(\theta,t) + \delta F (\theta,t)$ (with obvious meaning of the symbols). Now, averaging as in
the right hand side of Eq. (\ref{smooth}), an operation that can be denoted with angular brackets for brevity, one obtains
\begin{equation}
\label{klim_aver}
\frac{\partial f}{\partial t} + p \frac{\partial f}{\partial \theta} +F(\theta,t)
\frac{\partial f}{\partial p} = - \langle \delta F(\theta,t) \frac{\partial \delta f(\theta,p,t)}{\partial p} \rangle \, .
\end{equation}
This equation is still exact, and as such of little use, but it offers the possibility to obtain kinetic equations by suitable approximating
the right hand side. We note that by simply neglecting the right hand side, we have the Vlasov equation for $f(\theta,p,t)$. The Lenard-Balescu
equation concerns the corrections to the dynamical evolution of a homogeneous distribution function which is a stable stationary solution of
the Vlasov equation. Therefore, we apply Eq. (\ref{klim_aver}) to the case in which $f(\theta,p,t)$ actually does not depend on $\theta$,
and is moreover a stable stationary solution of the Vlasov equation. Since $f$ is homogeneous, $F(\theta,t)$ vanishes, and thus, we start
from\footnote{We can take the derivative sign outside the angular brackets, since the average that these brackets indicate is over the
Lagrangian coordinates of the particles, and not over the Eulerian coordinates of the distribution function.}
\begin{equation}
\label{klim_aver_st}
\frac{\partial f}{\partial t} = - \langle \delta F(\theta,t) \frac{\partial \delta f(\theta,p,t)}{\partial p} \rangle
= - \frac{\partial}{\partial p} \langle \delta F(\theta,t) \delta f(\theta,p,t) \rangle \, .
\end{equation}
Now, the approximation is introduced in which the time evolution of the fluctuations appearing on the right hand side is determined according
to the linearized Vlasov equation. At the end, one arrives at a closed equation for $f$. It will be seen that, in spite of the appearance of
the coordinate $\theta$, on averaging, the right hand side will depend only on $p$ and $t$. The linearized Vlasov equation for the
evolution of $\delta f(\theta,p,t)$ is
\begin{equation}
\label{evoldeltaf}
\frac{\partial}{\partial t}\delta f(\theta,p,t) + p \frac{\partial}{\partial \theta} \delta f(\theta,p,t)
+\delta F(\theta,t) \frac{\partial f}{\partial p} = 0 \, ,
\end{equation}
where, in this equation, the time evolution of $f(p,t)$ has to be considered frozen; this is in the spirit of the Lenard-Balescu equation,
where it is assumed that the time evolution of the fluctuations (and of the two-particle correlation function usually
denoted by $g_2$) is much faster than
that of $f$. In the same spirit, in the right hand side of Eq. (\ref{klim_aver_st}), we will consider its long-time behavior (physically,
the fluctuations reach practically asymptotic values before the function $f$ changes appreciably). Thus, in the following, we will
write explicitly only the dependence on $p$ of the function $f$. More comments on this point may be found later. In (\ref{evoldeltaf}),
$\delta F(\theta,t)$ is given by:
\begin{equation}
\label{deltaforce}
\delta F(\theta,t) = N \int_0^{2\pi} \dd \theta' \int_{-\infty}^{+\infty} \dd p \, {\cal F}(\theta - \theta') \delta f(\theta',p,t) \, ,
\end{equation}
We consider a generic $2\pi$-periodic function ${\cal F}(\theta)$ developed in Fourier series as:
\begin{equation}
{\cal F}(\theta)=\sum_{k=-\infty}^{+\infty} c_k e^{\ii k \theta} \, ;
\label{expanforce}
\end{equation}
in the main text the Fourier expansion of the force has only $c_{\pm 1}$ different from zero and of order
$1/N$. The force ${\cal F}(\theta)$ being real requires that $c_{-k}=c_k^*$, where, as usual, the star denotes complex conjugation.
Furthermore, we remind that we are assuming the absence of a constant term in ${\cal F}(\theta)$, meaning that $c_0=0$.

In order to have $\delta f(\theta,p,t)$, we are going to solve the linearized Vlasov equation as an initial value problem, using the
Fourier-Laplace transformation defined by:
\begin{equation}
\label{fltranf}
\widetilde{\delta f}(k,p,\omega) = \frac{1}{2\pi} \int_0^\infty \dd t \, \int_0^{2\pi} \dd \theta \,
e^{-\ii (k\theta - \omega t)} \delta f(\theta,p,t) \, .
\end{equation}
As we know, this transformation is defined for ${\rm Im} (\omega)$ sufficiently large, and by its analytic continuation for the rest
of the complex-$\omega$ plane. The inversion formula is:
\begin{equation}
\label{ifltranf}
\delta f(\theta,p,t) = \frac{1}{2\pi} \sum_{k=-\infty}^{+\infty} \int_{{\cal C}} \dd \omega \, e^{\ii (k\theta - \omega t)}
\widetilde{\delta f}(k,p,\omega) \, ,
\end{equation}
where the path ${\cal C}$ of integration in the complex-$\omega$ plane is a line parallel to the real axis that passes above
all singularities of $\widetilde{\delta f}(k,p,\omega)$ (or, any other path that can be obtained by this by deforming it and
without crossing any of the singularities of $\widetilde{\delta f}(k,p,\omega)$). The Fourier-Laplace transform of
Eq. (\ref{evoldeltaf}) is
\begin{equation}
\label{linvlatran}
-\ii \omega \widetilde{\delta f}(k,p,\omega) + \ii k p \widetilde{\delta f}(k,p,\omega)
+\widetilde{\delta F}(k,\omega)\frac{\partial f}{\partial p} = \widehat{\delta f}(k,p,0) \,
\end{equation}
where on the right hand side, we have the Fourier transform of $\delta f(\theta,p,t)$ at $t=0$:
\begin{equation}
\label{fouinit}
\widehat{\delta f}(k,p,0) = \frac{1}{2\pi} \int_0^{2\pi} \dd \theta \,
e^{-\ii k\theta} \delta f(\theta,p,0) \, .
\end{equation}
Using the Fourier-Laplace transform of Eq. (\ref{deltaforce}), i.e.,
\begin{equation}
\label{reldforcedf}
\widetilde{\delta F}(k,\omega) = 2\pi N c_k \int \dd p \, \widetilde{\delta f}(k,p,\omega) \,
\end{equation}
(where the expansion (\ref{expanforce}) has been exploited), the solution of Eq. (\ref{linvlatran}) is given by
\begin{equation}
\label{sol2linvlatran}
\widetilde{\delta f}(k,p,\omega) = \frac{\widehat{\delta f}(k,p,0)}{\ii (kp -\omega)} -2\pi N \frac{c_k f'(p)}
{\ii (kp - \omega )} \int \dd p' \, \widetilde{\delta f}(k,p',\omega) \, .
\end{equation}
where $f'(p)$ denotes the derivative of $f(p)$.
Integrating with respect to $p$ and defining the dielectric function $\epsilon(k,\omega)$ for ${\rm Im}(\omega) >0$ by
\begin{equation}
\label{dielPR}
\epsilon(k,\omega) = 1 + 2\pi \ii N c_k \int \dd p \, \frac{f'(p)}{\omega - k p} \, ,
\end{equation}
and by the analytic continuation of this expression for ${\rm Im}(\omega)\le 0$,
we obtain for the solution of Eq. (\ref{linvlatran}) the expression
\begin{equation}
\label{reldffluc}
\widetilde{\delta f}(k,p,\omega) = \frac{\widehat{\delta f}(k,p,0)}{\ii (kp -\omega)} -2\pi N \frac{c_kf'(p)}
{\ii (kp - \omega )} \frac{1}{\epsilon(k,\omega)} \int \dd p' \, \frac{\widehat{\delta f}(k,p',0)}{\ii (kp' - \omega)} \, ,
\end{equation}
i.e., $\widetilde{\delta f}(k,p,\omega)$ as a function of the Fourier transform of the initial time fluctuation.
Analogously, we find that $\widetilde{\delta F}(k,\omega)$ is given by
\begin{equation}
\label{reldforcefluc}
\widetilde{\delta F}(k,\omega) = 2\pi N c_k\frac{1}{\epsilon(k,\omega)} \int \dd p' \, \frac{\widehat{\delta f}(k,p',0)}
{\ii (kp' - \omega)} \, ,
\end{equation}
With these expressions, we can now evaluate the right hand side of Eq. (\ref{klim_aver_st}). Inverting the Fourier-Laplace
transform we have, for the quantity inside the angular brackets in this equation, the expression
\begin{eqnarray}
\label{invdeltadelta}
&&\delta F(\theta,t) \delta f(\theta,p,t) \\ &=& \frac{1}{(2\pi)^2} \sum_k \sum_{k'} \int_{{\cal C}_1} \dd \omega_1 \, \int_{{\cal C}_2}
\dd \omega_2 \, e^{\ii (k\theta - \omega_1 t)} e^{\ii (k'\theta - \omega_2 t)} \widetilde{\delta F}(k,\omega_1)
\widetilde{\delta f}(k',p,\omega_2) \, , \nonumber
\end{eqnarray}
where the paths of integration ${\cal C}_1$ and ${\cal C}_2$ must pass above the singularities of $\widetilde{\delta F}(k,\omega_1)$
and of $\widetilde{\delta f}(k',p,\omega_2)$, respectively. We can argue that there are no such singularities in the upper-half plane
with ${\rm Im}(\omega)>0$. In fact, from Eqs. (\ref{reldffluc}) and (\ref{reldforcefluc}), we see that $\widetilde{\delta f}(k,p,\omega)$
and $\widetilde{\delta F}(k,\omega)$ have singularities on the real $\omega$ axis; besides, the dielectric function $\epsilon(k,\omega)$
does not have zeros for ${\rm Im}(\omega) >0$, since we are studying the slow-time evolution of a Vlasov stable $f(p)$. Therefore, the paths
${\cal C}_1$ and ${\cal C}_2$ can be, in the corresponding complex plane, lines parallel to the real axes and with any positive imaginary
part. It is now useful to make in Eq. (\ref{invdeltadelta}) a change of integration variables from $(\omega_1,\omega_2)$ to
$(\omega_1,\omega=\omega_1 +\omega_2)$, obtaining:
\begin{eqnarray}
\label{invdeltadeltab}
&&\delta F(\theta,t) \delta f(\theta,p,t) \\ &=& \frac{1}{(2\pi)^2} \sum_k \sum_{k'} \int_{{\cal C}} e^{-\ii \omega t} \dd \omega \,
\int_{{\cal C}_1} \dd \omega_1 \, e^{\ii (k+k')\theta} \widetilde{\delta F}(k,\omega_1)
\widetilde{\delta f}(k',v_1,\omega - \omega_1) \, . \nonumber
\end{eqnarray}
Since we have ${\rm Im}(\omega_2)>0$, the path ${\cal C}$ is above the path ${\cal C}_1$. As remarked above, we are interested in the
long-time behavior of the quantity on the left hand side. This sought asymptotic value at $t=+\infty$ implies that the function
multiplying $e^{-\ii \omega t}$ in the above expression, (let us call it $G(\omega)$) has a pole for $\omega =0$; this pole has a residue
equal to the asymptotic value we want to obtain. There will be other poles (with corresponding residues) for $\omega$ values with
${\rm Im}(\omega)<0$, corresponding to transient exponential decays in time. The residue at $\omega =0$, the one in which we are interested,
will be given by the limit $(-2\pi \ii)\lim_{\omega \to 0} [\omega G(\omega)]$ (the residue must be obtained by closing the integration
over $\omega$ with a circle in the lower half plane). Then, we can write:
\begin{eqnarray}
\label{invdeltadeltac}
&&\delta F(\theta,t) \delta f(\theta,v_1,t) (t \to \infty) \\  &=& -(2\pi \ii) \lim_{\omega \to 0} \left\{ \frac{1}{(2\pi)^2} \sum_k \sum_{k'}
\int_{{\cal C}_1} \dd \omega_1 \, e^{\ii (k+k')\theta} \omega \widetilde{\delta F}(k,\omega_1)
\widetilde{\delta f}(k',v_1,\omega -\omega_1)\right\} \, . \nonumber
\end{eqnarray}
The limit in $\omega$ must be approached from the upper half plane. Since we also require that ${\rm Im}(\omega)>{\rm Im}(\omega_1)>0$,
performing this limit also ${\rm Im}(\omega_1)$ must go to zero from above, remaining always smaller than ${\rm Im}(\omega)$.

Before performing the $\omega$ limit, we perform the averaging $\langle \cdot \rangle$ required in Eq. (\ref{klim_aver_st}). From
Eqs. (\ref{reldffluc}) and (\ref{reldforcefluc}), we have:
\begin{eqnarray}
\label{averdforcedf}
&& \langle \widetilde{\delta F}(k,\omega_1) \widetilde{\delta f}(k',p,\omega - \omega_1) \rangle \\
&=& \frac{2\pi N c_k}{\epsilon(k,\omega_1) \ii (k'p -\omega +\omega_1)} \int \dd p' \,
\frac{\langle \widehat{\delta f}(k,p',0) \widehat{\delta f}(k',p,0) \rangle}{\ii (kp'-\omega_1)} \nonumber \\
&-& \! \frac{(2\pi N)^2 c_k c_{k'}f'(p)}{\epsilon(k,\omega_1)\epsilon(k',\omega-\omega_1)\ii (k'p-\omega +\omega_1)}
\int \dd p' \!\!\! \int \dd p'' \frac{\langle \widehat{\delta f}(k,p',0)\widehat{\delta f}(k',p'',0)\rangle}{\ii (kp'-\omega_1)
\ii (k'p''-\omega+\omega_1)} \, . \nonumber
\end{eqnarray}
Then, we need the autocorrelation of the fluctuation $\delta f$ at time $t=0$. The term that gives rise to a contribution that does not decay
exponentially in time is expressed by \cite{landaukin}
\begin{equation}
\label{flucaver}
\langle \widehat{\delta f}(k,p,0) \widehat{\delta f}(k',p',0) \rangle = \frac{1}{2\pi N}\delta_{k,-k'}f(p)\delta (p-p') \, .
\end{equation}
Since we have $c_0=0$, from Eqs. (\ref{averdforcedf}) and (\ref{flucaver}), we see that performing the averaging $\langle \cdot \rangle$, we
can assume that in Eq. (\ref{invdeltadeltac}), the terms with $k=0$ and $k'=0$ are absent.
Plugging Eq. (\ref{flucaver}) in Eq. (\ref{averdforcedf}), we obtain:
\begin{eqnarray}
\label{averdforcedfc}
&& \langle \widetilde{\delta F}(k,\omega_1) \widetilde{\delta f}(k',p,\omega - \omega_1) \rangle 
= \frac{ c_k\delta_{k,-k'}f(p)}{\epsilon(k,\omega_1) (kp +\omega -\omega_1) (kp-\omega_1)} \\
&+& \frac{2\pi N |c_k|^2\delta_{k,-k'}f'(p)}{\ii \epsilon(k,\omega_1)\epsilon(-k,\omega-\omega_1) (kp+\omega -\omega_1)}
\int \dd p' \,  \frac{f(p')}{(kp'-\omega_1) (kp'+\omega-\omega_1)} \, . \nonumber
\end{eqnarray}
The limit of this expression for $\omega \to 0$ and ${\rm Im}(\omega_1) \to 0$ is evaluated by using the Plemelj formula
(\ref{pleme}),  which for convenience we rewrite here:
\begin{equation}
\label{pleme_app}
\lim_{\eta \to 0^+} \frac{1}{x\pm \ii \eta} = P \frac{1}{x} \mp \ii \pi \delta(x) \, ,
\end{equation}
We also use the decomposition, inside the integral in Eq. (\ref{averdforcedfc}),
\begin{equation}
\label{separdelta}
\frac{1}{(kp'-\omega_1)(kp'+\omega-\omega_1)} = \frac{1}{\omega}\left[\frac{1}{kp'-\omega_1} -\frac{1}{kp'+\omega-\omega_1}\right] \, .
\end{equation}
We therefore obtain, using also the property $\delta(ax)=\delta(x)/|a|$, that
\begin{eqnarray}
\label{averdforcedfd}
&\lim_{\omega \to 0}& \left[ \omega \langle \widetilde{\delta F}(k,\omega_1) \widetilde{\delta f}(k',p,\omega - \omega_1) \rangle \right] 
= \frac{c_k\delta_{k,-k'}f(p)}{\epsilon(k,\omega_{1R})}2\pi \ii \delta (kp -\omega_{1R}) \\
&+& \!\! \frac{2\pi N |c_k|^2\delta_{k,-k'}f'(p)}{\ii \epsilon(k,\omega_{1R})\epsilon(-k,-\omega_{1R})}
\left[ P\frac{1}{kp-\omega_{1R}} -\ii \pi \delta(kp-\omega_{1R})\right] \frac{1}{|k|}2\pi \ii f\left(\frac{\omega_{1R}}{k}\right) \, ,
\nonumber
\end{eqnarray}
where the subscript $R$ denotes that now $\omega_1$ has become real, and the path of integration ${\cal C}_1$ has become the real axis.
We can finally substitute in Eq. (\ref{invdeltadeltac}), finding the expression
\begin{eqnarray}
\label{invdeltadeltad}
&&\langle \delta F(\theta,t) \delta f(\theta,p,t) (t \to \infty) \rangle \\
&=& \sum_k \left\{ \frac{c_k f(p)}{\epsilon(k,kp)} - 2\pi^2 N \frac{|c_k|^2f'(p)}{|k||\epsilon(k,kp)|^2}f(p)
\right. \nonumber \\ && \left. - 2\pi \ii N P\int \dd \omega_{1R} \, \frac{|c_k|^2 f'(p)}{|k||\epsilon(k,\omega_{1R})|^2}
\frac{f\left(\frac{\omega_{1R}}{k}\right)}{kp-\omega_{1R}} \right\} \, . \nonumber
\end{eqnarray}
To arrive at the final expression, we have to use the limit of the dielectric function $\epsilon(k,\omega)$ for real $\omega$. From
Eqs. (\ref{dielPR}) and (\ref{pleme}), we obtain
\begin{equation}
\label{dielrealPR}
\epsilon(k,\omega_R) = 1 + 2\pi \ii N c_k P \int \dd p \, \frac{f'(p)}{\omega_R - k p}
+2\pi^2 N \frac{c_k}{|k|} f'(\frac{\omega_R}{k}) \, ,
\end{equation}
From this expression, one finds that $\epsilon(-k,-\omega_R) = \epsilon^*(k,\omega_R)$. Furthermore, since $c_{-k} = c^*_k$, we deduce
that the last term on the right hand side of Eq. (\ref{invdeltadeltad}), the one with the integral, is odd in $k$, and therefore,
when summed over $k$, it gives a vanishing contribution (we remind that the term with $k=0$ is absent). In the first term, we
use Eq. (\ref{dielrealPR}) to write:
\begin{equation}
\label{oneover}
\frac{1}{\epsilon(k,kp)} = \frac{1}{|\epsilon(k,kp)|^2}\left[ 1 -2\pi \ii N c^*_k P \int \dd p' \, \frac{f'(p')}
{kp-kp'} +2\pi^2 N \frac{c^*_k}{|k|}f'(p)\right] \, .
\end{equation}
Plugged into (\ref{invdeltadeltad}), the second term in the square brackets gives rise to a term odd in $k$,  which therefore
vanishes on summing over $k$; the third term in square brackets, on the other hand, cancels with the second term in (\ref{invdeltadeltad}).
Thus, at the end, we get the final expression for Eq. (\ref{klim_aver_st}), i.e.;
\begin{equation}
\label{finalkl}
\frac{\partial f}{\partial t} = - \frac{\partial}{\partial p} \sum_k \frac{c_k f(p)}
{|\epsilon(k,kp)|^2} \, .
\end{equation}

Some remarks are in order. In the case of a Hamiltonian system, the force ${\cal F}(\theta)$ in Eq. (\ref{expanforce}) derives from
a potential, a real and even function of $\theta$. The Fourier coefficients $u_k$ of the potential are, consequently, real and
even in $k$ (again with $u_0=0$). Then, we would have $c_k = -\ii k u_k$, meaning that $c_k$ is purely imaginary and odd in $k$.
In this case, the right hand side of Eq. (\ref{finalkl}) vanishes, in agreement with the known result that the Lenard-Balescu evolution
operator vanishes for a 1D Hamiltonian system. On the other hand, for a fully non-Hamiltonian system, in which the Fourier
expansion of the force ${\cal F}(\theta)$ contains only the cosine terms, the coefficients $c_k$ are real and even in $k$; in this
case, the right hand side of Eq. (\ref{finalkl}) is in general nonzero. For the mixed case, with the presence of both Hamiltonian
and non-Hamiltonian terms in the force, it is easy to see that the real part of $c_k$, even in $k$, comes from the non-Hamiltonian
part, while the imaginary part of $c_k$, odd in $k$, comes from the Hamiltonian part. Therefore, the only non-vanishing contribution
in the right hand side of (\ref{finalkl}) is due to the non-Hamiltonian part of the force.

Another remark concerns the time dependence of $f(p)$, an issue that we have already mentioned above. Equation (\ref{finalkl}) clearly
determines a variation in time of $f(p)$. On the other hand, we had assumed, in the solution of the linearized Vlasov equation
to obtain $\delta f(\theta,p,t)$, that this dependence is frozen. As already emphasized above, this is a consequence of the
approximation, at the core of the derivation of the Lenard-Balescu kinetic equation, in which it is assumed that the time scale of
variation of the fluctuations is much smaller than that of $f(p)$ (the Bogoliubov hypothesis \cite{Nicholson:1992}). Thus, in
the derivation of the dynamics of the fluctuations, one can assume that $f(p)$ is constant in time, but then the result can be
used to obtaine the much slower time variation of $f(p)$; physically, this is a consistent procedure. We underline that in
Eq. (\ref{finalkl}), the dielectric function $\epsilon(k,kp)$ is computed from (\ref{dielrealPR}) by using the instantaneous value
of $f(p)$ as determined by Eq. (\ref{finalkl}) itself \cite{Nicholson:1992}.

We end this appendix by considering the caveat mentioned above, concerning the presence or absence, in the second equation
of motion in (\ref{gen_equ_non}), of the term with $j=i$. Given the definition (\ref{def_klim}) of the one-particle density
function $f_{{\rm d}}(\theta,p,t)$, one obtains the Klimontovich equation (\ref{klim_eq_non}) by using the equations of motion
(\ref{gen_equ_non}), with $F_{{\rm d}}(\theta,t)$ defined by Eq. (\ref{force_klim}). This is correct when in the second equation
of motion in (\ref{gen_equ_non}), the term with $j=i$ is present. The latter term represents a self-interaction of the $i$-th particle
with itself. We have seen in section \ref{sec:avermom} that, for the study done in this work, the choice between the exclusion
or the inclusion of the self-interaction term is a matter of convenience. Here we just want to show how the Lenard-Balescu equation would be
modified if we exclude it, i.e., when we exclude the term with $j=i$ in
(\ref{gen_equ_non}). In this case, we see that the factor multiplying $\frac{\partial f_{{\rm d}}}{\partial p}$ in the
Klimontovich equation should be given by $F_{{\rm d}}(\theta,t) - {\cal F}(0)$, with $F_{{\rm d}}(\theta,t)$ still defined by
(\ref{force_klim}). Then, the Klimontovich equation becomes
\begin{equation}
\label{klim2}
\frac{\partial f_d}{\partial t} + p\frac{\partial f_d}{\partial \theta} + F(\theta,t)\frac{\partial f_d}{\partial p} - {\cal F}(0)
\frac{\partial f_d}{\partial p} = 0 \, .
\end{equation}
All this is irrelevant in the Hamiltonian case, since then we have ${\cal F}(0)=0$. When we perform the averaging procedure as above,
the last term on the right hand side remains with just the substitution of $f_{{\rm d}}$ with $f(p)$. So, at the end, the Lenard-Balescu
equation (\ref{finalkl}) will become
\begin{equation}
\label{finalkl2}
\frac{\partial f}{\partial t} = \frac{\partial}{\partial p} \left\{ {\cal F}(0)f(p) - \sum_k \frac{c_k f(p)}
{|\epsilon(k,kp)|^2} \right\} \, .
\end{equation}




\end{document}